\documentclass[aps,pra,reprint,superscriptaddress,showkeys,nofootinbib]{revtex4-2}
\usepackage[dvipsnames]{xcolor}
\usepackage[colorlinks=true,linkcolor=OrangeRed,citecolor=Cerulean,urlcolor=ProcessBlue]{hyperref}
\usepackage{amsmath}
\usepackage{amsfonts}
\usepackage{amssymb}
\usepackage{bbm}
\usepackage{dsfont}
\usepackage{mathrsfs}
\usepackage{array}
\usepackage{dcolumn}
\usepackage{url}
\usepackage{textcomp}
\usepackage{graphicx}
\usepackage{color}
\usepackage{braket}
\usepackage[mathscr]{euscript}
\usepackage{bbm}
\usepackage{enumerate}
\usepackage{appendix}
\usepackage{lipsum}
\usepackage{float}

\def\lab{\label}
\def\alp{\alpha}

\def\sg{\sigma}

\def\Om{\Omega}

\def\lo{\lambda}

\def\gm{\gamma}

\def\ep{\epsilon}

\def\8{\phi}
\def\7{\Phi}
\def\9{\psi}

\def\fr{\frac}

\def\td{\tilde}

\def\b{\bra}

\def\K{\Ket}
\def\bk{\braket}
\def\BK{\Braket}
\def\h{\hat}
\def\dg{\dagger}

\def\ra{\rightarrow}

\def\ua{\uparrow}
\def\da{\downarrow}

\def\V{\vec}

\def\lsim{\lesssim}

\def\tens{\otimes}

\def\Cm{\mathcal{C}}
\def\Dm{\mathcal{D}}
\def\Em{\mathcal{E}}

\def\Hm{\mathcal{H}}
\def\Ie{\mathscr{I}}

\def\Oe{\mathcal{O}}
\def\Pm{\mathcal{P}}
\def\Rm{\mathcal{R}}

\def\iD{\mathbbm{1}}

\def\3{\big}
\def\4{\Big}
\def\5{\Bigg}
\definecolor{ogreen}{rgb}{0.2, 0.5, 0.1}
\definecolor{Orange}{rgb}{1, 0.6, 0.4}

\def\ti{\textit}

\def\tt{\text}
\newcommand{\tus}[1]{\underline{\smash{#1}}}
\def\sz{\small}
\def\fsz{\footnotesize}
\newcommand{\tsz}[1]{\text{\small #1}}

\newcommand{\fsi}[1]{\text{\footnotesize #1}}
\def\nsz{\normalsize}
\def\aref{\autoref}
\newcommand{\appref}[1]{\hyperref[#1]{Appendix~\ref{#1}}}
\usepackage[english]{babel}
\addto\extrasenglish{%
}

\addto\captionsenglish{}

\def\vh{\vspace{0.5mm}}
\def\vo{\vspace{1mm}}
\def\vt{\vspace{2mm}}
\def\vth{\vspace{3mm}}
\def\vfo{\vspace{4mm}}

\def\vTo{\vspace{-1mm}}
\def\vTt{\vspace{-2mm}}
\def\vTh{\vspace{-3mm}}
\def\vFo{\vspace{-4mm}}
\def\hs{\kern 0.16667em}
\def\Hs{\kern 0.5em}
\def\hsh{\kern 0.1em}
\def\hshh{\kern 0.07em}
\def\hsn{\kern -0.3em}
\def\hsN{\kern -0.6em}
\def\hseN{\kern -0.55em}
\def\hsnh{\kern -0.15em}
\def\hsnhh{\kern -0.1em}

\newcommand*{\Scale}[2][4]{\scalebox{#1}{$#2$}}

\newcommand{\SDrop}[2]{\fontdimen16\textfont2=#1
\fontdimen17\textfont2=#1 \tt{$#2$} \fontdimen16\textfont2=2pt
\fontdimen17\textfont2=2pt}
\newcommand{\SRaise}[2]{\fontdimen13\textfont2=#1
\fontdimen14\textfont2=#1 \fontdimen15\textfont2=#1 \tt{$#2$} \fontdimen13\textfont2=2pt
\fontdimen14\textfont2=2pt \fontdimen15\textfont2=2pt}

\usepackage{fancyhdr}
\def\nl{\noindent}

\numberwithin{equation}{section}

\def\tr{\text{Tr}}
	
\def\ths{\text{\tiny th}}

\def\its{\Ie_n}

\def\q16{IBMQ$_{16}$\hs}

\newcommand\qpu{\textsc{QPU}}

\def\uchiX{\h U_{\hsnhh \Scale[0.6]{\rchi}}}

\def\nisqT{\small NISQ\normalsize}

\def\qpUT{\sz QPU\normalsize}
\def\qpUTO{\sz QPU \normalsize}

\def\zneT{\sz $\text{ZNE}$\normalsize}
\def\zneTO{\sz $\text{ZNE}$ \normalsize}

\def\cnotTO{\sz CNOT \normalsize}

\def\vigoT{\sz IBMQ-Vigo\normalsize}
\def\vigoTO{\sz IBMQ-Vigo \normalsize}
\def\saintT{\sz IBMQ-Santiago\normalsize}

\def\NTt{\sz NT\normalsize}
\def\NTto{\sz NT \normalsize}

\def\claweA{\fsz CLAWE\sz}
\def\claweT{\sz CLAWE\normalsize}
\def\claweTO{\sz CLAWE \normalsize}

\def\vITO{\sz Variant I \normalsize}

\def\vIITO{\sz Variant II \normalsize}

\def\clawefTO{\sz{CLA}\nsz ssical \sz{W}\nsz hite-noise \sz{E}xtrapolation \nsz}
\def\fermiHT{\sz{F}\nsz ermi-\sz{H}\nsz ubbard \sz{M}odel\nsz}
\def\fermiHTO{\sz{F}\nsz ermi-\sz{H}\nsz ubbard \sz{M}odel \nsz}

\def\qpUFTO{\sz{Q}\nsz uantum \sz{P}\nsz rocessing \sz{U}nit \nsz}

\def\reynT{R\'{e}nyi }

\def\productFATO{\sz{P}\nsz roduct \sz{F}\nsz ormula \sz{A}lgorithm \nsz}

\def\parecT{\sz PAREC\nsz}

\def\rcoT{\sz RCo\nsz}
\def\rcoTO{\sz RCo \nsz}

\def\qs{\text{\tiny q}}

\def\qpus{\text{\tiny \qpu}}

\def\itsT{\text{\sz{ITS}}\normalsize}

\def\pnrT{\text{\sz{PNR}}\normalsize}
\def\pnrTO{\text{\sz{PNR}} \normalsize}

\def\qcnaT{\text{\sz{QCNA}}\normalsize}
\def\qcnaTO{\text{\sz{QCNA}} \normalsize}

\def\pfaT{\sz PFA\normalsize}
\def\pfaTO{\sz PFA \normalsize}

\def\ptasT{\sz PTAs\normalsize}
\def\ptasTO{\sz PTAs \normalsize}
\def\bqpT{\sz BQP\normalsize}
\def\bqpTO{\sz BQP \normalsize}
\def\bppT{\sz BPP\normalsize}
\def\bppTO{\sz BPP \normalsize}
\def\bbaT{\sz BBA\normalsize}
\def\bbaTO{\sz BBA \normalsize}

\def\cptpT{\sz CPTP\nsz}

\def\qiskT{\sz Q\nsz is\sz K\nsz it\nsz}

\def\hq{\Hm_{\text{\tiny q}}}
\def\henv{\Hm_{\text{\tiny env}}}
\def\happ{\Hm_{\text{\tiny app}}}
\def\hqpu{\Hm_{\text{\tiny \qpu}}}

\def\rq{\rho_{\qs}}
\def\rqpu{\rho_{\qpus}}
\DeclareRobustCommand{\rchi}{{\mathpalette\irchi\relax}}
\newcommand{\irchi}[2]{\raisebox{\depth}{$#1\chi$}}

\def\cxnavp{\h{\mathds{C}}^{ij}_{\rchi'}}
\def\cxid{\h C^{ij}_x}

\def\cxidO{C^{ij}_x}

\def\Enavp{\h \Em_{\rchi'}}

\def\VNavp{\h V_{\rchi'}}
\def\texp{t_{\text{\tiny exp}}}
\def\expe{\text{\tiny exp}}

\def\mape{\text{\tiny map}}
\def\targ{\text{\tiny targ}}
\def\acc{\text{\tiny acc}}
\def\idea{\text{\tiny{ideal}}}

\usepackage[all]{hypcap}

\usepackage[shortlabels]{enumitem}

\begin{document}

\title{Classical-Quantum Noise Mitigation for \text{\large NISQ} Hardware}
\date{\today}

\author{Andrew Shaw}
\email[Electronic Address: ]{ashaw12@umd.edu}
\affiliation{University of Maryland, College Park, MD 20742, USA}

\begin{abstract}

In this work, the \ti{global white-noise model} is proved from first principles. The adherence of NISQ hardware to the global white-noise model is used to perform noise mitigation using \ti{CLAssical White-noise Extrapolation} (\claweA). 

\end{abstract}

\maketitle

\section{Global White-Noise Model}\label{sec:gwn}

\vTh
Universal quantum computation involves the encoding of algorithms into sequences of \ti{local} gates. The noise processes in \ti{Noisy Intermediate-Scale Quantum} (\nisqT) hardware do not exhibit such locality, resulting in \ti{qubit crosstalk} \cite{crosst0,crosst1,crosst2,crosst3,crosst4,crosst5,crosst6,crosst7,crosst8,crosst9,crosst10,crosst11,crosst12,crosst13,crosst14,crosst15,crosst16,crosst17,crosst18,crosst19,crosst20,crosst21,crosst22,crosst23,crosst24}. 

\vo

\begin{figure}
\includegraphics[scale=0.048]{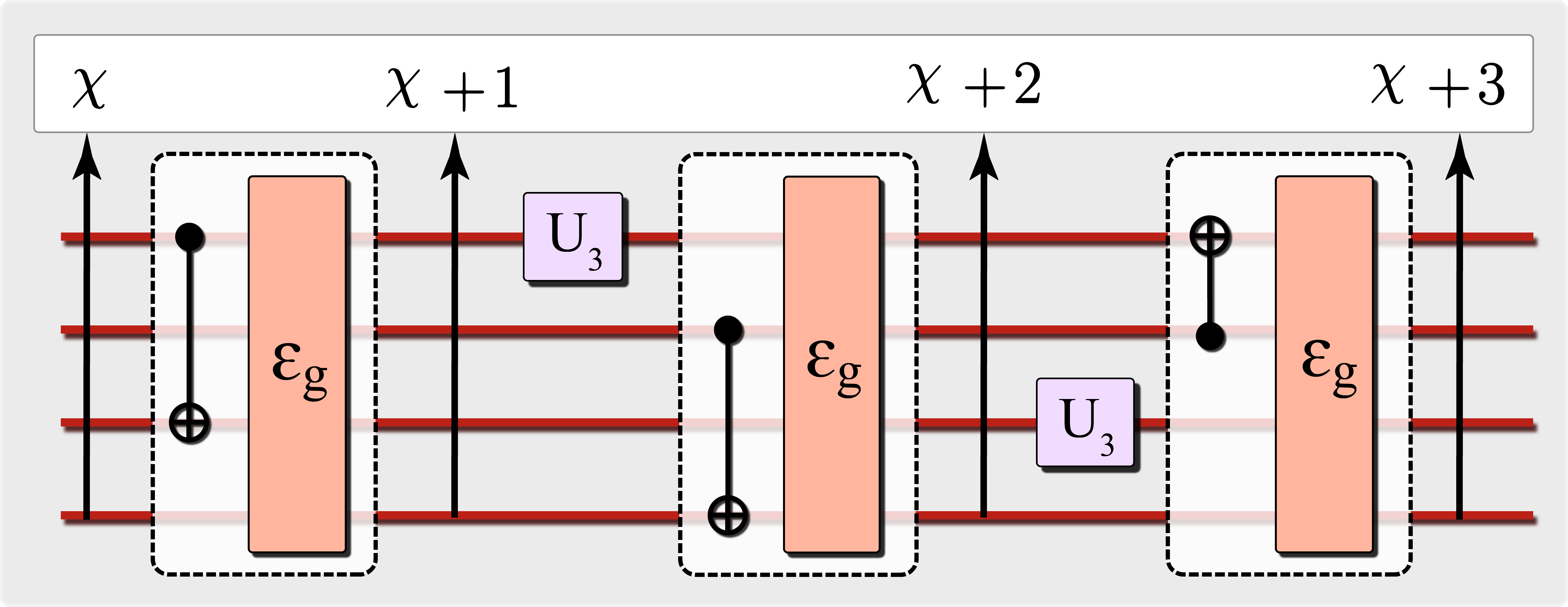}
\caption{Depolarizing channels follow CNOT gates.}\vspace{-5mm}
\label{fig:gwnm}
\end{figure}

This non-locality suggests a global description of the average noise dynamics. The \ti{global white-noise model} \cite{CLAWEPost0,CLAWEConference0,CLAWEPost1,CLAWEPresent1} presumes that the noise dynamics associated with \ti{entangling operations} \cite{
transmEnt0,transmEnt1,transmEnt2,transmEnt3,transmEnt4,transmEnt5,transmEnt6,transmEnt7,transmEnt8,transmEnt9,transmEnt10,transmEnt11,transmEnt12,transmEnt13,transmEnt14,transmEnt15,transmEnt16,transmEnt17,transmEnt18,transmEnt19,transmEnt20,transmEnt21,transmEnt22,transmEnt23,transmEnt24,transmEnt25,transmEnt26,transmEnt27,transmEnt28,transmEnt29,transmEnt30,transmEnt31,transmEnt32,transmEnt33,transmEnt34,transmEnt35,transmEnt36} can be approximated by depolarizing events that span the entire Hilbert space (\aref{fig:gwnm}). The global white-noise model is now derived from first principles using \ti{quantum channel technology}.

\vspace{-2mm}
\subsection{Introduction to Superoperators}\label{ssc:supop}
\vTt

Rigorously treating noise dynamics requires a density matrix representation \cite{densitymatI,densitymatII,densitymatIII}, because the evolution of open quantum systems generates \ti{mixed quantum states}. Vector representations cannot describe such processes, as they only characterize \ti{pure quantum states}: \tsz{$\K{\9}$}.
\vo

Mixed quantum states have interacted with unknown degrees of freedom. They are composed of an ensemble of pure states \tsz{$\{\K{\9_k}\}$}, with observational probabilities \tsz{$\{\alp_k\}$}. The corresponding density matrix is the following:
\vTo
\begin{equation}
\rho_{\tt{\tiny mixed}}=\sum_k \alp_k \K{\9_k}\hsnh \b{\9_k}
\end{equation}

\vTt
In the density matrix representation, \tsz{$d\hsnh\times\hsnh d$}-dimensional operators that act on wavefunctions, generalize to \tsz{$d^2\hsnh\times \hsnh d^2$}-dimensional \ti{quantum channels} known as \ti{superoperators}. Each superoperator can be represented by sets of \ti{Kraus operators} \tsz{\{$M_k$\}} \cite{superOp}:
\vTt
\begin{equation}
\h \Em =\sum^{\Rm}_{k=1} M_k \tens M_k^{\dg}	
\end{equation}

\vTo

Meaningful density matrices are positive semi-definite, with \tsz{$\tr(\rho)=1$}. To map such density matrices onto one another, superoperators must be \ti{completely positive and trace-preserving} (\cptpT): \tsz{$\sum_k M_k M_k^{\dg}=1$} \cite{densitymatIV}.

\vt

%\nl
In this work, hat-notation is reserved for the action of superoperators in matrix representation:

\vTh

\begin{align}
\begin{split}
&\tt{Matrix} \\[-2mm] &\tt{Representation:}
\end{split} \hspace{5em} \h \Em \rho  \\[2mm]
\begin{split}
&\tt{Operator-Sum \vTh} \\[-2mm] &\tt{Representation:}
\end{split} \ \   \hspace{2em} \Em(\rho)=\sum_k \hs M_k\hsh \rho \hsh M_k^{\dg}
\end{align}

\subsection{The Depolarizing Channel}\label{ssc:depolChan}
\vTt

Depolarization results in the mixing of a quantum state with the infinite temperature Gibbs state, one for which all micro-states are equally likely \cite{gibbsstate}. Such processes are described by the \tsz{$n$}-qubit \ti{depolarizing channel}:

\vTh

\begin{equation}
\h \Dm_{\ep,n} \hs \rho =(1-\ep)\hsh \rho+\ep \hs \its
\end{equation}
where \tsz{$\its=\iD/2^n$} is the \ti{infinite temperature state} (\itsT) and \tsz{$\ep$} is the \ti{noise strength}.

\vo

 The repeated action of the depolarizing channel is the following:

\vTh

\begin{align}
\h \Dm^k_{\ep,n} \hsh \rho &=(1-\ep)^k\hsh \rho +f(\ep,k) \hsh \its \\[0.3em]
f&(\ep,k)=\ep\sum_{n=0}^{k-1}(1-\ep)^n \label{recursiveM}
\end{align}

\vo

\begin{figure}
\includegraphics[scale=0.1005]{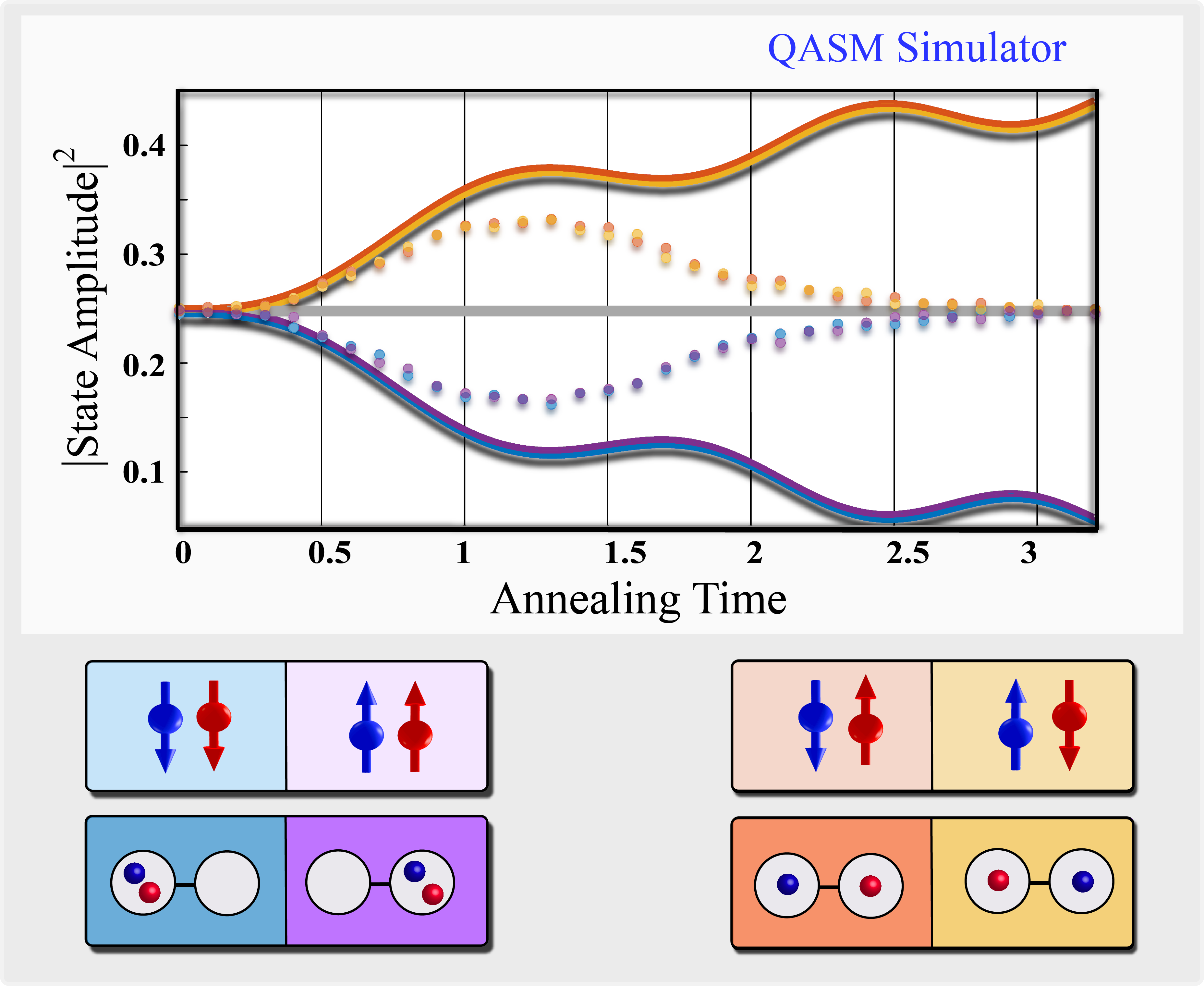}
\caption{The time evolution of the Fermi-Hubbard Model is computed numerically (solid). Depolarizing noise dynamics (circles) cause saturation with the \text{\fsz ITS\sz} (grey). \vspace{-3mm}}
\label{fig:nisqual}
\end{figure}

The input state is suppressed exponentially, indicating a \ti{signal-to-noise problem}. After its qubits are saturated by the \itsT, a \ti{\qpUFTO}(\qpUT) can no longer perform meaningful computation (\aref{fig:nisqual}).

\subsection{Proving the Global White-Noise Model}

\vo

A \qpUTO is composed of qubits coupled to the \ti{environment} and a \ti{measurement apparatus}:

\vTh

\begin{equation}
\hqpu=\henv\tens\hq\tens \happ
 \end{equation} 

\vo

\nl
In the \qpUT's expanded Hilbert space, noise dynamics map onto unitary time evolution \cite{qpuMOD1}.

\vt

Consider a \qpUTO that performs digital computations with \cnotTO gates. It is examined throughout a quantum computation obeying the \ti{computational cycle}:

\vt

(I.) \tus{\ti{state preparation}}: The qubits are prepared in a pure initial state.

\vt

(II.)  \tus{\ti{quantum computation}}: The \ti{target computation} is performed by applying \cnotTO gates to the qubits.

\nl

\vt

(III.) \tus{\ti{measurement}}: An observable on \tsz{$\hq$} is measured.

\vth

 \vo
 
 The computational cycle is repeated \tsz{$N_m$} times. Within the \qpUT, the entire computation is described  by unitary evolution for time \tsz{$\texp={\tiny{N_m}\hsh t_{\text{\tiny cycle}}}$}:
 
 \vTh
 
 \begin{equation}
 \rqpu^{(t)}=\h U_{\qpus}^{(t)} \hsh \rqpu^{(0)}
 \end{equation}

\vo

 A location metric is required to identify the occurrence of \cnotTO gates in a computation. In this work, the \ti{scalar depth} (\tsz{$\rchi$}) is used: the number of entangling operations applied after state preparation.

 \vo
 
 The state of the \qpUTO \hsnh during the computation can be parametrized by the scalar depth \tsz{$\rchi(t)$}:
 
 \vTh
 
 \begin{equation}
 \rqpu^{(t)}\ra \rqpu[\rchi(t)]
 \end{equation}

The qubits will be acted on by the \tsz{$\rchi'^{\hsh \ths}$} \cnotTO gate at times \tsz{$\{t_{\sg}\}$} satisfying \tsz{$\rchi(t_{\sg})=\rchi'$}. The state of the \qpUTO at times \tsz{$\{t_{\sg}\}$} defines a set of \ti{encountering states}:

\vTh

\begin{equation}
 \rqpu^{(t_k)}[\rchi']=\rqpu[\rchi'(t_k)]
 \end{equation}

Encountering states at times \tsz{$t_l$} and \tsz{$t_k$} are related by unitary evolution on \tsz{$\hqpu$}, represented by \ti{encountering transformations}:

\vTh

\begin{align}
\h U_{lk}&=\h U_{\qpus}^{(t_l)} \hs \h U_{\qpus}^{\dg(t_k)} \\ \label{encountT}
\rqpu^{(t_l)}[\rchi']&=\h U_{lk}\hs \rqpu^{(t_k)}[\rchi']
\end{align}

 The action of the \tsz{$\rchi'^{\ths}$} \cnotTO gate on the encountering states is treated with quantum channel technology. The native \cnotTO gate coupling qubits \tsz{$\sz{\{i,j}\}$} is the following:

\vTh

\begin{equation} \label{navCNOT}
 \cxnavp  =\Enavp\hsh \cxid
 \end{equation}

\tsz{$\Enavp$} is a superoperator describing the noise dynamics of the native gate. \tsz{$\cxidO$} is the ideal \cnotTO gate.

\vo

To treat the noise dynamics in a unitary fashion, a \ti{Stinespring dilation} \cite{stinespring} is performed:

\vTh

\begin{equation}
\hq\ra\hqpu, \ \  \Enavp\ra \VNavp \hs \h \xi
\end{equation}

\tsz{$\h \xi$} is a \ti{unital homomorphism} that takes \tsz{$\hq\hsnh \ra \hqpu$}. \tsz{$\VNavp$} is a unitary transformation acting on \tsz{$\hqpu$}.

\vo

The action of \hsh \tsz{$\cxnavp$} on the \tsz{$\SRaise{3pt}{k^{\tt{th}}}$} encountering state is as follows:

\begin{equation}
\rqpu^{(t_k)}[\rchi'\sz{+}1]=\VNavp \hsh \hsh\cxid \hs \rqpu^{(t_k)}[\rchi']
\end{equation}

\vo

The expectation value of \tsz{$O_{\qs}$} is obtained by averaging the contributions of the encountering states:

\vTh

\begin{equation}\label{ohaarsum}
\begin{split}
\bk{O_{\qs}}&=\fr{1}{N_m}\sum_{k=1}^{N_m}\tr_{\qpus}\4[\4\{\VNavp \hsh \hsh\cxid \hs \rqpu^{(t_k)}[\rchi']\4\} \hs O_{\qs}\4] \\
&=\fr{1}{N_m}\sum_{k=1}^{N_m}\tr_{\qpus}\4[\4\{\VNavp \hsh \hsh\cxid \hs \h U_{k1}\hsh \rqpu^{(t_1)}[\rchi'] \4\}\hs O_{\qs}\4]\raisetag{3\baselineskip}
\end{split}
\end{equation}

\nl
In the second line, \tsz{$\rqpu^{(t_k)}[\rchi']$} is expressed in terms of an encountering transformation on \tsz{$\rqpu^{(t_1)}[\rchi']$}.

\vo

In the ideal limit, the target computation is applied identically for all cycles. This implies the following:

\vTh 

\begin{equation}
\3[O_{\qs},U^{\idea}_{lk}\3]= 0	
\end{equation}

\vo

\tsz{$U_{_{k1}}U^{\text{\tiny$\dg$}}_{k1}$} is now inserted into \aref{ohaarsum}:

\vTh

\begin{equation}\label{ambientP} 
\begin{split}
\bk{O_{\qs}}=\fr{1}{N_m}\sum_{k=1}^{N_m}\tr_{\qpus}\4[\hsh U^{\dg}_{k1}\4\{ \VNavp \hsh \hsh\cxid \hs \h U_{k1}\hsh \rqpu^{(t_1)}[\rchi']\4\} \hsh O_{\qs} \hsh \SDrop{3pt}{U_{k1}}\4]\\
\approx\fr{1}{N_m}\sum_{k=1}^{N_m}\tr_{\qpus}\4[\hsh U^{\dg}_{k1}\4\{ \hsh \VNavp \hsh \SDrop{3pt}{\h U_{k1}} \hs \cxid\hsh \rqpu^{(t_1)}[\rchi']\4\} \hs \SDrop{3pt}{U_{k1}} \hs O_{\qs} \4]
\end{split}
\end{equation}

\vt

This allows \aref{ambientP} to be expressed as follows:

\vTh

\begin{equation}\label{Haarave}
\begin{split}
\bk{O_{\qs}}&\approx\fr{1}{N_m}\sum_{k=1}^{N_m}\tr_{\qpus}\4[\hsh \4\{ \h U^{\dg}_{k1} \hsh \VNavp \hsh\hs \SDrop{3pt}{\h U_{k1}} \ \hsh \cxid  \hs \rqpu^{(t_1)}[\rchi']\4\} \hs O_{\qs} \4]
\end{split}
\end{equation}

\vo

Note the following identity \cite{gwnm1}:

\vTh

\begin{equation}\label{eqFullTwirl}
\begin{split}
\Em^{\tt{\tiny ave}} \rho &=\int dU \hs \h U^{\dg} \hsh \h \Em \hsh \h U \hs \rho \\
&=(1-\ep)\hsh \rho+\ep\hsh \its
\end{split}
\end{equation}

\vo

Applying this relation yields the following:

\vTh

\begin{align}
\bk{O_{\qs}}&\sim \int dU \hs \tr_{\qpus}\4[\4\{\hs \h U^{\dg} \hs \VNavp \hs \h U \ \cxid \rqpu^{(t_1)}[\rchi']\4\} \hs O_{\qs}\4] \notag \\[0.4em]
&=\hs \tr_{\qs} \4[  \4\{(1-\ep_g) \hs \cxid \rq[\rchi']+\ep_g \hs \its \4\} \hs O_{\qs}\4]
\end{align}

\vo

The averaged noise dynamics of the \cnotTO gate approximate a global depolarizing channel with \ti{global noise strength} \tsz{$\ep_g$}.

\vo

The noise dynamics of the \cnotTO gate can be modified using \ti{noise tailoring} (\NTt) algorithms (\appref{appendix:Atail}).

\vspace{-3mm}

\section{CLAssical White-noise Extrapolation}\label{sec:CLAWEdius}
\vspace{-1mm}

 The objective of \clawefTO (\claweT) \cite{CLAWEPresent1}, is the extraction of ideal observables from their noisy counterparts. This extraction requires the noise dynamics to obey the global white-noise model.

\vTh 

\subsection{Model-Based Extrapolation}\label{subs:mbe}
\vTt

\claweTO is a \ti{model-based extrapolation} algorithm. Model-based extrapolation \cite{mbextrap0,mbextrap1,mbextrap2,mbextrap3} can be represented schematically in three stages:

\begin{enumerate}[I.]
\item \tus{Modeling}: Find a suitable model that describes a class of systems \tsz{$\{\Pm_{\sg}\}$}, using parameters \tsz{$\{\lo_{\alp}\}$}. The model must relate accessible observables to target observables: $\{\bk{O^{\acc}_{\mu}}\}\ra \{\bk{O_{\nu}^{\targ}}\}$.
\item \tus{Calibration}: Determine the model parameters \tsz{$\{\lo_{\alp}^{\expe}\}$} for a system \tsz{$\Pm_{\tt{phys}}$}, by measuring calibration observables \tsz{$\{\bk{ O_{\gm}^{\text{\tiny cal}}}\}$}.
\item \tus{Extrapolation}: Use the model to estimate target observables of \tsz{$\Pm_{\tt{phys}}$}. 
\end{enumerate}

\vTh

\subsubsection{Modeling}\label{ssc:param}
\vspace{-1mm}

\vo

\vspace{3mm}
 
\tus{\ti{Model Inversion}}: It is possible to exactly invert the global white-noise model due to the \ti{unitary invariance} of the depolarizing channel:

\vTh

\begin{equation}
\h \Dm_{\ep,n}\hsh \h U \hs \rho \hs =\hs \h U \hs\h \Dm_{\ep,n}\hs \rho
\end{equation}

 Consider a target computation of scalar depth \tsz{$\rchi$}: \tsz{$U_{_{\rchi}}$}. There are \tsz{$\rchi$} global depolarizing channels dressing the ideal gates, which can be moved to the end of the computation, such that they act on the perfect output state:
 
\vTh

\begin{equation}
\rho_{_{\hsnh n}}= (1-\ep_g)^{\rchi}\hs \hsh \SDrop{3pt}{\uchiX} \hsh \rho+f(\ep_g,\rchi)\hs \Ie_n.
\end{equation}

This relation can be used to express a noisy observable in terms of its \ti{infinite temperature value} \tsz{$\Om_{ITS}=\text{\nsz{Tr}}[\hsh \Ie_n\hsh O]$}:

\vTh

\begin{equation}
\begin{split}
\bk{O_{_{\hsnh n}}}&=(1-\ep_g)^{\rchi} \ \Scale[1.1]{\tr}\3 [\hsh \SDrop{3pt}{\uchiX} \hshh \rho\hsh \hshh O\hshh \3] +f(\ep_g,\rchi)\hs \Scale[1.1]{\tr} \3[\hsh\Ie_n O\hshh\3] \\
&=(1-\ep_g)^{\rchi} \ \Scale[1.1]{\tr} \3 [\hsh \rho \hsh \hshh \SDrop{3pt}{\uchiX}^{\hsn \hshh\dg} \hsnhh O \hshh\3] +f(\ep_g,\rchi)\hs \Scale[1.1]{\tr} \3[\hsh\Ie_n O\hshh\3]\\[0.4em]
&= (1-\ep_g)^{\rchi} \BK{ \SDrop{3pt}{\uchiX}^{\hsn \hshh \dg} O}+f(\ep_g,\rchi) \hs \Om_{ITS}\\[0.4em]
&=(1-\ep_g)^{\rchi}\bk{\hshh O_{\hsnhh u}\hshh}+f(\ep_g,\rchi) \hs \Om_{ITS}
\end{split}
\end{equation}

\vo

%\nl
Apply the geometric sum formula to \tsz{$f(\ep_g,\rchi)$} (\aref{recursiveM}), to isolate the ideal observable \tsz{$\bk{O_u}$}:

\vTh

\begin{equation}
\begin{split}
\bk{O_{_{\hsn n}}}=(1-\ep_g)^{\rchi}&\bk{\hshh O_{\hsnhh u}\hshh}+\3[1-(1-\ep_g)^{\rchi}\hshh\3]\hs \Om_{ITS} \\[0.6em]
\bk{O_{_{\hsn n}}}-\Om_{ITS}&=(1-\ep_g)^{\rchi}\hsh \3[\hsnhh \bk{\hshh O_{\hsnhh u}\hshh}-\Om_{ITS}\hshh \3]
\end{split}	
\end{equation}

\vh

To obtain a simple relation, the \ti{rescaled observable} is defined \tsz{$\Om\equiv O- \Om_{ITS}\hsnh \times \hsnh \Ie_n$}:

\vTh

\begin{equation}\label{claweinvert}
\bk{\Om_n}=\SRaise{5pt}{(1-\ep_g)^{\hshh \rchi}} \bk{\Om_u}
\end{equation}

\vTh
\subsubsection{Calibration}
\vTt

A \ti{calibrating unitary} is a secondary computation that mimics the noise dynamics of the primary computation. It must have a \ti{calibrating observable}: a quantity whose ideal observable \tsz{$\bk{O^c_u}$} is known for a \ti{calibration state} \tsz{$\rho_{_{\hsnh c}}$}.

\vo

To obtain \tsz{$\bk{O^c_n}$}, the calibrating observable is measured after applying the calibrating unitary to \tsz{$\rho_{_{\hsnh c}}$}. Taking the ratio of the noisy observable to the ideal observable after rescaling, yields the \ti{contamination} \tsz{$\Cm\tt{\fsz ($\rchi_c$)}$}:

\vTh

\begin{equation}\label{contamination}
\Cm\tt{\fsz ($\rchi_c$)}=\fr{\bk{\SRaise{4pt}{\Om_n^{c}}}}{\bk{\SRaise{4pt}{{\Om_u^{c}}}}}
\end{equation}

The \ti{secondary global noise strength} is given in terms of the contamination:

\vTh

\begin{equation}\label{epdet}
\ep^s_g=1-\SRaise{5pt}{\Cm\tt{\fsz ($\rchi_c$)}^{1/\rchi_c}}
\end{equation}

\vTh
\subsubsection{Extrapolation}
\vTt

The \ti{ideal map} is the estimate of the ideal observable obtained from \claweT:

\vTh

\begin{align}
\bk{\SRaise{5pt}{\Om^{\hshh \mape}_{u}}}&=(1-\ep^s_g\hsh \SRaise{5pt}{)^{\hsnhh-\rchi}} \bk{\Om_n} \label{idealmap} \\[0.4em]
\bk{\SRaise{5pt}{\Om^{\hshh \mape}_{u}}}&=(1-\ep^s_g\hsh \SRaise{5pt}{)^{\hsnhh-\rchi}}\hsh \hshh \SRaise{5pt}{(1-\ep_g)^{\rchi}} \bk{\Om_u}
\end{align}

\vo

The viability of \claweTO is determined by the ratio of the ideal map to the ideal observable:

\vTh

\begin{equation}
\fr{\bk{\SRaise{5pt}{\Om^{\hshh \mape}_{u}}}}{\bk{\Om_u}}=\Scale[0.9]{\5(}\fr{1-\ep_g}{1-\ep_g^s}\Scale[0.9]{\5)}^{\hshh \rchi}
\end{equation}

\vo

 The ratio scales \ti{super-polynomially} as the scalar depth increases. The ideal map becomes unstable beyond the \ti{gate cutoff}: \tsz{$\rchi_{\text{g}} \hs \tt{$\scriptstyle{\sim}$}\hs \Oe(1/\ep_g)$}. This comes as a result of the depolarizing channel's signal-to-noise problem.

\vTh

\subsection{Calibration Algorithms}\label{subs:calib}
\vTt

In the following algorithms, \ti{motion-reversal} (\tsz{$\h U^{\dg} \h U$}) of the target computation is used as the calibrating unitary \cite{CLAWEPost0}.

\vTh
\vTo 

\subsubsection{Variant I}

\vTo

\begin{figure}
\includegraphics[scale=0.058]{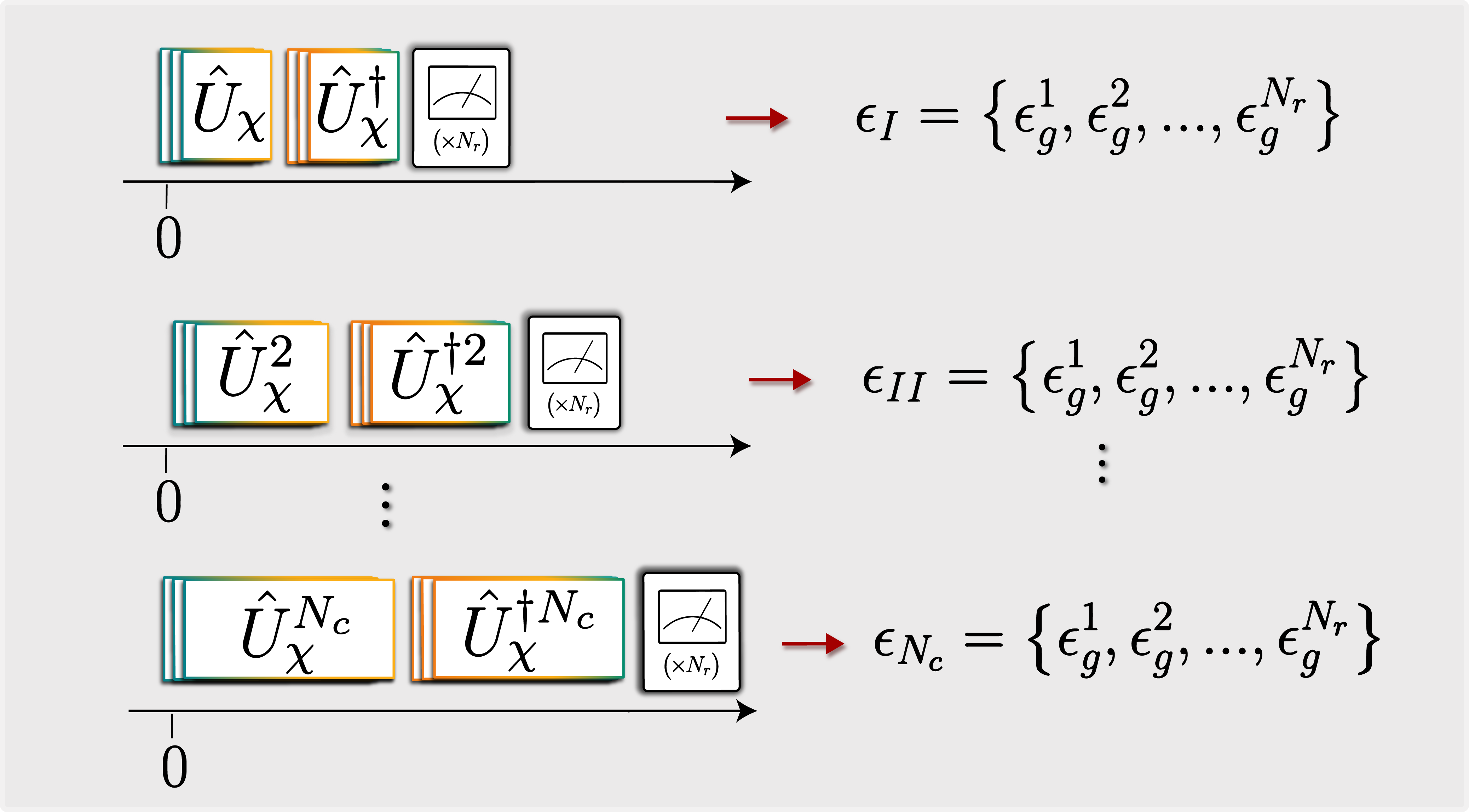}
\caption{\ti{Variant I}: A series of noise-tailored, motion-reversal experiments is performed. Measuring the contamination generates estimates of \tsz{$\ep_g$}.}
\label{fig:calibI}
\vspace{4mm}
\includegraphics[scale=0.039]{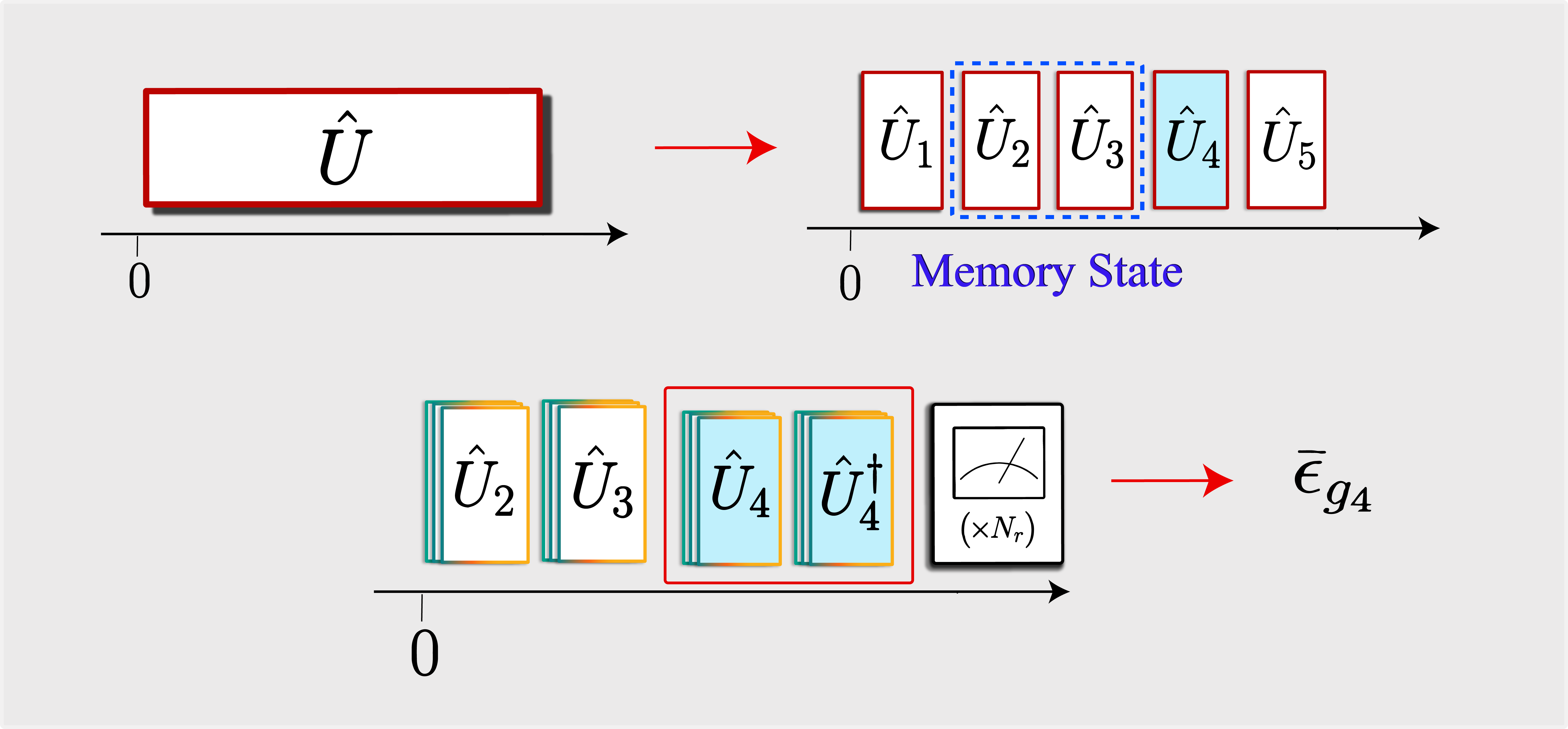}
\caption{\ti{Variant II}: [Upper panel] A target computation is partitioned into unitary fragments. A memory window (dashed) is used to generate the memory state.
[Lower panel] Applying motion-reversal to the memory state with \fsi{$U_4$} generates estimates of \tsz{$\ep_{g_4}$}.} \vspace{-3mm}
\label{fig:calibII}
\end{figure}

An implicit assumption of \aref{claweinvert} is that \tsz{$\ep_g$} is relatively constant during the course of the computation. The \vITO calibration algorithm is designed to extract this constant, by applying motion-reversal in powers of the target computation (\aref{fig:calibI}):

\vTh

\begin{equation}
\3\{\hs \h U_{\rchi}^{\dg k}\hs \hsh  \h U_{\rchi}^{k} \hs \hsh  \3| \scalebox{0.9}{\hs \hsh  for $\forall$ \hs $k$, from $1$ to $N_c$}\3\} \hsnh \hsN \hsN\hsN
\end{equation}

\vTh
\subsubsection{Variant II}

\vTo

The \vIITO calibration algorithm partitions the target computation into \ti{unitary fragments} of scalar depth \tsz{$\{\rchi_i\}$} \cite{CLAWEPost0}:

\vTh

\begin{equation}
U_{\rchi}\ra U_{\rchi_{N_p}} \hsnh \cdots \hs \hsh U_{\rchi_2} \hs U_{\rchi_{1}}
\end{equation}

\vTo

Throughout a computation, \tsz{$\ep_g$} will modulate across the unitary fragments. This allows \tsz{$\ep_g$} to be modeled as a \ti{global noise vector} \tsz{$\V \ep_g=\{\ep_{g_1}, \hsh ... \hs, \ep_{g_{N_p}}\}$}. 

\vt

\tus{\ti{Memory Window}}: Calibrating the unitary fragments in a vacuum will not capture the true noise dynamics of \tsz{$U_{\rchi}$}. The qubits retain memory of past interactions with the environment. The noise dynamics of a unitary fragment depend on those of its predecessors.

\vo

Due to \ti{quantum memory loss}, this dependence can be mimicked by a \ti{memory window}: a portion of the unitary fragment's predecessors. Applying the memory window to the calibration state will imprint noise dynamics onto a \ti{memory state} (\aref{fig:calibII}).

\vo 

After constructing the memory state, motion-reversal is performed to generate \tsz{$\ep^s_{g_s}$}. Repeating this procedure for every unitary fragment yields \tsz{$\V {\ep_g^s}$}.

\vo

%\nl
The ideal map is the following:

\vTh

\begin{equation}\label{piecewise}
\bk{\SRaise{5pt}{\Om^{\hshh \mape}_{u}}}= (1-\ep^s_{g_1})^{-\rchi_1} ...\hs\hsh (1-\ep^s_{g_{N_p}})^{-\rchi_{N_p}}\hs \bk{\Om_n}
\end{equation}

\vt

\vTo
 
\section{Hardware Applications}\label{sec:IBMQ}

\vTt

\claweTO and \ti{Zero-Noise Extrapolation} (\zneT) \cite{zeronoiseIII,zeronoise} are applied to two computations that lie squarely within the \ti{perturbative noise regime} (\pnrT) \cite{PNR0}:

\vTh
\vTo

\begin{equation}
\rchi_{\tt{\tiny PNR}} \hsh  \lsim \hs \hsh \fr{n}{2} \hsh \rchi_g
\end{equation}

\vTh

\vTh

\subsection{Quantum Simulation}

\vTh

A sensible choice for a benchmark computation is one that is expected to exhibit a \ti{quantum speedup}. Rigorizing computing advantages is a key concern of \ti{computational complexity theory}, which aims to classify computational problems by their difficulty \cite{complexi,complexii,complexAaron0,complexAaron1}.

\vo

Postulated by Alan Cobham \cite{cobham} and Jack Edmonds \cite{edmonds}, \ti{Cobham's thesis} states that feasible computations have known \hs \ti{polynomial-time algorithms} (\ptasT) \cite{complexOne}. \ptasTO are run on either classical \cite{cturing0,cturing1,cturing2} or quantum \cite{qturing0,qturing1,qturing2} \ti{Turing machines}. This distinction is used to organize feasible computations in \ti{complexity classes}. 

\vo

Classically feasible computations are placed in the class \ti{bounded-error probabilistic polynomial time} (\bppT) \cite{complexThree}. Quantumly feasible computations are placed in the class \ti{bounded-error quantum polynomial time} (\bqpT) \cite{complexFour}. Computations that exhibit quantum speedups are those \tsz{$ \in$} \bqpTO and \hsnh \text{ $\notin$} \bppTO  \cite{qspeedII}.

\vo

\ti{Local quantum simulation} is a computation thought to be \text{$\notin$} \bppTO \cite{feynmansupremacy,complexTwo,completeIII,completeIV,completeI,completeII}. Calculating the time evolution of an \tsz{$n$}-qubit system involves multiplying \tsz{$\SRaise{4pt}{2^n}\times \SRaise{4pt}{2^n}$} matrices. Because classical algorithms manipulate matrix elements, computing the dynamics will take exponential time:\footnote{Some theories possess a poly($n$) representation in the low-entanglement regime, enabling efficient computation \cite{MPSI,MPSII,MPSIII,MPSIV,MPSV,MPSVI,DMRGI,DMRGII,DMRGIII,DMRGIV}.\\[0.01em]}

\vTh

\begin{equation}
\ti{time-complexity} \hsh \sim \hsh  \Oe(\SRaise{4pt}{2^{2n}})\footnote{Time-complexity is the amount of computer time required to run an algorithm. It is usually approximated by counting the number of elementary operations.}
\end{equation}

\vo

Seth Lloyd proved that local quantum simulation is \tsz{$ \in$} \bqpTO by deriving the \ti{\productFATO} (\pfaT) \cite{complexTwo}. It is tailored to \ti{\tsz{$k$}-local theories} \cite{interI},  which have interactions coupling \tsz{$\leq k$} qubits:

\vFo

\begin{equation}
H=\sum_{\sg=1}^{N_i} \hsh H_{\sg}
\end{equation}

The \pfaTO \hs applies a \ti{Trotter-Suzuki expansion} \cite{trotterI,trotterII,trotterIII} to the evolution operator:

\vspace{-3mm}

\begin{equation}\label{pfaE}
\Scale[1.3]{e}^{^{\hsnh -\Scale[1]{i H t}}}\approx \5(\hsh \prod_{\sg=1}^{N_i} \hs \Scale[1.3]{e}^{^{\hsnh -\Scale[1.02]{i H_{\sg} t/n_t}}} \hsh \5)^{\hsnh \Scale[1.02]{n_t}}
\end{equation}

\vo

Approximating the dynamics to accuracy \tsz{$\ep$} will take the following time:

\vTh

\begin{equation}
\ti{time-complexity} \hsh \sim \hsh N_i \hsh \SRaise{4pt}{4^k} \hs \hsh \SRaise{4pt}{t^2}\3{/}\ep
\end{equation}

\vo

This will scale polynomially, provided \tsz{$N_i\sim \tt{poly}(n)$} and \tsz{$k\sim \tt{polylog}(n)$}. For such theories, the \pfaTO generates an exponential quantum speedup over classical approaches.

\subsection{Simulating the Fermi-Hubbard Model \\on a Quantum Computer}\label{ssc:fhm}

The benchmark computations involve digital quantum simulation using the \pfaT.

\vTo

\subsubsection{Simulation Prescription}\label{ssc:adiab}

\vTo

\vo

The benchmark computations require time-dependent quantum simulation of the \fermiHT:

\vTh

\begin{equation}\label{Hrun}
H_{f}(t)=-h(t)\hs \4\{\sg_x^1+\sg_x^2\4\}\hs +\hs \fr{u(t)}{2}\hs \sg_z^1\tens\sg_z^2
\end{equation}

\vth

%\nl 
The Hamiltonian becomes dimensionless when rescaled with \tsz{$h(t)$}. The dimensionless interaction is \tsz{$\td u(t)=u(t)/h(t)$}:

\vTh

\begin{equation}\lab{Fhresc}
\td H_{f}(t)=-\4\{\sg_x^1+\sg_x^2\4\}\hs +\hs \fr{\td u(t)}{2}\hs \sg_z^1\tens\sg_z^2
\end{equation}

\vt

\begin{figure*}
\includegraphics[scale=0.125]{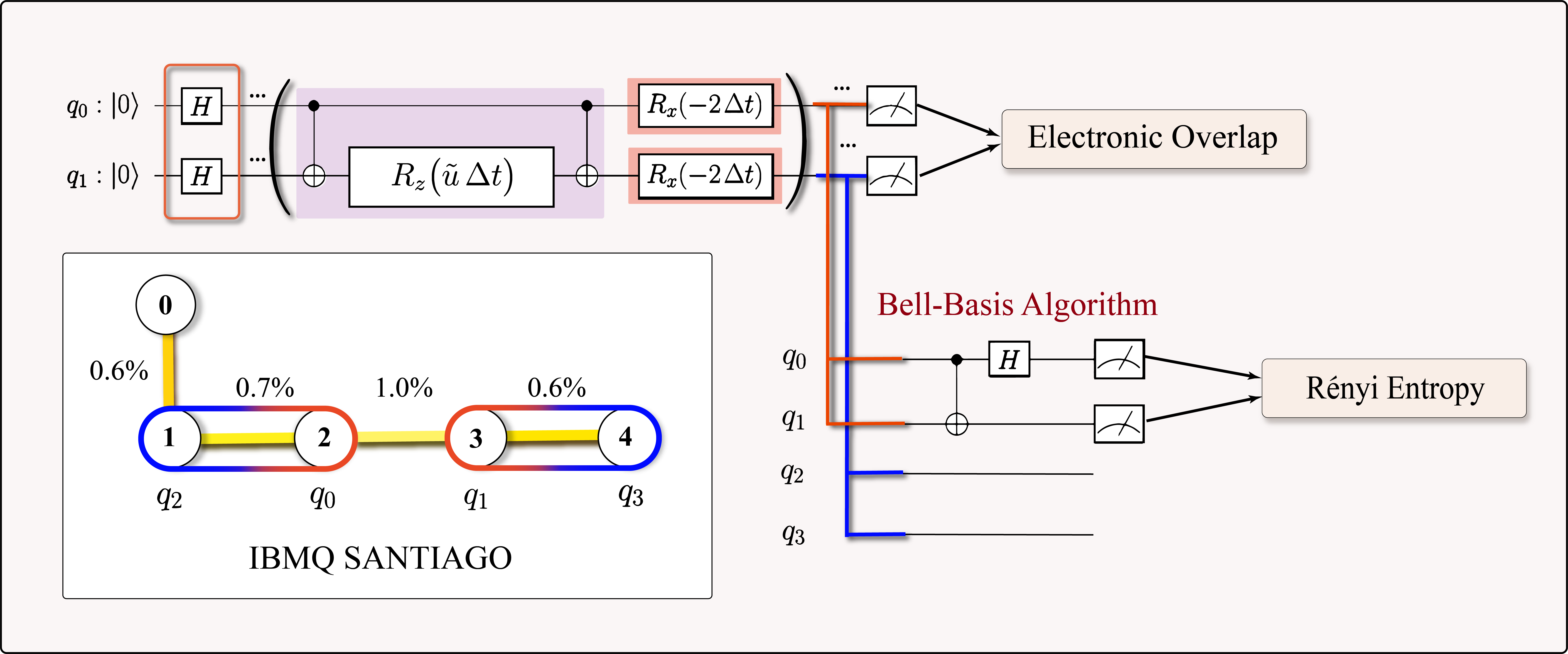}
\caption{[Upper Panel] The PFA is used to perform quantum simulation of the Fermi-Hubbard Model. [Lower Right Panel]
The \reynT entropy is computed using the Bell-Basis Algorithm.}
\vspace{-3mm}
\label{fig:FullCirc}
\end{figure*}

The initial state is prepared with a pair of Hadamard gates (\aref{fig:FullCirc}):

\vTh

\begin{equation}\label{fermtb}
\K{\9_{_\tt{init.}}}=\5\{\hsh\fr{\K{0}+\K{1}}{\sqrt{2}}\hsh \5\}\tens\5\{\hsh \fr{\K{0}+\K{1}}{\sqrt{2}} \hsh \5\}
\end{equation}

\vth

\vTt
\subsubsection{Computation Parameters}\label{ssc:nisqComp}

\vTt

\vt

\tus{\ti{Theoretical Result}}: To compute the theoretical result, quantum simulation is performed numerically.

\vo
The digitization error is estimated by permuting terms in the \pfaT. Each \pfaTO step contains \tsz{$3$} unitaries, of which two mutually commute (\aref{fig:FullCirc}). As such, there are \tsz{$\SRaise{3pt}{4^k}$} total permutations for the \tsz{$k^{\text{th}}$} \pfaTO step.

\vo

\vo

\vTt

\subsubsection{Electronic Overlap Dynamics}\label{ssc:rtnisqC}

The presence of electrons occupying shared lattice sites is indicated by the \ti{electronic overlap}:

\vTh

\begin{equation}\label{EsOp}
E_o=\K{00}\hsn\b{00}+\K{11}\hsn\b{11}
\end{equation}

\vo

The electronic overlap is measured during quantum simulation of the \fermiHT. 

\vfo

{\centering \paragraph{NISQ Computation} ~\\}

\vth

\tus{\ti{QPU Interfacing}}: The \vigoTO is accessed via \ti{\qiskT}, which enables preparation of quantum circuits within an intuitive framework \cite{QisKit}. Quantum circuits are packaged into a \ti{job object}, and sent to the \qpUT. Each job object can contain up to \tsz{$75$} individual  circuits.

\vt

\tus{\ti{Data Acquisition}}: The benchmark computation requires \tsz{$10$} quantum circuits. The circuits are evaluated with a single call to the \vigoT. A \ti{bootstrap algorithm} \cite{bootstrapI,bootstrapII,bootstrapIII} is run on the measurements to generate the electronic overlap and its uncertainty.

\vfo

{\centering \paragraph{ZNE Noise Mitigation} ~\\}\label{ssc:znedcond}

\vth

\zneTO is used to establish a performance baseline for \claweT. 

\vo

\zneTO is performed with a polynomial fit and a \ti{Richardson extrapolation}.

\vo

In both approaches, errors are modulated using the \ti{quartic-cycle noise amplification}  (\qcnaT) prescription. \qcnaTO amplifies the noise dynamics by injecting pairs of \cnotTO gates in four rounds of computation  (\aref{fig:ZNEdCondCalib}). In this work, a (\tsz{$\times 7$})-\qcnaTO is utilized: \tsz{$\rchi\ra 7\hsh \rchi$}.

\vo

\vth

\tus{\ti{ZNE Viability}}: \zneTO is designed to treat computations that remain within the \pnrTO after error amplification is performed. Applying (\tsz{$\times 7$})-\qcnaTO to the \pfaT-expansion probes noise dynamics outside of the \pnrTO (\tsz{$\rchi\hsnh :20\ra 140$}):

\vFo

\begin{equation}\label{eq:vigopnr}
\text{\small IBMQ-Vigo:} \ \rchi_{\tt{\tiny PNR}} \approx 111
\end{equation}

\nl
As such, \zneTO should reliably correct \tsz{$7/10$} time-steps.

\vt

\tus{\ti{Polynomial Fit}}: Performing \zneTO with a third-order polynomial fit demonstrates theoretical agreement for \tsz{$2/10$} points  (\aref{fig:ZNEdCondPoly}).

\vt

\tus{\ti{Richardson Extrapolation}}: Performing \zneTO with a second-order Richardson extrapolation demonstrates theoretical agreement for \tsz{$6/10$} points (\aref{fig:ZNEdCondRichard}).

\vfo

{\centering \paragraph{CLAWE Noise Mitigation} ~\\}\label{ssc:clawecond}

 \vth

\vITO and \vIITO are applied to the electronic overlap computation.

\vt

\vt\tus{\ti{Variant I}}: Performing \vITO demonstrates theoretical agreement for \tsz{$8/10$} points (\aref{fig:CLAWECondI}).

\vt

\tus{\ti{Variant II}}: Performing \vIITO demonstrates theoretical agreement for \tsz{$8/10$} points (\aref{fig:CLAWECondII}).

\vt

\tus{\ti{Variant I Calibration}}: The global noise strength is used to perform extrapolation (\aref{fig:CLAWEIEPG}).

\vt

\tus{\ti{Variant II Calibration}}: The global noise vector is used to perform extrapolation (\aref{fig:CLAWEIIEPG}).

\subsubsection{\reynT Entropy Dynamics}\label{subs:rtimerenyi}

The \ti{\reynT entropy} is measured during  quantum simulation of the \fermiHT.

\vfo

{\centering \paragraph{\reynT Entropy} ~\\}\label{ssc:renyiBBA}

 \vth

The Hilbert space of the \fermiHTO is bipartite in the electron spin:

\vTh

\begin{equation}
\Hm=\Hm_{\ua}\tens \Hm_{\da}	
\end{equation}

\vo

For a state \tsz{$\rho$}, the \ti{reduced density matrix} for each electron spin species is obtained by applying a \ti{partial trace}:

\vTh

\begin{align}
\rho_{\ua}=\tr_{\da}\3[\hsh \rho\hsh \3] \\[0.4em]
\rho_{\da}=\tr_{\ua}\3[\hsh \rho\hsh \3]
\end{align}

\vo

The \reynT entropy of \tsz{$\rho_{\ua}$} is the following:

\vTh

\begin{align}\label{eq:reyniDef}
S\hshh(\rho_{\ua}) &=-\fr{1}{2}\hsh \log{\4\{ \hsh \text{Tr} \3(\hsh \rho_{\ua}^2 \hsh \3) \hsh \4\}}
\end{align}

\vfo

{\centering \paragraph{NISQ Computation} ~\\}\label{ssc:nisqRenyi}

 \vth

The \reynT entropy can be computed using the \ti{Bell-Basis Algorithm} (\bbaT) \cite{reyniXI}.

\vt

\tus{\ti{Bell-Basis Algorithm}}: Quantum simulation is applied to two copies of the initial state. The quantum stage of the \bbaTO is applied using a \cnotTO gate between the spin-up qubits (\aref{fig:FullCirc}).

\vt

\tus{\ti{Data Acquisition}}: The \reynT entropy computation requires \tsz{$10$} quantum circuits, which are evaluated with a single call to the \saintT. A bootstrap algorithm is run to generate the \reynT entropy and its uncertainty.

\vt

\vfo

{\centering \paragraph{ZNE Noise Mitigation} ~\\}\label{ssc:znedRenyi}

\vth

\zneTO is applied to the \reynT entropy computation.

\vt

\tus{\ti{Polynomial Fit}}: Performing \zneTO with a third-order polynomial fit demonstrates theoretical agreement for \tsz{$1/10$} points (\aref{fig:ZNEdRenyiPoly}).

\vth

\tus{\ti{Richardson Extrapolation}}: Performing \zneTO with a second-order Richardson extrapolation demonstrates theoretical agreement for \tsz{$4/10$} points (\aref{fig:ZNEdRenyiRichard}).

 \vfo

{\centering \paragraph{CLAWE Noise Mitigation} ~\\}\label{ssc:claweRenyi}

 \vth

 \vITO and \vIITO are applied to the \reynT entropy computation.

\vt\tus{\ti{Variant I}}: Performing \vITO demonstrates theoretical agreement for \tsz{$3/10$} points (\aref{fig:CLAWERenyiI}).

\vt

\tus{\ti{Variant II}}: Performing \vIITO demonstrates theoretical agreement for \tsz{$7/10$} points (\aref{fig:CLAWERenyiII}).

\section{Acknowledgements}\label{sec:ack}

\begin{center}
The author is supported in part by the \ti{\href{http://bit.ly/MCFPII}{Maryland Center for Fundamental Physics}, University of Maryland, College Park} and by the \ti{U.S. Department of Energy (DOE), Office of Science, \href{http://bit.ly/OASCRII}{Office of Advanced Scientific Computing Research} (ASCR) Quantum Computing Applications Teams program}, under fieldwork proposal number \ti{ERKJ347}.

\end{center}

\begin{center}

\noindent\rule{3cm}{0.4pt}

\end{center}

\begin{center}
\ti{Those who wait on the Lord shall renew their strength;
they shall mount up with wings like eagles,
they shall run and not be weary,
they shall walk and not faint.}
\vh

-\ti{Isaiah 40:31}
\end{center}

\begin{center}
$ -AMDG - $
\end{center}

%\appendix\label{appendix:ref}
\renewcommand\thesubsection{\Roman{subsection}}

\begin{appendices}

\section*{Appendix}
\subsection{Noise Tailoring Algorithms}\label{appendix:Atail}

\vTh

\ti{Coarse-grained decoupling} \cite{PAREC,randomization,randcomp,randcompII} can be used to alter noise dynamics using gate-level control. Oliver Kern, Gernot Alber, and Dima Shepelyansky first demonstrated this using \ti{Pauli Random Error Correction} (\parecT) \cite{PAREC}.

\vo

In \parecT, the \ti{computational frame} is toggled during a computation by inserting randomized pairs of Pauli operations. This leaves the computation logically equivalent. Toggling the computational frame generates a \ti{dynamical decoupling}-like effect without the need for rapid time-modulation \cite{DDI}.

\vo

The \NTto algorithm used in this work is \ti{Randomized Compiling} (\rcoT) \cite{randcomp}. \rcoTO applies coarse-grained decoupling across a sequence of circuits.

\begin{figure}
\includegraphics[scale=0.18]{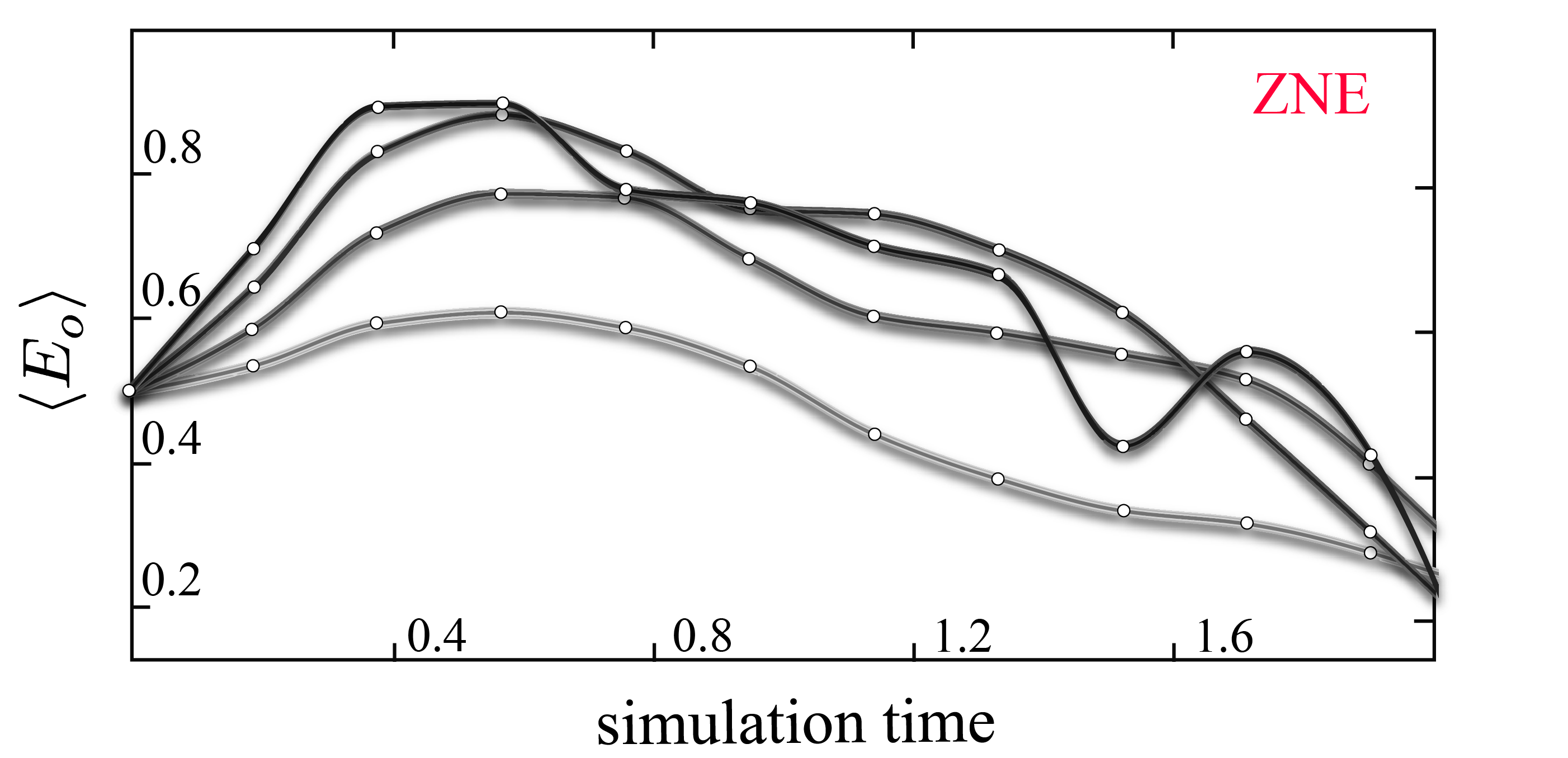}
\caption{Error-amplified dynamics (gray), are extrapolated to the zero-noise solution.}
\label{fig:ZNEdCondCalib}
\vspace{6mm}
\includegraphics[scale=0.18]{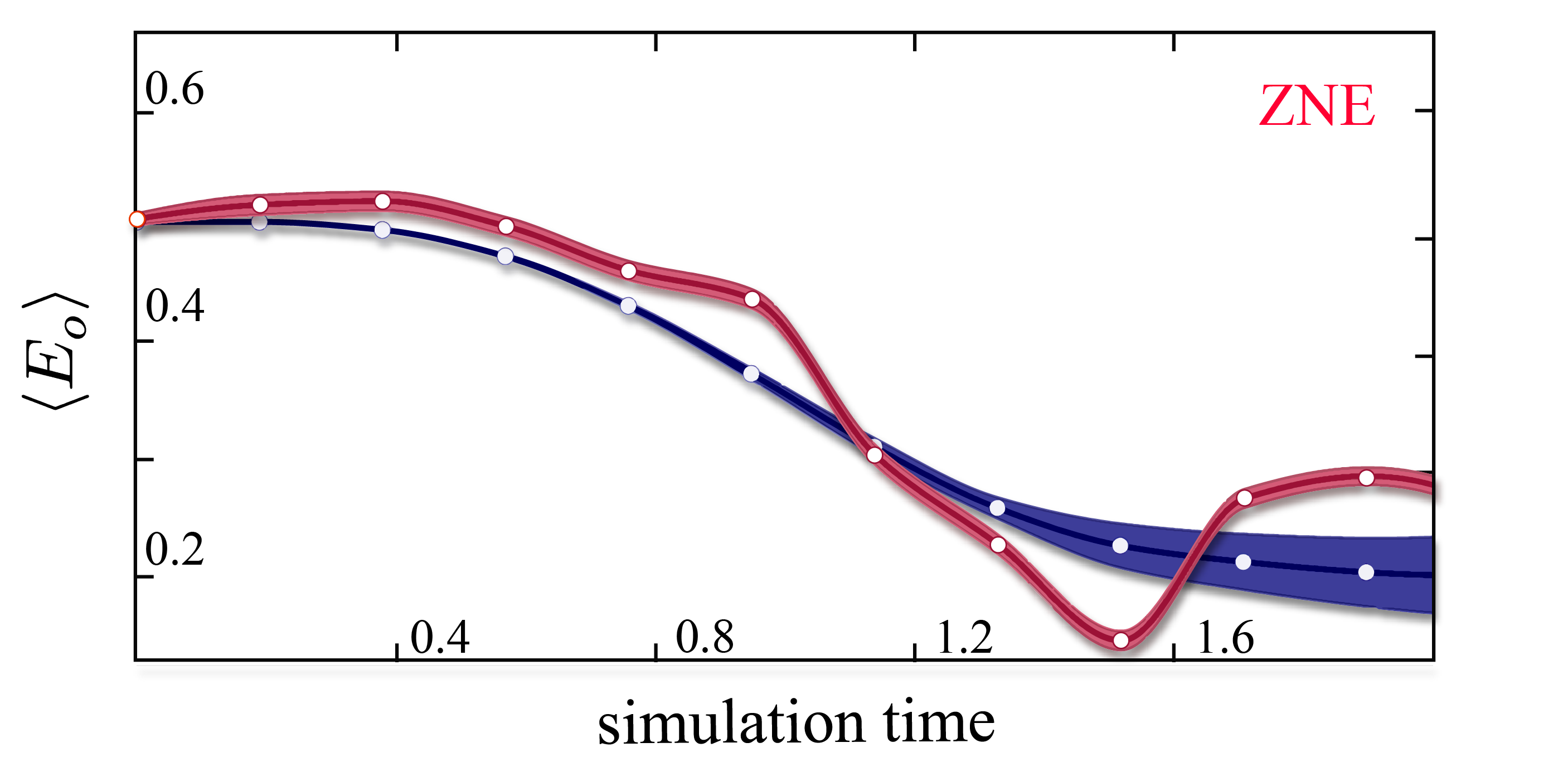}
\caption{The ZNE polynomial fit (red) is compared with the theoretical result (indigo) for the electronic overlap.}
\label{fig:ZNEdCondPoly}
\vspace{6mm}
\includegraphics[scale=0.18]{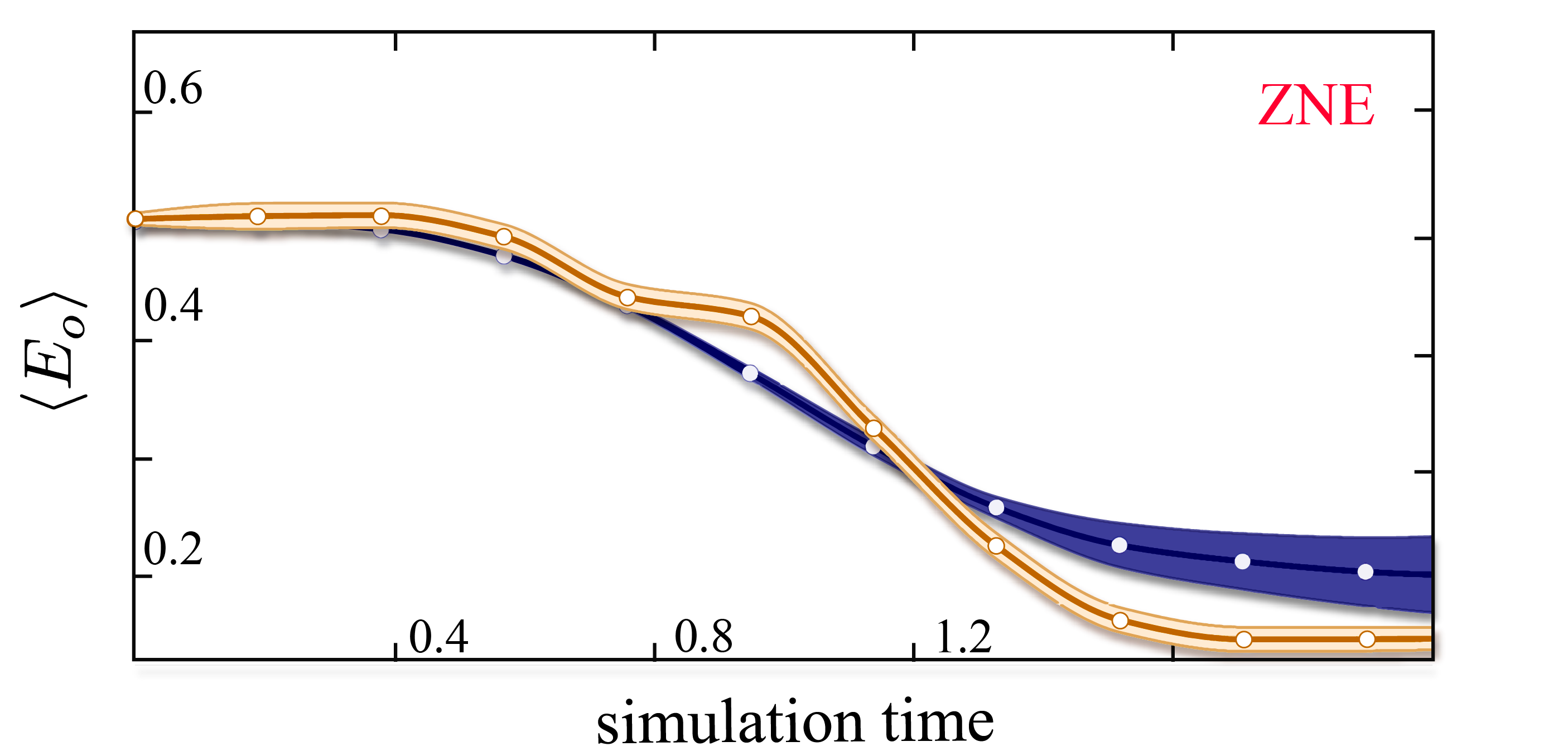}
\caption{The ZNE Richardson extrapolation (golden) is compared with the theoretical result (indigo) for the electronic overlap.}
\label{fig:ZNEdCondRichard}
\end{figure}

\clearpage

\rhead{}

\begin{figure}
\includegraphics[scale=0.18]{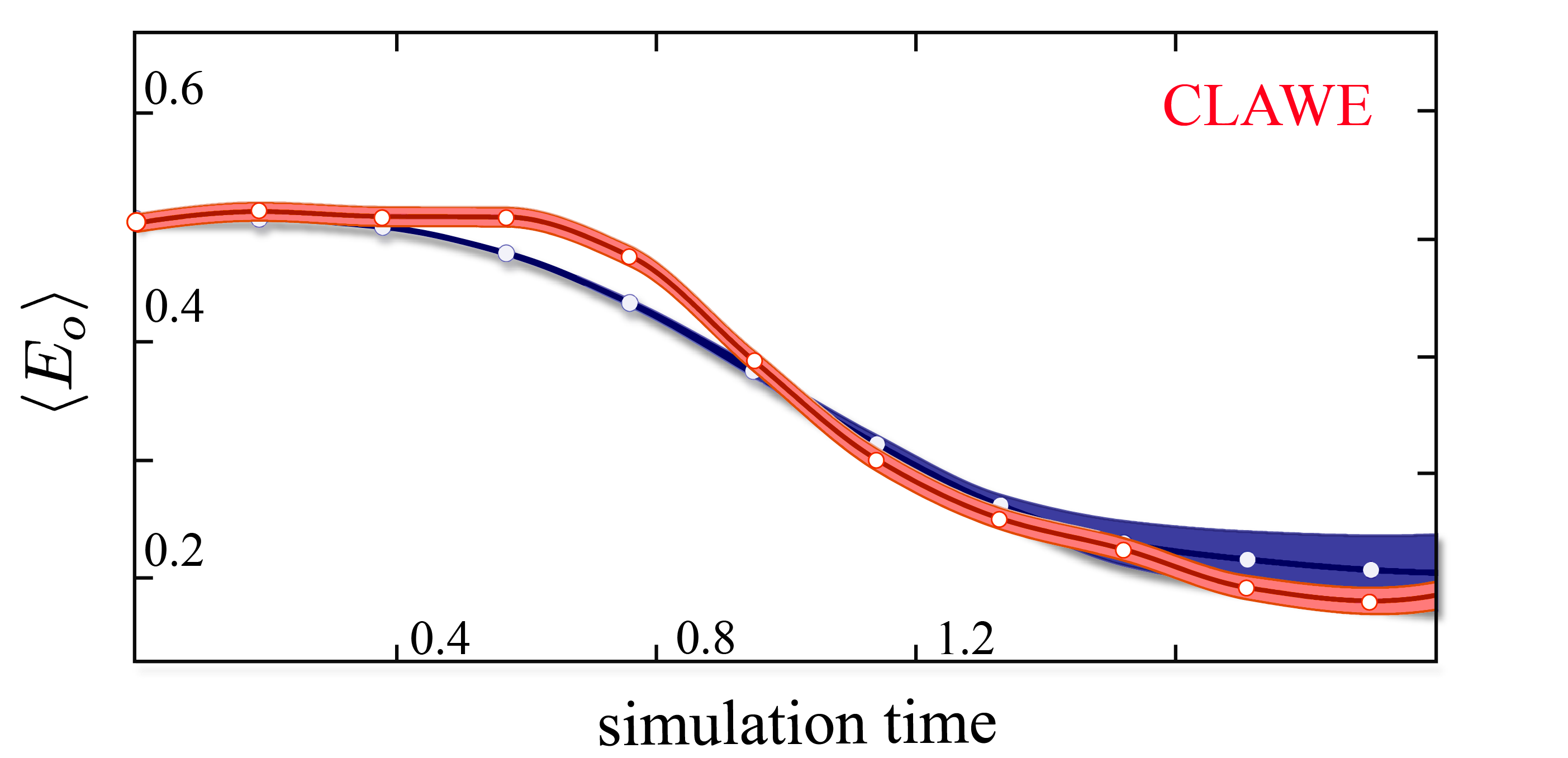}
\caption{Variant I (red) is compared with the theoretical result (indigo) for the electronic overlap.}
\label{fig:CLAWECondI}
\vspace{4mm}
\setcounter{figure}{10}
\includegraphics[scale=0.18]{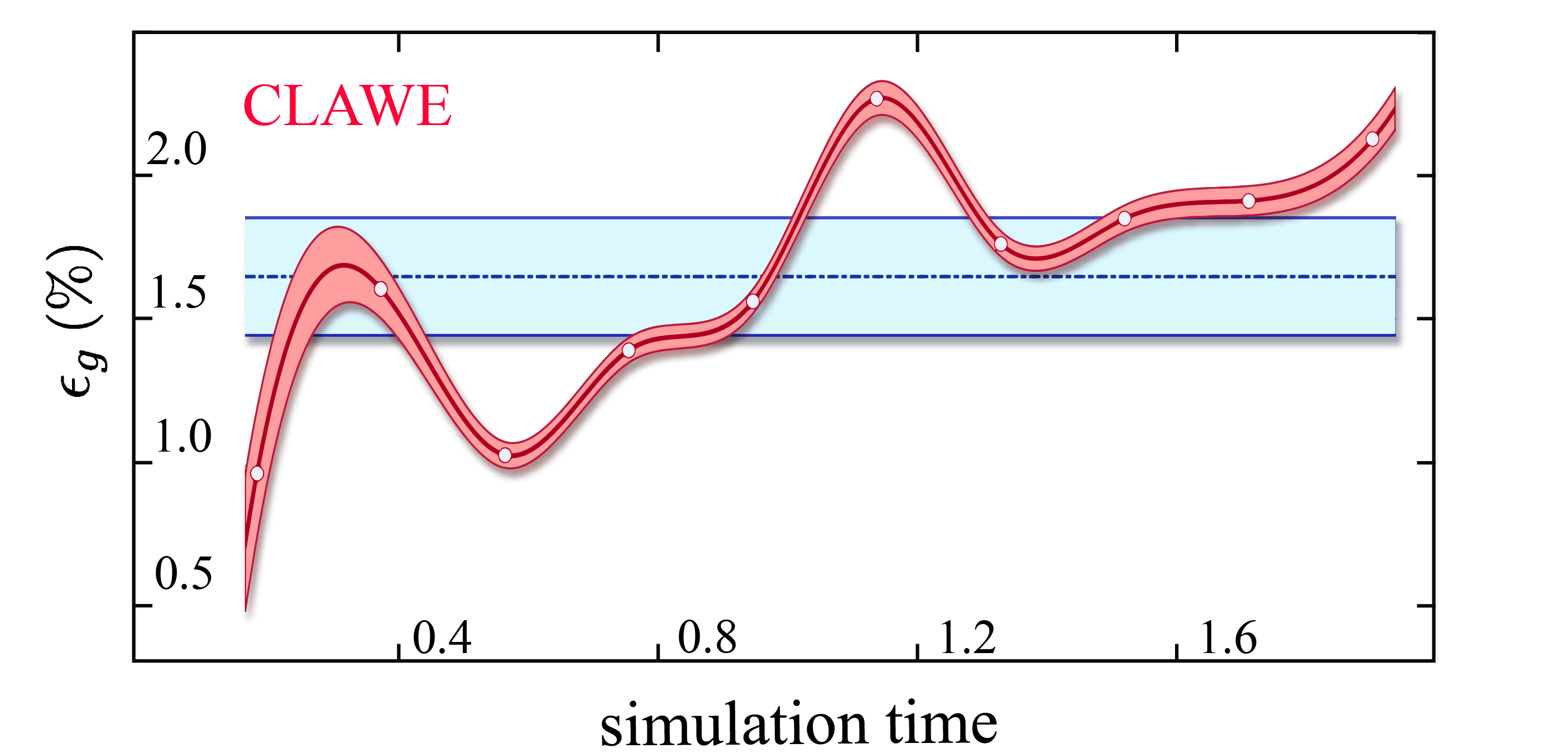}
\caption{The global noise strength (red) is shown alongside its mean value (blue).}
\label{fig:CLAWEIEPG}
\vspace{4mm}
\setcounter{figure}{12}
\includegraphics[scale=0.08]{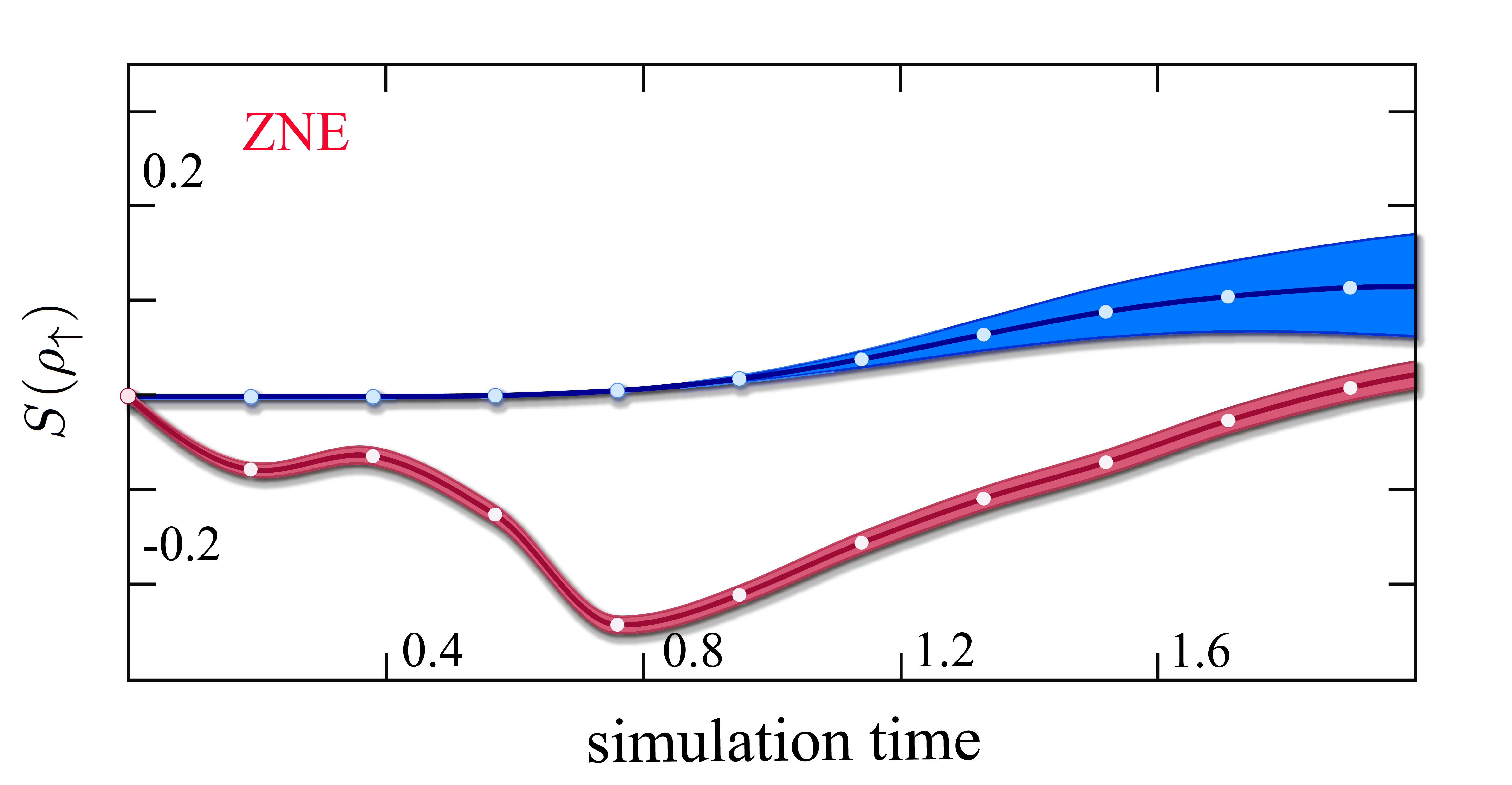}
\caption{The ZNE polynomial fit (red) is compared with the theoretical result (blue) for the \reynT entropy.}
\label{fig:ZNEdRenyiPoly}
\vspace{4mm}
\setcounter{figure}{14}
\includegraphics[scale=0.08]{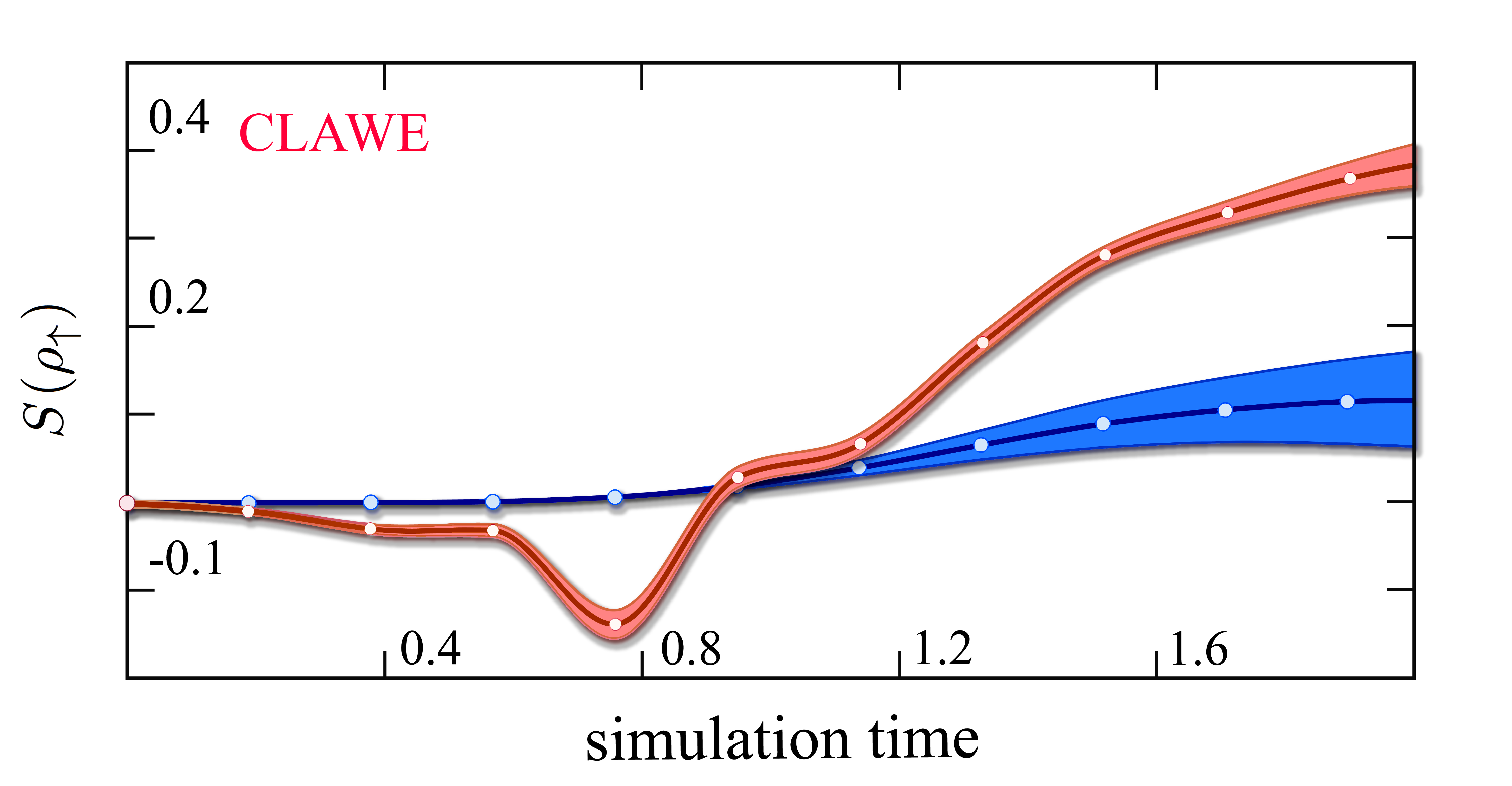}
\caption{Variant I (red) is compared with the theoretical result (blue) for the \reynT entropy.}
\label{fig:CLAWERenyiI}
\end{figure}

\begin{figure}
\setcounter{figure}{9}
\includegraphics[scale=0.18]{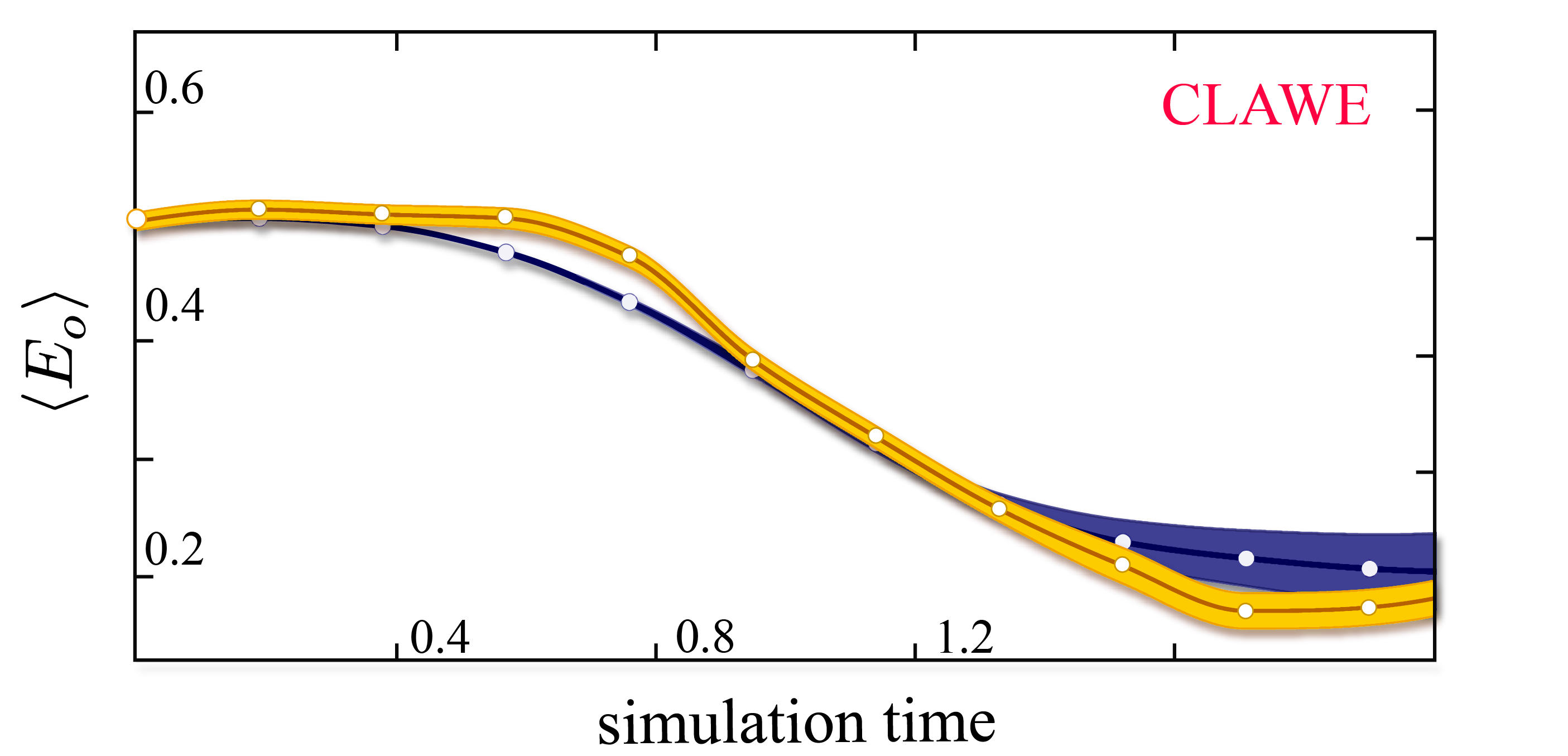}
\caption{Variant II (golden) is compared with the theoretical result (indigo) for the electronic overlap.}
\label{fig:CLAWECondII}
\vspace{4mm}
\setcounter{figure}{11}
\includegraphics[scale=0.18]{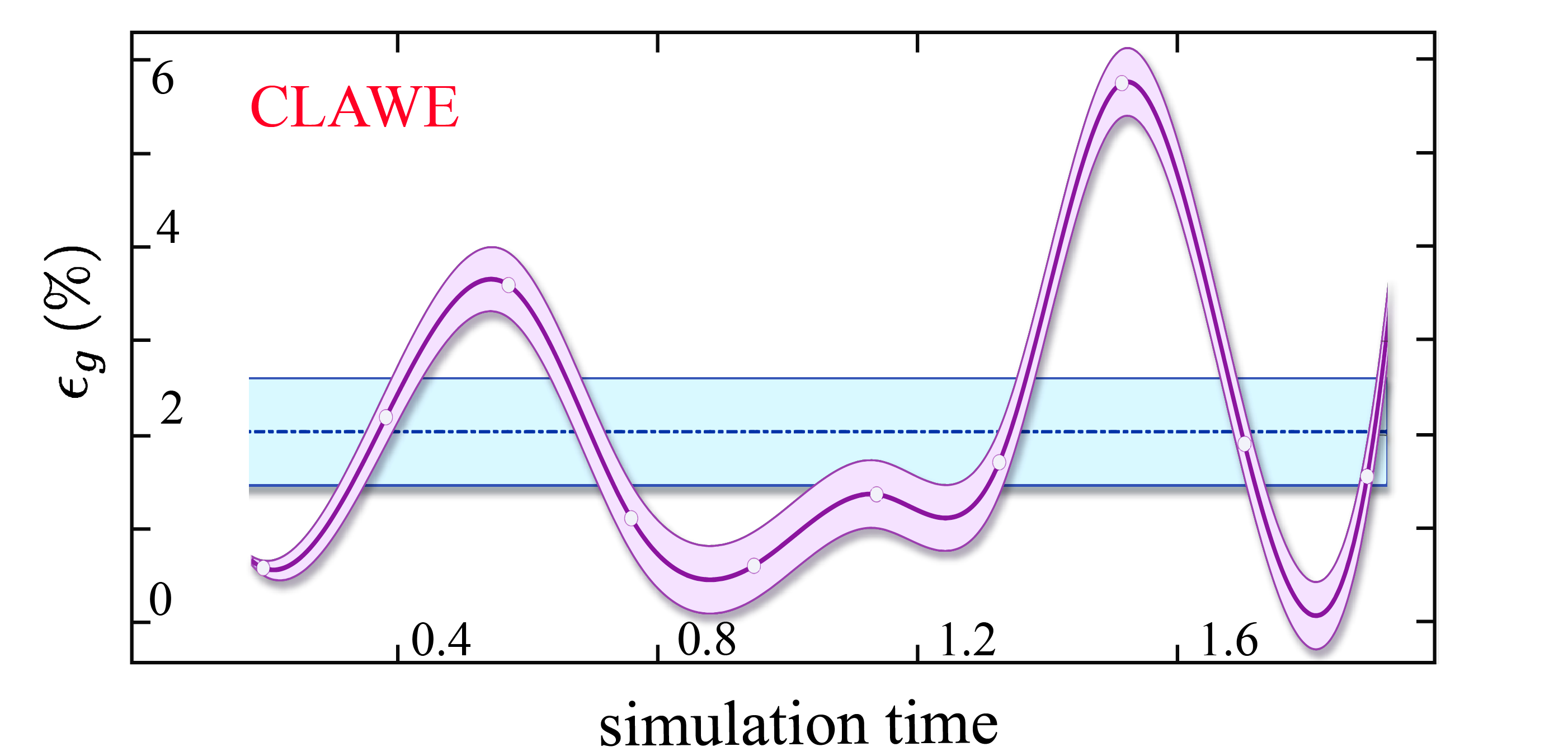}
\caption{The global noise vector (purple) is shown alongside its mean value (blue).}
\label{fig:CLAWEIIEPG}
\vspace{4mm}
\setcounter{figure}{13}
\includegraphics[scale=0.08]{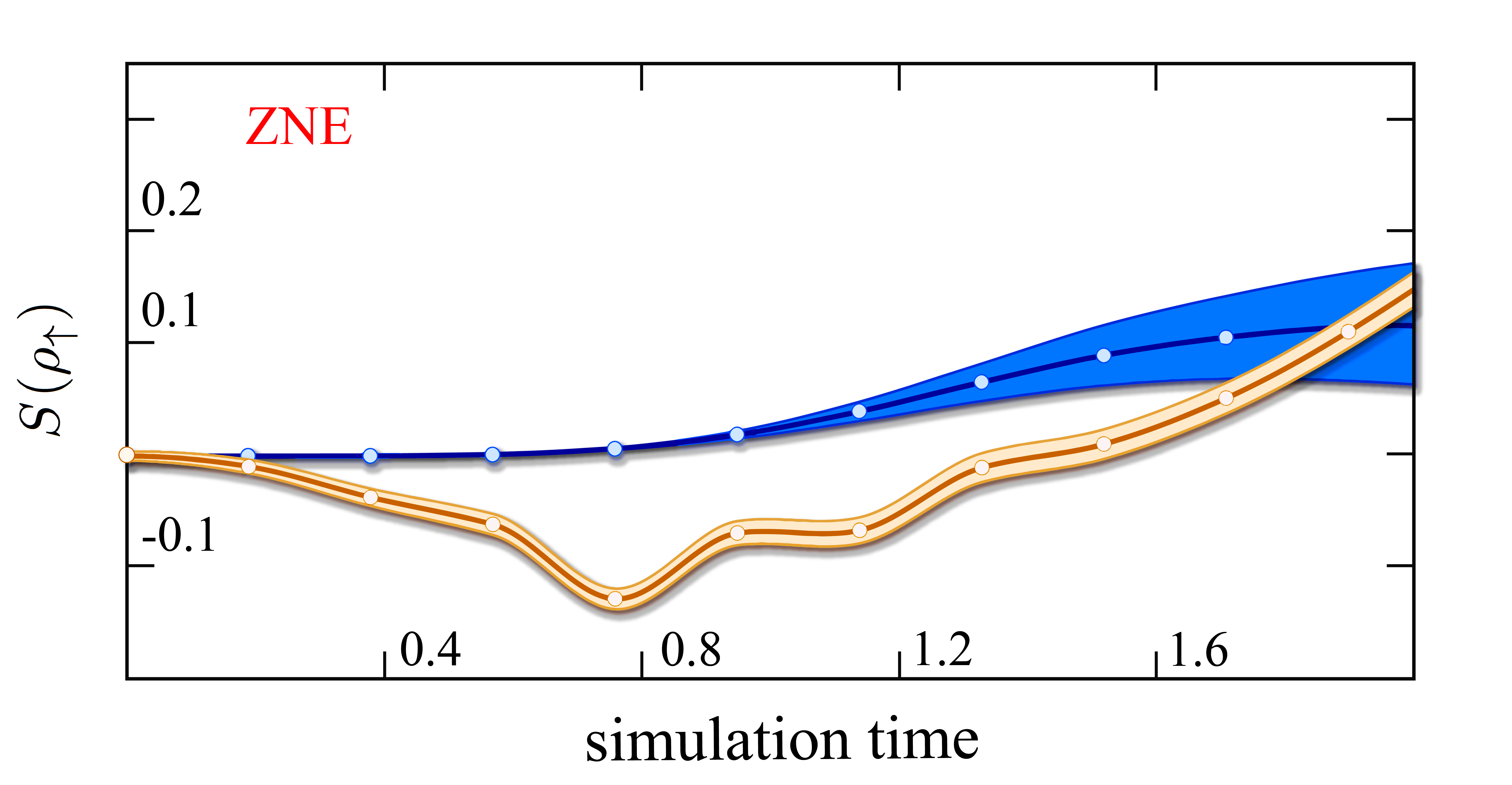}
\caption{The ZNE Richardson extrapolation (golden) is compared with the theoretical result (blue) for the \reynT entropy.}
\label{fig:ZNEdRenyiRichard}
\vspace{4mm}
\setcounter{figure}{15}
\includegraphics[scale=0.08]{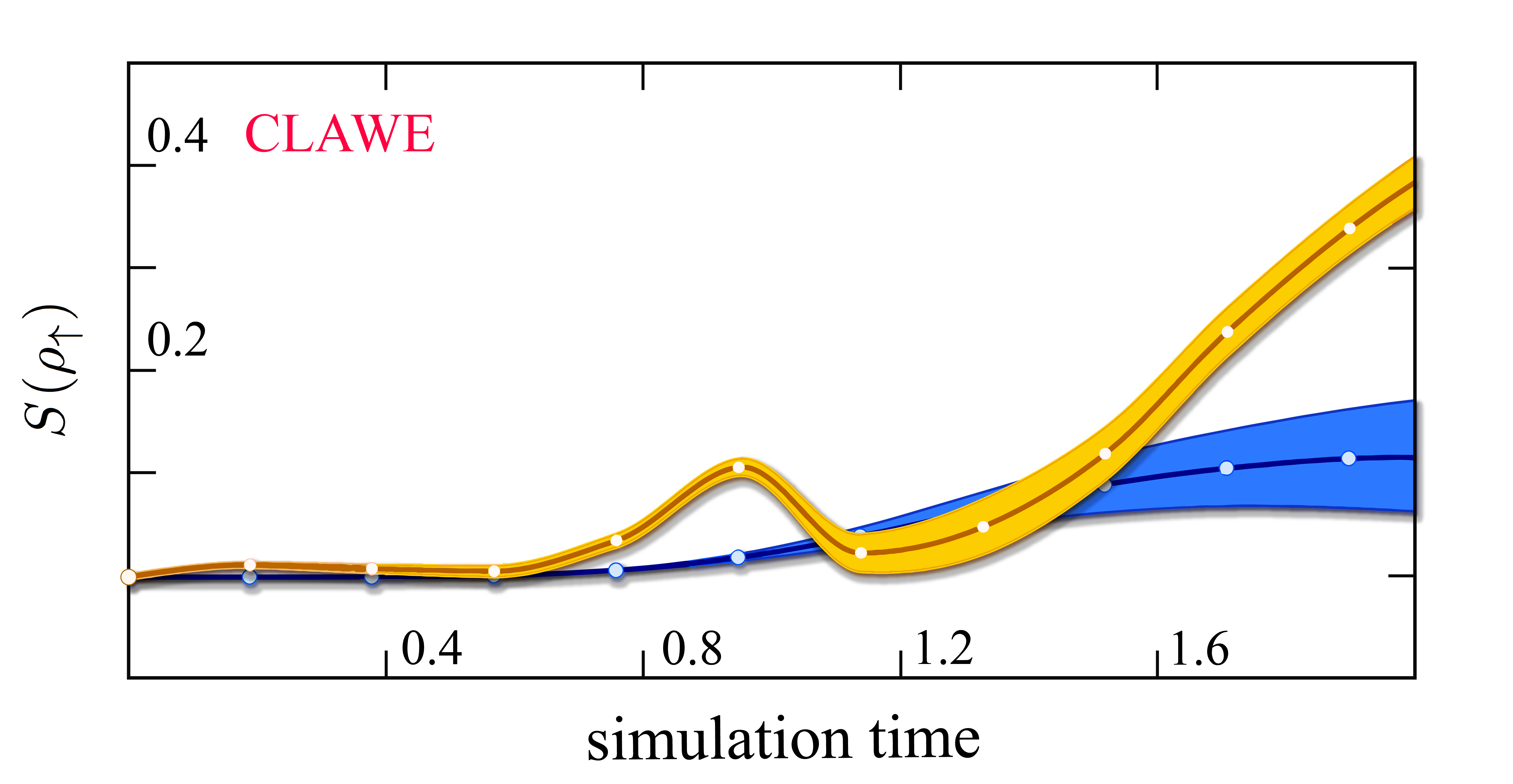}
\caption{Variant II (golden) is compared with the theoretical result (blue) for the \reynT entropy.}
\label{fig:CLAWERenyiII}
\end{figure}

\clearpage

\end{appendices}

\clearpage

\rhead{}

\renewcommand*{\bibfont}{\scriptsize}

\bibliography{CLAWE}

\providecommand{\noopsort}[1]{}\providecommand{\singleletter}[1]{#1}%
\begin{thebibliography}{128}%
\makeatletter
\providecommand \@ifxundefined [1]{%
 \@ifx{#1\undefined}
}%
\providecommand \@ifnum [1]{%
 \ifnum #1\expandafter \@firstoftwo
 \else \expandafter \@secondoftwo
 \fi
}%
\providecommand \@ifx [1]{%
 \ifx #1\expandafter \@firstoftwo
 \else \expandafter \@secondoftwo
 \fi
}%
\providecommand \natexlab [1]{#1}%
\providecommand \enquote  [1]{``#1''}%
\providecommand \bibnamefont  [1]{#1}%
\providecommand \bibfnamefont [1]{#1}%
\providecommand \citenamefont [1]{#1}%
\providecommand \href@noop [0]{\@secondoftwo}%
\providecommand \href [0]{\begingroup \@sanitize@url \@href}%
\providecommand \@href[1]{\@@startlink{#1}\@@href}%
\providecommand \@@href[1]{\endgroup#1\@@endlink}%
\providecommand \@sanitize@url [0]{\catcode `\\12\catcode `\$12\catcode
  `\&12\catcode `\#12\catcode `\^12\catcode `\_12\catcode `\%12\relax}%
\providecommand \@@startlink[1]{}%
\providecommand \@@endlink[0]{}%
\providecommand \url  [0]{\begingroup\@sanitize@url \@url }%
\providecommand \@url [1]{\endgroup\@href {#1}{\urlprefix }}%
\providecommand \urlprefix  [0]{URL }%
\providecommand \Eprint [0]{\href }%
\providecommand \doibase [0]{https://doi.org/}%
\providecommand \selectlanguage [0]{\@gobble}%
\providecommand \bibinfo  [0]{\@secondoftwo}%
\providecommand \bibfield  [0]{\@secondoftwo}%
\providecommand \translation [1]{[#1]}%
\providecommand \BibitemOpen [0]{}%
\providecommand \bibitemStop [0]{}%
\providecommand \bibitemNoStop [0]{.\EOS\space}%
\providecommand \EOS [0]{\spacefactor3000\relax}%
\providecommand \BibitemShut  [1]{\csname bibitem#1\endcsname}%
\let\auto@bib@innerbib\@empty
%</preamble>
\bibitem [{\citenamefont {Abrams}\ \emph {et~al.}(2019)\citenamefont {Abrams},
  \citenamefont {Didier}, \citenamefont {Caldwell}, \citenamefont {Johnson},\
  and\ \citenamefont {Ryan}}]{crosst0}%
  \BibitemOpen
  \bibfield  {author} {\bibinfo {author} {\bibfnamefont {D.~M.}\ \bibnamefont
  {Abrams}}, \bibinfo {author} {\bibfnamefont {N.}~\bibnamefont {Didier}},
  \bibinfo {author} {\bibfnamefont {S.~A.}\ \bibnamefont {Caldwell}}, \bibinfo
  {author} {\bibfnamefont {B.~R.}\ \bibnamefont {Johnson}},\ and\ \bibinfo
  {author} {\bibfnamefont {C.~A.}\ \bibnamefont {Ryan}},\ }\bibfield  {title}
  {\bibinfo {title} {Methods for measuring magnetic flux crosstalk between
  tunable transmons},\ }\href
  {https://doi.org/10.1103/PhysRevApplied.12.064022} {\bibfield  {journal}
  {\bibinfo  {journal} {Phys. Rev. Applied}\ }\textbf {\bibinfo {volume}
  {12}},\ \bibinfo {pages} {064022} (\bibinfo {year} {2019})}\BibitemShut
  {NoStop}%
\bibitem [{\citenamefont {Hashim}\ \emph {et~al.}(2020)\citenamefont {Hashim},
  \citenamefont {Naik}, \citenamefont {Morvan}, \citenamefont {Ville},
  \citenamefont {Mitchell}, \citenamefont {Kreikebaum}, \citenamefont {Davis},
  \citenamefont {Smith}, \citenamefont {Iancu}, \citenamefont {O'Brien},
  \citenamefont {Hincks}, \citenamefont {Wallman}, \citenamefont {Emerson},\
  and\ \citenamefont {Siddiqi}}]{crosst1}%
  \BibitemOpen
  \bibfield  {author} {\bibinfo {author} {\bibfnamefont {A.}~\bibnamefont
  {Hashim}}, \bibinfo {author} {\bibfnamefont {R.~K.}\ \bibnamefont {Naik}},
  \bibinfo {author} {\bibfnamefont {A.}~\bibnamefont {Morvan}}, \bibinfo
  {author} {\bibfnamefont {J.-L.}\ \bibnamefont {Ville}}, \bibinfo {author}
  {\bibfnamefont {B.}~\bibnamefont {Mitchell}}, \bibinfo {author}
  {\bibfnamefont {J.~M.}\ \bibnamefont {Kreikebaum}}, \bibinfo {author}
  {\bibfnamefont {M.}~\bibnamefont {Davis}}, \bibinfo {author} {\bibfnamefont
  {E.}~\bibnamefont {Smith}}, \bibinfo {author} {\bibfnamefont
  {C.}~\bibnamefont {Iancu}}, \bibinfo {author} {\bibfnamefont {K.~P.}\
  \bibnamefont {O'Brien}}, \bibinfo {author} {\bibfnamefont {I.}~\bibnamefont
  {Hincks}}, \bibinfo {author} {\bibfnamefont {J.~J.}\ \bibnamefont {Wallman}},
  \bibinfo {author} {\bibfnamefont {J.}~\bibnamefont {Emerson}},\ and\ \bibinfo
  {author} {\bibfnamefont {I.}~\bibnamefont {Siddiqi}},\ }\bibfield  {title}
  {\bibinfo {title} {\href{https://arxiv.org/abs/2010.00215}{Randomized
  compiling for scalable quantum computing on a noisy superconducting quantum
  processor}},\ }\href@noop {} {\  (\bibinfo {year} {2020})},\ \Eprint
  {https://arxiv.org/abs/2010.00215} {arXiv:2010.00215 [quant-ph]} \BibitemShut
  {NoStop}%
\bibitem [{\citenamefont {{Ding}}\ \emph {et~al.}(2020)\citenamefont {{Ding}},
  \citenamefont {{Gokhale}}, \citenamefont {{Fuhui Lin}}, \citenamefont
  {{Rines}}, \citenamefont {{Propson}},\ and\ \citenamefont
  {{Chong}}}]{crosst2}%
  \BibitemOpen
  \bibfield  {author} {\bibinfo {author} {\bibfnamefont {Y.}~\bibnamefont
  {{Ding}}}, \bibinfo {author} {\bibfnamefont {P.}~\bibnamefont {{Gokhale}}},
  \bibinfo {author} {\bibfnamefont {S.}~\bibnamefont {{Fuhui Lin}}}, \bibinfo
  {author} {\bibfnamefont {R.}~\bibnamefont {{Rines}}}, \bibinfo {author}
  {\bibfnamefont {T.}~\bibnamefont {{Propson}}},\ and\ \bibinfo {author}
  {\bibfnamefont {F.~T.}\ \bibnamefont {{Chong}}},\ }\bibfield  {title}
  {\bibinfo {title} {\href{https://arxiv.org/abs/2008.09503}{Systematic
  Crosstalk Mitigation for Superconducting Qubits via Frequency-Aware
  Compilation}},\ }\href@noop {} {\bibfield  {journal} {\bibinfo  {journal}
  {arXiv e-prints}\ ,\ \bibinfo {eid} {arXiv:2008.09503}} (\bibinfo {year}
  {2020})},\ \Eprint {https://arxiv.org/abs/2008.09503} {arXiv:2008.09503
  [quant-ph]} \BibitemShut {NoStop}%
\bibitem [{\citenamefont {Yousefjani}\ and\ \citenamefont
  {Bayat}(2020)}]{crosst3}%
  \BibitemOpen
  \bibfield  {author} {\bibinfo {author} {\bibfnamefont {R.}~\bibnamefont
  {Yousefjani}}\ and\ \bibinfo {author} {\bibfnamefont {A.}~\bibnamefont
  {Bayat}},\ }\bibfield  {title} {\bibinfo {title}
  {\href{https://arxiv.org/abs/2008.12771}{Parallel entangling gate operations
  and two-way quantum communication in spin chains}},\ }\href@noop {} {\
  (\bibinfo {year} {2020})},\ \Eprint {https://arxiv.org/abs/2008.12771}
  {arXiv:2008.12771 [quant-ph]} \BibitemShut {NoStop}%
\bibitem [{\citenamefont {{Willsch}}(2020)}]{crosst4}%
  \BibitemOpen
  \bibfield  {author} {\bibinfo {author} {\bibfnamefont {D.}~\bibnamefont
  {{Willsch}}},\ }\bibfield  {title} {\bibinfo {title}
  {\href{https://arxiv.org/abs/2008.13490}{Supercomputer simulations of
  transmon quantum computers}},\ }\href@noop {} {\bibfield  {journal} {\bibinfo
   {journal} {arXiv e-prints}\ ,\ \bibinfo {eid} {arXiv:2008.13490}} (\bibinfo
  {year} {2020})},\ \Eprint {https://arxiv.org/abs/2008.13490}
  {arXiv:2008.13490 [quant-ph]} \BibitemShut {NoStop}%
\bibitem [{\citenamefont {Winick}\ \emph {et~al.}(2020)\citenamefont {Winick},
  \citenamefont {Wallman},\ and\ \citenamefont {Emerson}}]{crosst5}%
  \BibitemOpen
  \bibfield  {author} {\bibinfo {author} {\bibfnamefont {A.}~\bibnamefont
  {Winick}}, \bibinfo {author} {\bibfnamefont {J.~J.}\ \bibnamefont
  {Wallman}},\ and\ \bibinfo {author} {\bibfnamefont {J.}~\bibnamefont
  {Emerson}},\ }\bibfield  {title} {\bibinfo {title}
  {\href{https://arxiv.org/abs/2006.09596}{Simulating and mitigating
  crosstalk}},\ }\href@noop {} {\  (\bibinfo {year} {2020})},\ \Eprint
  {https://arxiv.org/abs/2006.09596} {arXiv:2006.09596 [quant-ph]} \BibitemShut
  {NoStop}%
\bibitem [{\citenamefont {Xu}\ \emph {et~al.}(2020)\citenamefont {Xu},
  \citenamefont {Chu}, \citenamefont {Yuan}, \citenamefont {Qiu}, \citenamefont
  {Zhou}, \citenamefont {Zhang}, \citenamefont {Tan}, \citenamefont {Yu},
  \citenamefont {Liu}, \citenamefont {Li}, \citenamefont {Yan},\ and\
  \citenamefont {Yu}}]{crosst6}%
  \BibitemOpen
  \bibfield  {author} {\bibinfo {author} {\bibfnamefont {Y.}~\bibnamefont
  {Xu}}, \bibinfo {author} {\bibfnamefont {J.}~\bibnamefont {Chu}}, \bibinfo
  {author} {\bibfnamefont {J.}~\bibnamefont {Yuan}}, \bibinfo {author}
  {\bibfnamefont {J.}~\bibnamefont {Qiu}}, \bibinfo {author} {\bibfnamefont
  {Y.}~\bibnamefont {Zhou}}, \bibinfo {author} {\bibfnamefont {L.}~\bibnamefont
  {Zhang}}, \bibinfo {author} {\bibfnamefont {X.}~\bibnamefont {Tan}}, \bibinfo
  {author} {\bibfnamefont {Y.}~\bibnamefont {Yu}}, \bibinfo {author}
  {\bibfnamefont {S.}~\bibnamefont {Liu}}, \bibinfo {author} {\bibfnamefont
  {J.}~\bibnamefont {Li}}, \bibinfo {author} {\bibfnamefont {F.}~\bibnamefont
  {Yan}},\ and\ \bibinfo {author} {\bibfnamefont {D.}~\bibnamefont {Yu}},\
  }\bibfield  {title} {\bibinfo {title}
  {\href{https://arxiv.org/abs/2006.11860}{High-fidelity, high-scalability
  two-qubit gate scheme for superconducting qubits}},\ }\href@noop {} {\
  (\bibinfo {year} {2020})},\ \Eprint {https://arxiv.org/abs/2006.11860}
  {arXiv:2006.11860 [quant-ph]} \BibitemShut {NoStop}%
\bibitem [{\citenamefont {Collodo}\ \emph {et~al.}(2020)\citenamefont
  {Collodo}, \citenamefont {Herrmann}, \citenamefont {Lacroix}, \citenamefont
  {Andersen}, \citenamefont {Remm}, \citenamefont {Lazar}, \citenamefont
  {Besse}, \citenamefont {Walter}, \citenamefont {Wallraff},\ and\
  \citenamefont {Eichler}}]{crosst7}%
  \BibitemOpen
  \bibfield  {author} {\bibinfo {author} {\bibfnamefont {M.~C.}\ \bibnamefont
  {Collodo}}, \bibinfo {author} {\bibfnamefont {J.}~\bibnamefont {Herrmann}},
  \bibinfo {author} {\bibfnamefont {N.}~\bibnamefont {Lacroix}}, \bibinfo
  {author} {\bibfnamefont {C.~K.}\ \bibnamefont {Andersen}}, \bibinfo {author}
  {\bibfnamefont {A.}~\bibnamefont {Remm}}, \bibinfo {author} {\bibfnamefont
  {S.}~\bibnamefont {Lazar}}, \bibinfo {author} {\bibfnamefont {J.-C.}\
  \bibnamefont {Besse}}, \bibinfo {author} {\bibfnamefont {T.}~\bibnamefont
  {Walter}}, \bibinfo {author} {\bibfnamefont {A.}~\bibnamefont {Wallraff}},\
  and\ \bibinfo {author} {\bibfnamefont {C.}~\bibnamefont {Eichler}},\
  }\href@noop {} {\bibinfo {title}
  {\href{https://arxiv.org/abs/2005.08863}{Implementation of Conditional-Phase
  Gates based on tunable ZZ-Interactions}}} (\bibinfo {year} {2020}),\ \Eprint
  {https://arxiv.org/abs/2005.08863} {arXiv:2005.08863 [quant-ph]} \BibitemShut
  {NoStop}%
\bibitem [{\citenamefont {Krinner}\ \emph {et~al.}(2020)\citenamefont
  {Krinner}, \citenamefont {Lazar}, \citenamefont {Remm}, \citenamefont
  {Andersen}, \citenamefont {Lacroix}, \citenamefont {Norris}, \citenamefont
  {Hellings}, \citenamefont {Gabureac}, \citenamefont {Eichler},\ and\
  \citenamefont {Wallraff}}]{crosst8}%
  \BibitemOpen
  \bibfield  {author} {\bibinfo {author} {\bibfnamefont {S.}~\bibnamefont
  {Krinner}}, \bibinfo {author} {\bibfnamefont {S.}~\bibnamefont {Lazar}},
  \bibinfo {author} {\bibfnamefont {A.}~\bibnamefont {Remm}}, \bibinfo {author}
  {\bibfnamefont {C.}~\bibnamefont {Andersen}}, \bibinfo {author}
  {\bibfnamefont {N.}~\bibnamefont {Lacroix}}, \bibinfo {author} {\bibfnamefont
  {G.}~\bibnamefont {Norris}}, \bibinfo {author} {\bibfnamefont
  {C.}~\bibnamefont {Hellings}}, \bibinfo {author} {\bibfnamefont
  {M.}~\bibnamefont {Gabureac}}, \bibinfo {author} {\bibfnamefont
  {C.}~\bibnamefont {Eichler}},\ and\ \bibinfo {author} {\bibfnamefont
  {A.}~\bibnamefont {Wallraff}},\ }\bibfield  {title} {\bibinfo {title}
  {Benchmarking coherent errors in controlled-phase gates due to spectator
  qubits},\ }\href {https://doi.org/10.1103/PhysRevApplied.14.024042}
  {\bibfield  {journal} {\bibinfo  {journal} {Phys. Rev. Applied}\ }\textbf
  {\bibinfo {volume} {14}},\ \bibinfo {pages} {024042} (\bibinfo {year}
  {2020})}\BibitemShut {NoStop}%
\bibitem [{\citenamefont {Throckmorton}\ and\ \citenamefont
  {Das~Sarma}(2020)}]{crosst9}%
  \BibitemOpen
  \bibfield  {author} {\bibinfo {author} {\bibfnamefont {R.~E.}\ \bibnamefont
  {Throckmorton}}\ and\ \bibinfo {author} {\bibfnamefont {S.}~\bibnamefont
  {Das~Sarma}},\ }\bibfield  {title} {\bibinfo {title} {Fidelity of a sequence
  of swap operations on a spin chain},\ }\href
  {https://doi.org/10.1103/PhysRevB.102.035439} {\bibfield  {journal} {\bibinfo
   {journal} {Phys. Rev. B}\ }\textbf {\bibinfo {volume} {102}},\ \bibinfo
  {pages} {035439} (\bibinfo {year} {2020})}\BibitemShut {NoStop}%
\bibitem [{\citenamefont {McKay}\ \emph {et~al.}(2020)\citenamefont {McKay},
  \citenamefont {Cross}, \citenamefont {Wood},\ and\ \citenamefont
  {Gambetta}}]{crosst10}%
  \BibitemOpen
  \bibfield  {author} {\bibinfo {author} {\bibfnamefont {D.~C.}\ \bibnamefont
  {McKay}}, \bibinfo {author} {\bibfnamefont {A.~W.}\ \bibnamefont {Cross}},
  \bibinfo {author} {\bibfnamefont {C.~J.}\ \bibnamefont {Wood}},\ and\
  \bibinfo {author} {\bibfnamefont {J.~M.}\ \bibnamefont {Gambetta}},\
  }\bibfield  {title} {\bibinfo {title}
  {\href{https://arxiv.org/abs/2003.02354}{Correlated Randomized
  Benchmarking}},\ }\href@noop {} {\  (\bibinfo {year} {2020})},\ \Eprint
  {https://arxiv.org/abs/2003.02354} {arXiv:2003.02354 [quant-ph]} \BibitemShut
  {NoStop}%
\bibitem [{\citenamefont {Ku}\ \emph {et~al.}(2020)\citenamefont {Ku},
  \citenamefont {Xu}, \citenamefont {Brink}, \citenamefont {McKay},
  \citenamefont {Hertzberg}, \citenamefont {Ansari},\ and\ \citenamefont
  {Plourde}}]{crosst11}%
  \BibitemOpen
  \bibfield  {author} {\bibinfo {author} {\bibfnamefont {J.}~\bibnamefont
  {Ku}}, \bibinfo {author} {\bibfnamefont {X.}~\bibnamefont {Xu}}, \bibinfo
  {author} {\bibfnamefont {M.}~\bibnamefont {Brink}}, \bibinfo {author}
  {\bibfnamefont {D.~C.}\ \bibnamefont {McKay}}, \bibinfo {author}
  {\bibfnamefont {J.~B.}\ \bibnamefont {Hertzberg}}, \bibinfo {author}
  {\bibfnamefont {M.~H.}\ \bibnamefont {Ansari}},\ and\ \bibinfo {author}
  {\bibfnamefont {B.~L.~T.}\ \bibnamefont {Plourde}},\ }\bibfield  {title}
  {\bibinfo {title} {\href{https://arxiv.org/abs/2003.02775}{Suppression of
  Unwanted $ZZ$ Interactions in a Hybrid Two-Qubit System}},\ }\href@noop {} {\
   (\bibinfo {year} {2020})},\ \Eprint {https://arxiv.org/abs/2003.02775}
  {arXiv:2003.02775 [quant-ph]} \BibitemShut {NoStop}%
\bibitem [{\citenamefont {Han}\ \emph {et~al.}(2020)\citenamefont {Han},
  \citenamefont {Cai}, \citenamefont {Li}, \citenamefont {Wu}, \citenamefont
  {Ma}, \citenamefont {Ma}, \citenamefont {Wang}, \citenamefont {Zhang},
  \citenamefont {Song},\ and\ \citenamefont {Duan}}]{crosst12}%
  \BibitemOpen
  \bibfield  {author} {\bibinfo {author} {\bibfnamefont {X.~Y.}\ \bibnamefont
  {Han}}, \bibinfo {author} {\bibfnamefont {T.~Q.}\ \bibnamefont {Cai}},
  \bibinfo {author} {\bibfnamefont {X.~G.}\ \bibnamefont {Li}}, \bibinfo
  {author} {\bibfnamefont {Y.~K.}\ \bibnamefont {Wu}}, \bibinfo {author}
  {\bibfnamefont {Y.~W.}\ \bibnamefont {Ma}}, \bibinfo {author} {\bibfnamefont
  {Y.~L.}\ \bibnamefont {Ma}}, \bibinfo {author} {\bibfnamefont {J.~H.}\
  \bibnamefont {Wang}}, \bibinfo {author} {\bibfnamefont {H.~Y.}\ \bibnamefont
  {Zhang}}, \bibinfo {author} {\bibfnamefont {Y.~P.}\ \bibnamefont {Song}},\
  and\ \bibinfo {author} {\bibfnamefont {L.~M.}\ \bibnamefont {Duan}},\
  }\bibfield  {title} {\bibinfo {title} {Error analysis in suppression of
  unwanted qubit interactions for a parametric gate in a tunable
  superconducting circuit},\ }\href
  {https://doi.org/10.1103/PhysRevA.102.022619} {\bibfield  {journal} {\bibinfo
   {journal} {Phys. Rev. A}\ }\textbf {\bibinfo {volume} {102}},\ \bibinfo
  {pages} {022619} (\bibinfo {year} {2020})}\BibitemShut {NoStop}%
\bibitem [{\citenamefont {Huang}\ \emph {et~al.}(2020)\citenamefont {Huang},
  \citenamefont {Ni}, \citenamefont {Zhang}, \citenamefont {Newman},
  \citenamefont {Ding}, \citenamefont {Gao}, \citenamefont {Wang},
  \citenamefont {Zhao}, \citenamefont {Wu}, \citenamefont {Zhang},
  \citenamefont {Deng}, \citenamefont {Ku}, \citenamefont {Chen},\ and\
  \citenamefont {Shi}}]{crosst13}%
  \BibitemOpen
  \bibfield  {author} {\bibinfo {author} {\bibfnamefont {C.}~\bibnamefont
  {Huang}}, \bibinfo {author} {\bibfnamefont {X.}~\bibnamefont {Ni}}, \bibinfo
  {author} {\bibfnamefont {F.}~\bibnamefont {Zhang}}, \bibinfo {author}
  {\bibfnamefont {M.}~\bibnamefont {Newman}}, \bibinfo {author} {\bibfnamefont
  {D.}~\bibnamefont {Ding}}, \bibinfo {author} {\bibfnamefont {X.}~\bibnamefont
  {Gao}}, \bibinfo {author} {\bibfnamefont {T.}~\bibnamefont {Wang}}, \bibinfo
  {author} {\bibfnamefont {H.-H.}\ \bibnamefont {Zhao}}, \bibinfo {author}
  {\bibfnamefont {F.}~\bibnamefont {Wu}}, \bibinfo {author} {\bibfnamefont
  {G.}~\bibnamefont {Zhang}}, \bibinfo {author} {\bibfnamefont
  {C.}~\bibnamefont {Deng}}, \bibinfo {author} {\bibfnamefont {H.-S.}\
  \bibnamefont {Ku}}, \bibinfo {author} {\bibfnamefont {J.}~\bibnamefont
  {Chen}},\ and\ \bibinfo {author} {\bibfnamefont {Y.}~\bibnamefont {Shi}},\
  }\bibfield  {title} {\bibinfo {title}
  {\href{https://arxiv.org/abs/2002.08918}{Alibaba Cloud Quantum Development
  Platform: Surface Code Simulations with Crosstalk}},\ }\href@noop {} {\
  (\bibinfo {year} {2020})},\ \Eprint {https://arxiv.org/abs/2002.08918}
  {arXiv:2002.08918 [quant-ph]} \BibitemShut {NoStop}%
\bibitem [{\citenamefont {Kono}\ \emph {et~al.}(2020)\citenamefont {Kono},
  \citenamefont {Koshino}, \citenamefont {Lachance-Quirion}, \citenamefont {van
  Loo}, \citenamefont {Tabuchi}, \citenamefont {Noguchi},\ and\ \citenamefont
  {Nakamura}}]{crosst14}%
  \BibitemOpen
  \bibfield  {author} {\bibinfo {author} {\bibfnamefont {S.}~\bibnamefont
  {Kono}}, \bibinfo {author} {\bibfnamefont {K.}~\bibnamefont {Koshino}},
  \bibinfo {author} {\bibfnamefont {D.}~\bibnamefont {Lachance-Quirion}},
  \bibinfo {author} {\bibfnamefont {A.~F.}\ \bibnamefont {van Loo}}, \bibinfo
  {author} {\bibfnamefont {Y.}~\bibnamefont {Tabuchi}}, \bibinfo {author}
  {\bibfnamefont {A.}~\bibnamefont {Noguchi}},\ and\ \bibinfo {author}
  {\bibfnamefont {Y.}~\bibnamefont {Nakamura}},\ }\bibfield  {title} {\bibinfo
  {title} {Breaking the trade-off between fast control and long lifetime of a
  superconducting qubit},\ }\href {https://doi.org/10.1038/s41467-020-17511-y}
  {\bibfield  {journal} {\bibinfo  {journal} {Nature Communications}\ }\textbf
  {\bibinfo {volume} {11}},\ \bibinfo {pages} {3683} (\bibinfo {year}
  {2020})}\BibitemShut {NoStop}%
\bibitem [{\citenamefont {Murali}\ \emph {et~al.}(2020)\citenamefont {Murali},
  \citenamefont {Mckay}, \citenamefont {Martonosi},\ and\ \citenamefont
  {Javadi-Abhari}}]{crosst15}%
  \BibitemOpen
  \bibfield  {author} {\bibinfo {author} {\bibfnamefont {P.}~\bibnamefont
  {Murali}}, \bibinfo {author} {\bibfnamefont {D.~C.}\ \bibnamefont {Mckay}},
  \bibinfo {author} {\bibfnamefont {M.}~\bibnamefont {Martonosi}},\ and\
  \bibinfo {author} {\bibfnamefont {A.}~\bibnamefont {Javadi-Abhari}},\
  }\bibfield  {title} {\bibinfo {title} {Software mitigation of crosstalk on
  noisy intermediate-scale quantum computers}\ }\href
  {https://doi.org/10.1145/3373376.3378477} {10.1145/3373376.3378477} (\bibinfo
  {year} {2020})\BibitemShut {NoStop}%
\bibitem [{\citenamefont {Sun}\ \emph {et~al.}(2020)\citenamefont {Sun},
  \citenamefont {Yuan}, \citenamefont {Tsunoda}, \citenamefont {Vedral},
  \citenamefont {Bejamin},\ and\ \citenamefont {Endo}}]{crosst16}%
  \BibitemOpen
  \bibfield  {author} {\bibinfo {author} {\bibfnamefont {J.}~\bibnamefont
  {Sun}}, \bibinfo {author} {\bibfnamefont {X.}~\bibnamefont {Yuan}}, \bibinfo
  {author} {\bibfnamefont {T.}~\bibnamefont {Tsunoda}}, \bibinfo {author}
  {\bibfnamefont {V.}~\bibnamefont {Vedral}}, \bibinfo {author} {\bibfnamefont
  {S.~C.}\ \bibnamefont {Bejamin}},\ and\ \bibinfo {author} {\bibfnamefont
  {S.}~\bibnamefont {Endo}},\ }\bibfield  {title} {\bibinfo {title}
  {\href{https://arxiv.org/abs/2001.04891}{Mitigating realistic noise in
  practical noisy intermediate-scale quantum devices}},\ }\href@noop {} {\
  (\bibinfo {year} {2020})},\ \Eprint {https://arxiv.org/abs/2001.04891}
  {arXiv:2001.04891 [quant-ph]} \BibitemShut {NoStop}%
\bibitem [{\citenamefont {G\"ung\"ord\"u}\ and\ \citenamefont
  {Kestner}(2020)}]{crosst17}%
  \BibitemOpen
  \bibfield  {author} {\bibinfo {author} {\bibfnamefont {U.}~\bibnamefont
  {G\"ung\"ord\"u}}\ and\ \bibinfo {author} {\bibfnamefont {J.~P.}\
  \bibnamefont {Kestner}},\ }\bibfield  {title} {\bibinfo {title} {Robust
  implementation of quantum gates despite always-on exchange coupling in
  silicon double quantum dots},\ }\href
  {https://doi.org/10.1103/PhysRevB.101.155301} {\bibfield  {journal} {\bibinfo
   {journal} {Phys. Rev. B}\ }\textbf {\bibinfo {volume} {101}},\ \bibinfo
  {pages} {155301} (\bibinfo {year} {2020})}\BibitemShut {NoStop}%
\bibitem [{\citenamefont {Qi}\ and\ \citenamefont {Jing}(2020)}]{crosst18}%
  \BibitemOpen
  \bibfield  {author} {\bibinfo {author} {\bibfnamefont {S.-f.}\ \bibnamefont
  {Qi}}\ and\ \bibinfo {author} {\bibfnamefont {J.}~\bibnamefont {Jing}},\
  }\bibfield  {title} {\bibinfo {title} {Generating noon states in circuit qed
  using a multiphoton resonance in the presence of counter-rotating
  interactions},\ }\href {https://doi.org/10.1103/PhysRevA.101.033809}
  {\bibfield  {journal} {\bibinfo  {journal} {Phys. Rev. A}\ }\textbf {\bibinfo
  {volume} {101}},\ \bibinfo {pages} {033809} (\bibinfo {year}
  {2020})}\BibitemShut {NoStop}%
\bibitem [{\citenamefont {Debroy}\ \emph {et~al.}(2020)\citenamefont {Debroy},
  \citenamefont {Li}, \citenamefont {Huang},\ and\ \citenamefont
  {Brown}}]{crosst19}%
  \BibitemOpen
  \bibfield  {author} {\bibinfo {author} {\bibfnamefont {D.~M.}\ \bibnamefont
  {Debroy}}, \bibinfo {author} {\bibfnamefont {M.}~\bibnamefont {Li}}, \bibinfo
  {author} {\bibfnamefont {S.}~\bibnamefont {Huang}},\ and\ \bibinfo {author}
  {\bibfnamefont {K.~R.}\ \bibnamefont {Brown}},\ }\bibfield  {title} {\bibinfo
  {title} {\href{https://arxiv.org/abs/1910.08495}{Logical Performance of 9
  Qubit Compass Codes in Ion Traps with Crosstalk Errors}},\ }\href@noop {} {\
  (\bibinfo {year} {2020})},\ \Eprint {https://arxiv.org/abs/1910.08495}
  {arXiv:1910.08495 [quant-ph]} \BibitemShut {NoStop}%
\bibitem [{\citenamefont {Sarovar}\ \emph {et~al.}(2020)\citenamefont
  {Sarovar}, \citenamefont {Proctor}, \citenamefont {Rudinger}, \citenamefont
  {Young}, \citenamefont {Nielsen},\ and\ \citenamefont
  {Blume-Kohout}}]{crosst20}%
  \BibitemOpen
  \bibfield  {author} {\bibinfo {author} {\bibfnamefont {M.}~\bibnamefont
  {Sarovar}}, \bibinfo {author} {\bibfnamefont {T.}~\bibnamefont {Proctor}},
  \bibinfo {author} {\bibfnamefont {K.}~\bibnamefont {Rudinger}}, \bibinfo
  {author} {\bibfnamefont {K.}~\bibnamefont {Young}}, \bibinfo {author}
  {\bibfnamefont {E.}~\bibnamefont {Nielsen}},\ and\ \bibinfo {author}
  {\bibfnamefont {R.}~\bibnamefont {Blume-Kohout}},\ }\bibfield  {title}
  {\bibinfo {title} {Detecting crosstalk errors in quantum information
  processors},\ }\href {https://doi.org/10.22331/q-2020-09-11-321} {\bibfield
  {journal} {\bibinfo  {journal} {{Quantum}}\ }\textbf {\bibinfo {volume}
  {4}},\ \bibinfo {pages} {321} (\bibinfo {year} {2020})}\BibitemShut {NoStop}%
\bibitem [{\citenamefont {Chamberland}\ \emph {et~al.}(2020)\citenamefont
  {Chamberland}, \citenamefont {Zhu}, \citenamefont {Yoder}, \citenamefont
  {Hertzberg},\ and\ \citenamefont {Cross}}]{crosst21}%
  \BibitemOpen
  \bibfield  {author} {\bibinfo {author} {\bibfnamefont {C.}~\bibnamefont
  {Chamberland}}, \bibinfo {author} {\bibfnamefont {G.}~\bibnamefont {Zhu}},
  \bibinfo {author} {\bibfnamefont {T.~J.}\ \bibnamefont {Yoder}}, \bibinfo
  {author} {\bibfnamefont {J.~B.}\ \bibnamefont {Hertzberg}},\ and\ \bibinfo
  {author} {\bibfnamefont {A.~W.}\ \bibnamefont {Cross}},\ }\bibfield  {title}
  {\bibinfo {title} {Topological and subsystem codes on low-degree graphs with
  flag qubits},\ }\href {https://doi.org/10.1103/PhysRevX.10.011022} {\bibfield
   {journal} {\bibinfo  {journal} {Phys. Rev. X}\ }\textbf {\bibinfo {volume}
  {10}},\ \bibinfo {pages} {011022} (\bibinfo {year} {2020})}\BibitemShut
  {NoStop}%
\bibitem [{\citenamefont {Majumder}\ \emph {et~al.}(2020)\citenamefont
  {Majumder}, \citenamefont {Andreta~de Castro},\ and\ \citenamefont
  {Brown}}]{crosst22}%
  \BibitemOpen
  \bibfield  {author} {\bibinfo {author} {\bibfnamefont {S.}~\bibnamefont
  {Majumder}}, \bibinfo {author} {\bibfnamefont {L.}~\bibnamefont {Andreta~de
  Castro}},\ and\ \bibinfo {author} {\bibfnamefont {K.~R.}\ \bibnamefont
  {Brown}},\ }\bibfield  {title} {\bibinfo {title} {Real-time calibration with
  spectator qubits},\ }\href {https://doi.org/10.1038/s41534-020-0251-y}
  {\bibfield  {journal} {\bibinfo  {journal} {npj Quantum Information}\
  }\textbf {\bibinfo {volume} {6}},\ \bibinfo {pages} {19} (\bibinfo {year}
  {2020})}\BibitemShut {NoStop}%
\bibitem [{\citenamefont {Harper}\ \emph {et~al.}(2020)\citenamefont {Harper},
  \citenamefont {Flammia},\ and\ \citenamefont {Wallman}}]{crosst23}%
  \BibitemOpen
  \bibfield  {author} {\bibinfo {author} {\bibfnamefont {R.}~\bibnamefont
  {Harper}}, \bibinfo {author} {\bibfnamefont {S.~T.}\ \bibnamefont
  {Flammia}},\ and\ \bibinfo {author} {\bibfnamefont {J.~J.}\ \bibnamefont
  {Wallman}},\ }\bibfield  {title} {\bibinfo {title} {Efficient learning of
  quantum noise},\ }\bibfield  {journal} {\bibinfo  {journal} {Nature Physics}\
  }\href {https://doi.org/10.1038/s41567-020-0992-8}
  {10.1038/s41567-020-0992-8} (\bibinfo {year} {2020})\BibitemShut {NoStop}%
\bibitem [{\citenamefont {Ash-Saki}\ \emph {et~al.}(2020)\citenamefont
  {Ash-Saki}, \citenamefont {Alam},\ and\ \citenamefont {Ghosh}}]{crosst24}%
  \BibitemOpen
  \bibfield  {author} {\bibinfo {author} {\bibfnamefont {A.}~\bibnamefont
  {Ash-Saki}}, \bibinfo {author} {\bibfnamefont {M.}~\bibnamefont {Alam}},\
  and\ \bibinfo {author} {\bibfnamefont {S.}~\bibnamefont {Ghosh}},\ }\bibfield
   {title} {\bibinfo {title} {Analysis of crosstalk in nisq devices and
  security implications in multi-programming regime},\ }\href
  {https://doi.org/10.1145/3370748.3406570} {\ ,\ \bibinfo {pages} {25}
  (\bibinfo {year} {2020})}\BibitemShut {NoStop}%
\bibitem [{\citenamefont {Shaw}(2019{\natexlab{a}})}]{CLAWEPost0}%
  \BibitemOpen
  \bibfield  {author} {\bibinfo {author} {\bibfnamefont {A.}~\bibnamefont
  {Shaw}},\ }\bibfield  {title} {\bibinfo {title}
  {\href{https://drive.google.com/file/d/1pKUFua_Z1Zohec2ZpWlCD9N7OuLRaG6B/view?usp=sharing}{A
  New Noise Mitigation Scheme for Computations on NISQ-era Devices}}} (\bibinfo
  {year} {June 19, 2019}{\natexlab{a}}),\ \bibinfo {note} {{UMD} NISQ/TQC
  Conference}\BibitemShut {NoStop}%
\bibitem [{\citenamefont {Shaw}(2019{\natexlab{b}})}]{CLAWEConference0}%
  \BibitemOpen
  \bibfield  {author} {\bibinfo {author} {\bibfnamefont {A.}~\bibnamefont
  {Shaw}},\ }\bibfield  {title} {\bibinfo {title}
  {\href{https://drive.google.com/file/d/1CRecVS1jPfQJJEhp96PXMGmOXs-9RgWZ/view?usp=sharing}{Extending
  the Range of Near-Term Quantum Simulation with Classical-Quantum Hybrid
  Methods}}} (\bibinfo {year} {October 13, 2019}{\natexlab{b}}),\ \bibinfo
  {note} {{Fall 2019} DNP-APS Meeting}\BibitemShut {NoStop}%
\bibitem [{\citenamefont {Shaw}(2019{\natexlab{c}})}]{CLAWEPost1}%
  \BibitemOpen
  \bibfield  {author} {\bibinfo {author} {\bibfnamefont {A.}~\bibnamefont
  {Shaw}},\ }\bibfield  {title} {\bibinfo {title}
  {\href{https://drive.google.com/file/d/110NYehQ7rORmzqZeom_nPaPUim89VM2T/view?usp=sharing}{A
  Noise Mitigation Scheme for NISQ Hardware}}} (\bibinfo {year} {December 04,
  2019}{\natexlab{c}}),\ \bibinfo {note} {{FAR-QC} Conference}\BibitemShut
  {NoStop}%
\bibitem [{\citenamefont {Shaw}(2020{\natexlab{a}})}]{CLAWEPresent1}%
  \BibitemOpen
  \bibfield  {author} {\bibinfo {author} {\bibfnamefont {A.}~\bibnamefont
  {Shaw}},\ }\bibfield  {title} {\bibinfo {title}
  {\href{http://bit.ly/June2020CLAWETalk}{Classical-Quantum Noise Mitigation
  for NISQ Devices}}} (\bibinfo {year} {June 23, 2020}{\natexlab{a}}),\
  \bibinfo {note} {{Quantum} Computing Applications Team Meeting}\BibitemShut
  {NoStop}%
\bibitem [{\citenamefont {Paraoanu}(2006)}]{transmEnt0}%
  \BibitemOpen
  \bibfield  {author} {\bibinfo {author} {\bibfnamefont {G.~S.}\ \bibnamefont
  {Paraoanu}},\ }\bibfield  {title} {\bibinfo {title} {Microwave-induced
  coupling of superconducting qubits},\ }\href
  {https://doi.org/10.1103/PhysRevB.74.140504} {\bibfield  {journal} {\bibinfo
  {journal} {Phys. Rev. B}\ }\textbf {\bibinfo {volume} {74}},\ \bibinfo
  {pages} {140504} (\bibinfo {year} {2006})}\BibitemShut {NoStop}%
\bibitem [{\citenamefont {Li}\ \emph {et~al.}(2008)\citenamefont {Li},
  \citenamefont {Chalapat},\ and\ \citenamefont {Paraoanu}}]{transmEnt1}%
  \BibitemOpen
  \bibfield  {author} {\bibinfo {author} {\bibfnamefont {J.}~\bibnamefont
  {Li}}, \bibinfo {author} {\bibfnamefont {K.}~\bibnamefont {Chalapat}},\ and\
  \bibinfo {author} {\bibfnamefont {G.~S.}\ \bibnamefont {Paraoanu}},\
  }\bibfield  {title} {\bibinfo {title} {Entanglement of superconducting qubits
  via microwave fields: Classical and quantum regimes},\ }\href
  {https://doi.org/10.1103/PhysRevB.78.064503} {\bibfield  {journal} {\bibinfo
  {journal} {Phys. Rev. B}\ }\textbf {\bibinfo {volume} {78}},\ \bibinfo
  {pages} {064503} (\bibinfo {year} {2008})}\BibitemShut {NoStop}%
\bibitem [{\citenamefont {Li}\ \emph {et~al.}(2009)\citenamefont {Li},
  \citenamefont {Chalapat},\ and\ \citenamefont {Paraoanu}}]{transmEnt2}%
  \BibitemOpen
  \bibfield  {author} {\bibinfo {author} {\bibfnamefont {J.}~\bibnamefont
  {Li}}, \bibinfo {author} {\bibfnamefont {K.}~\bibnamefont {Chalapat}},\ and\
  \bibinfo {author} {\bibfnamefont {G.~S.}\ \bibnamefont {Paraoanu}},\
  }\bibfield  {title} {\bibinfo {title} {Measurement-induced entanglement of
  two superconducting qubits},\ }\href
  {https://doi.org/10.1088/1742-6596/150/2/022051} {\bibfield  {journal}
  {\bibinfo  {journal} {Journal of Physics: Conference Series}\ }\textbf
  {\bibinfo {volume} {150}},\ \bibinfo {pages} {022051} (\bibinfo {year}
  {2009})}\BibitemShut {NoStop}%
\bibitem [{\citenamefont {Kurpas}\ \emph {et~al.}(2009)\citenamefont {Kurpas},
  \citenamefont {Dajka},\ and\ \citenamefont {Zipper}}]{transmEnt3}%
  \BibitemOpen
  \bibfield  {author} {\bibinfo {author} {\bibfnamefont {M.}~\bibnamefont
  {Kurpas}}, \bibinfo {author} {\bibfnamefont {J.}~\bibnamefont {Dajka}},\ and\
  \bibinfo {author} {\bibfnamefont {E.}~\bibnamefont {Zipper}},\ }\bibfield
  {title} {\bibinfo {title} {Entanglement of qubits via a nonlinear
  resonator},\ }\href {https://doi.org/10.1088/0953-8984/21/23/235602}
  {\bibfield  {journal} {\bibinfo  {journal} {Journal of Physics: Condensed
  Matter}\ }\textbf {\bibinfo {volume} {21}},\ \bibinfo {pages} {235602}
  (\bibinfo {year} {2009})}\BibitemShut {NoStop}%
\bibitem [{\citenamefont {Li}\ and\ \citenamefont
  {Paraoanu}(2009)}]{transmEnt4}%
  \BibitemOpen
  \bibfield  {author} {\bibinfo {author} {\bibfnamefont {J.}~\bibnamefont
  {Li}}\ and\ \bibinfo {author} {\bibfnamefont {G.~S.}\ \bibnamefont
  {Paraoanu}},\ }\bibfield  {title} {\bibinfo {title} {Generation and
  propagation of entanglement in driven coupled-qubit systems},\ }\href
  {https://doi.org/10.1088/1367-2630/11/11/113020} {\bibfield  {journal}
  {\bibinfo  {journal} {New Journal of Physics}\ }\textbf {\bibinfo {volume}
  {11}},\ \bibinfo {pages} {113020} (\bibinfo {year} {2009})}\BibitemShut
  {NoStop}%
\bibitem [{\citenamefont {Rigetti}\ and\ \citenamefont
  {Devoret}(2010)}]{transmEnt5}%
  \BibitemOpen
  \bibfield  {author} {\bibinfo {author} {\bibfnamefont {C.}~\bibnamefont
  {Rigetti}}\ and\ \bibinfo {author} {\bibfnamefont {M.}~\bibnamefont
  {Devoret}},\ }\bibfield  {title} {\bibinfo {title} {Fully microwave-tunable
  universal gates in superconducting qubits with linear couplings and fixed
  transition frequencies},\ }\href {https://doi.org/10.1103/PhysRevB.81.134507}
  {\bibfield  {journal} {\bibinfo  {journal} {Phys. Rev. B}\ }\textbf {\bibinfo
  {volume} {81}},\ \bibinfo {pages} {134507} (\bibinfo {year}
  {2010})}\BibitemShut {NoStop}%
\bibitem [{\citenamefont {Leandro}\ \emph {et~al.}()\citenamefont {Leandro},
  \citenamefont {de~Castro}, \citenamefont {Munhoz},\ and\ \citenamefont
  {Semi{\~a}o}}]{transmEnt6}%
  \BibitemOpen
  \bibfield  {author} {\bibinfo {author} {\bibfnamefont {J.~F.}\ \bibnamefont
  {Leandro}}, \bibinfo {author} {\bibfnamefont {A.~S.~M.}\ \bibnamefont
  {de~Castro}}, \bibinfo {author} {\bibfnamefont {P.~P.}\ \bibnamefont
  {Munhoz}},\ and\ \bibinfo {author} {\bibfnamefont {F.~L.}\ \bibnamefont
  {Semi{\~a}o}},\ }\bibfield  {title} {\bibinfo {title} {Active control of
  qubit-qubit entanglement evolution},\ }\href
  {https://doi.org/https://doi.org/10.1016/j.physleta.2010.08.035} {\bibinfo
  {journal} {Physics Letters A}\ ,\ \bibinfo {pages} {4199}}\BibitemShut
  {NoStop}%
\bibitem [{\citenamefont {Chow}\ \emph {et~al.}(2011)\citenamefont {Chow},
  \citenamefont {C\'orcoles}, \citenamefont {Gambetta}, \citenamefont
  {Rigetti}, \citenamefont {Johnson}, \citenamefont {Smolin}, \citenamefont
  {Rozen}, \citenamefont {Keefe}, \citenamefont {Rothwell}, \citenamefont
  {Ketchen},\ and\ \citenamefont {Steffen}}]{transmEnt7}%
  \BibitemOpen
\bibfield  {journal} {  }\bibfield  {author} {\bibinfo {author} {\bibfnamefont
  {J.~M.}\ \bibnamefont {Chow}}, \bibinfo {author} {\bibfnamefont {A.~D.}\
  \bibnamefont {C\'orcoles}}, \bibinfo {author} {\bibfnamefont {J.~M.}\
  \bibnamefont {Gambetta}}, \bibinfo {author} {\bibfnamefont {C.}~\bibnamefont
  {Rigetti}}, \bibinfo {author} {\bibfnamefont {B.~R.}\ \bibnamefont
  {Johnson}}, \bibinfo {author} {\bibfnamefont {J.~A.}\ \bibnamefont {Smolin}},
  \bibinfo {author} {\bibfnamefont {J.~R.}\ \bibnamefont {Rozen}}, \bibinfo
  {author} {\bibfnamefont {G.~A.}\ \bibnamefont {Keefe}}, \bibinfo {author}
  {\bibfnamefont {M.~B.}\ \bibnamefont {Rothwell}}, \bibinfo {author}
  {\bibfnamefont {M.~B.}\ \bibnamefont {Ketchen}},\ and\ \bibinfo {author}
  {\bibfnamefont {M.}~\bibnamefont {Steffen}},\ }\bibfield  {title} {\bibinfo
  {title} {Simple all-microwave entangling gate for fixed-frequency
  superconducting qubits},\ }\href
  {https://doi.org/10.1103/PhysRevLett.107.080502} {\bibfield  {journal}
  {\bibinfo  {journal} {Phys. Rev. Lett.}\ }\textbf {\bibinfo {volume} {107}},\
  \bibinfo {pages} {080502} (\bibinfo {year} {2011})}\BibitemShut {NoStop}%
\bibitem [{\citenamefont {Chaudhry}\ and\ \citenamefont
  {Gong}(2012)}]{transmEnt8}%
  \BibitemOpen
  \bibfield  {author} {\bibinfo {author} {\bibfnamefont {A.~Z.}\ \bibnamefont
  {Chaudhry}}\ and\ \bibinfo {author} {\bibfnamefont {J.}~\bibnamefont
  {Gong}},\ }\bibfield  {title} {\bibinfo {title} {Decoherence control:
  Universal protection of two-qubit states and two-qubit gates using continuous
  driving fields},\ }\href {https://doi.org/10.1103/PhysRevA.85.012315}
  {\bibfield  {journal} {\bibinfo  {journal} {Phys. Rev. A}\ }\textbf {\bibinfo
  {volume} {85}},\ \bibinfo {pages} {012315} (\bibinfo {year}
  {2012})}\BibitemShut {NoStop}%
\bibitem [{\citenamefont {Lü}\ \emph {et~al.}(2012)\citenamefont {Lü},
  \citenamefont {Ashhab}, \citenamefont {Cui}, \citenamefont {Wu},\ and\
  \citenamefont {Nori}}]{transmEnt9}%
  \BibitemOpen
  \bibfield  {author} {\bibinfo {author} {\bibfnamefont {X.-Y.}\ \bibnamefont
  {Lü}}, \bibinfo {author} {\bibfnamefont {S.}~\bibnamefont {Ashhab}},
  \bibinfo {author} {\bibfnamefont {W.}~\bibnamefont {Cui}}, \bibinfo {author}
  {\bibfnamefont {R.}~\bibnamefont {Wu}},\ and\ \bibinfo {author}
  {\bibfnamefont {F.}~\bibnamefont {Nori}},\ }\bibfield  {title} {\bibinfo
  {title} {Two-qubit gate operations in superconducting circuits with strong
  coupling and weak anharmonicity},\ }\href
  {https://doi.org/10.1088/1367-2630/14/7/073041} {\bibfield  {journal}
  {\bibinfo  {journal} {New Journal of Physics}\ }\textbf {\bibinfo {volume}
  {14}},\ \bibinfo {pages} {073041} (\bibinfo {year} {2012})}\BibitemShut
  {NoStop}%
\bibitem [{\citenamefont {de~Groot}\ \emph {et~al.}(2012)\citenamefont
  {de~Groot}, \citenamefont {Ashhab}, \citenamefont {Lupa{\c{s}}cu},
  \citenamefont {DiCarlo}, \citenamefont {Nori}, \citenamefont {Harmans},\ and\
  \citenamefont {Mooij}}]{transmEnt10}%
  \BibitemOpen
  \bibfield  {author} {\bibinfo {author} {\bibfnamefont {P.~C.}\ \bibnamefont
  {de~Groot}}, \bibinfo {author} {\bibfnamefont {S.}~\bibnamefont {Ashhab}},
  \bibinfo {author} {\bibfnamefont {A.}~\bibnamefont {Lupa{\c{s}}cu}}, \bibinfo
  {author} {\bibfnamefont {L.}~\bibnamefont {DiCarlo}}, \bibinfo {author}
  {\bibfnamefont {F.}~\bibnamefont {Nori}}, \bibinfo {author} {\bibfnamefont
  {C.~J. P.~M.}\ \bibnamefont {Harmans}},\ and\ \bibinfo {author}
  {\bibfnamefont {J.~E.}\ \bibnamefont {Mooij}},\ }\bibfield  {title} {\bibinfo
  {title} {Selective darkening of degenerate transitions for implementing
  quantum controlled-{NOT} gates},\ }\href
  {https://doi.org/10.1088/1367-2630/14/7/073038} {\bibfield  {journal}
  {\bibinfo  {journal} {New Journal of Physics}\ }\textbf {\bibinfo {volume}
  {14}},\ \bibinfo {pages} {073038} (\bibinfo {year} {2012})}\BibitemShut
  {NoStop}%
\bibitem [{\citenamefont {Chow}\ \emph {et~al.}(2012)\citenamefont {Chow},
  \citenamefont {Gambetta}, \citenamefont {C\'orcoles}, \citenamefont {Merkel},
  \citenamefont {Smolin}, \citenamefont {Rigetti}, \citenamefont {Poletto},
  \citenamefont {Keefe}, \citenamefont {Rothwell}, \citenamefont {Rozen},
  \citenamefont {Ketchen},\ and\ \citenamefont {Steffen}}]{transmEnt11}%
  \BibitemOpen
  \bibfield  {author} {\bibinfo {author} {\bibfnamefont {J.~M.}\ \bibnamefont
  {Chow}}, \bibinfo {author} {\bibfnamefont {J.~M.}\ \bibnamefont {Gambetta}},
  \bibinfo {author} {\bibfnamefont {A.~D.}\ \bibnamefont {C\'orcoles}},
  \bibinfo {author} {\bibfnamefont {S.~T.}\ \bibnamefont {Merkel}}, \bibinfo
  {author} {\bibfnamefont {J.~A.}\ \bibnamefont {Smolin}}, \bibinfo {author}
  {\bibfnamefont {C.}~\bibnamefont {Rigetti}}, \bibinfo {author} {\bibfnamefont
  {S.}~\bibnamefont {Poletto}}, \bibinfo {author} {\bibfnamefont {G.~A.}\
  \bibnamefont {Keefe}}, \bibinfo {author} {\bibfnamefont {M.~B.}\ \bibnamefont
  {Rothwell}}, \bibinfo {author} {\bibfnamefont {J.~R.}\ \bibnamefont {Rozen}},
  \bibinfo {author} {\bibfnamefont {M.~B.}\ \bibnamefont {Ketchen}},\ and\
  \bibinfo {author} {\bibfnamefont {M.}~\bibnamefont {Steffen}},\ }\bibfield
  {title} {\bibinfo {title} {Universal quantum gate set approaching
  fault-tolerant thresholds with superconducting qubits},\ }\href
  {https://doi.org/10.1103/PhysRevLett.109.060501} {\bibfield  {journal}
  {\bibinfo  {journal} {Phys. Rev. Lett.}\ }\textbf {\bibinfo {volume} {109}},\
  \bibinfo {pages} {060501} (\bibinfo {year} {2012})}\BibitemShut {NoStop}%
\bibitem [{\citenamefont {Poletto}\ \emph {et~al.}(2012)\citenamefont
  {Poletto}, \citenamefont {Gambetta}, \citenamefont {Merkel}, \citenamefont
  {Smolin}, \citenamefont {Chow}, \citenamefont {C\'orcoles}, \citenamefont
  {Keefe}, \citenamefont {Rothwell}, \citenamefont {Rozen}, \citenamefont
  {Abraham}, \citenamefont {Rigetti},\ and\ \citenamefont
  {Steffen}}]{transmEnt12}%
  \BibitemOpen
  \bibfield  {author} {\bibinfo {author} {\bibfnamefont {S.}~\bibnamefont
  {Poletto}}, \bibinfo {author} {\bibfnamefont {J.~M.}\ \bibnamefont
  {Gambetta}}, \bibinfo {author} {\bibfnamefont {S.~T.}\ \bibnamefont
  {Merkel}}, \bibinfo {author} {\bibfnamefont {J.~A.}\ \bibnamefont {Smolin}},
  \bibinfo {author} {\bibfnamefont {J.~M.}\ \bibnamefont {Chow}}, \bibinfo
  {author} {\bibfnamefont {A.~D.}\ \bibnamefont {C\'orcoles}}, \bibinfo
  {author} {\bibfnamefont {G.~A.}\ \bibnamefont {Keefe}}, \bibinfo {author}
  {\bibfnamefont {M.~B.}\ \bibnamefont {Rothwell}}, \bibinfo {author}
  {\bibfnamefont {J.~R.}\ \bibnamefont {Rozen}}, \bibinfo {author}
  {\bibfnamefont {D.~W.}\ \bibnamefont {Abraham}}, \bibinfo {author}
  {\bibfnamefont {C.}~\bibnamefont {Rigetti}},\ and\ \bibinfo {author}
  {\bibfnamefont {M.}~\bibnamefont {Steffen}},\ }\bibfield  {title} {\bibinfo
  {title} {Entanglement of two superconducting qubits in a waveguide cavity via
  monochromatic two-photon excitation},\ }\href
  {https://doi.org/10.1103/PhysRevLett.109.240505} {\bibfield  {journal}
  {\bibinfo  {journal} {Phys. Rev. Lett.}\ }\textbf {\bibinfo {volume} {109}},\
  \bibinfo {pages} {240505} (\bibinfo {year} {2012})}\BibitemShut {NoStop}%
\bibitem [{\citenamefont {Ghosh}\ \emph {et~al.}(2013)\citenamefont {Ghosh},
  \citenamefont {Galiautdinov}, \citenamefont {Zhou}, \citenamefont {Korotkov},
  \citenamefont {Martinis},\ and\ \citenamefont {Geller}}]{transmEnt13}%
  \BibitemOpen
  \bibfield  {author} {\bibinfo {author} {\bibfnamefont {J.}~\bibnamefont
  {Ghosh}}, \bibinfo {author} {\bibfnamefont {A.}~\bibnamefont {Galiautdinov}},
  \bibinfo {author} {\bibfnamefont {Z.}~\bibnamefont {Zhou}}, \bibinfo {author}
  {\bibfnamefont {A.~N.}\ \bibnamefont {Korotkov}}, \bibinfo {author}
  {\bibfnamefont {J.~M.}\ \bibnamefont {Martinis}},\ and\ \bibinfo {author}
  {\bibfnamefont {M.~R.}\ \bibnamefont {Geller}},\ }\bibfield  {title}
  {\bibinfo {title} {High-fidelity controlled-${\ensuremath{\sigma}}^{Z}$ gate
  for resonator-based superconducting quantum computers},\ }\href
  {https://doi.org/10.1103/PhysRevA.87.022309} {\bibfield  {journal} {\bibinfo
  {journal} {Phys. Rev. A}\ }\textbf {\bibinfo {volume} {87}},\ \bibinfo
  {pages} {022309} (\bibinfo {year} {2013})}\BibitemShut {NoStop}%
\bibitem [{\citenamefont {C\'orcoles}\ \emph {et~al.}(2013)\citenamefont
  {C\'orcoles}, \citenamefont {Gambetta}, \citenamefont {Chow}, \citenamefont
  {Smolin}, \citenamefont {Ware}, \citenamefont {Strand}, \citenamefont
  {Plourde},\ and\ \citenamefont {Steffen}}]{transmEnt14}%
  \BibitemOpen
  \bibfield  {author} {\bibinfo {author} {\bibfnamefont {A.~D.}\ \bibnamefont
  {C\'orcoles}}, \bibinfo {author} {\bibfnamefont {J.~M.}\ \bibnamefont
  {Gambetta}}, \bibinfo {author} {\bibfnamefont {J.~M.}\ \bibnamefont {Chow}},
  \bibinfo {author} {\bibfnamefont {J.~A.}\ \bibnamefont {Smolin}}, \bibinfo
  {author} {\bibfnamefont {M.}~\bibnamefont {Ware}}, \bibinfo {author}
  {\bibfnamefont {J.}~\bibnamefont {Strand}}, \bibinfo {author} {\bibfnamefont
  {B.~L.~T.}\ \bibnamefont {Plourde}},\ and\ \bibinfo {author} {\bibfnamefont
  {M.}~\bibnamefont {Steffen}},\ }\bibfield  {title} {\bibinfo {title} {Process
  verification of two-qubit quantum gates by randomized benchmarking},\ }\href
  {https://doi.org/10.1103/PhysRevA.87.030301} {\bibfield  {journal} {\bibinfo
  {journal} {Phys. Rev. A}\ }\textbf {\bibinfo {volume} {87}},\ \bibinfo
  {pages} {030301} (\bibinfo {year} {2013})}\BibitemShut {NoStop}%
\bibitem [{\citenamefont {Leghtas}\ \emph {et~al.}(2013)\citenamefont
  {Leghtas}, \citenamefont {Vool}, \citenamefont {Shankar}, \citenamefont
  {Hatridge}, \citenamefont {Girvin}, \citenamefont {Devoret},\ and\
  \citenamefont {Mirrahimi}}]{transmEnt15}%
  \BibitemOpen
  \bibfield  {author} {\bibinfo {author} {\bibfnamefont {Z.}~\bibnamefont
  {Leghtas}}, \bibinfo {author} {\bibfnamefont {U.}~\bibnamefont {Vool}},
  \bibinfo {author} {\bibfnamefont {S.}~\bibnamefont {Shankar}}, \bibinfo
  {author} {\bibfnamefont {M.}~\bibnamefont {Hatridge}}, \bibinfo {author}
  {\bibfnamefont {S.~M.}\ \bibnamefont {Girvin}}, \bibinfo {author}
  {\bibfnamefont {M.~H.}\ \bibnamefont {Devoret}},\ and\ \bibinfo {author}
  {\bibfnamefont {M.}~\bibnamefont {Mirrahimi}},\ }\bibfield  {title} {\bibinfo
  {title} {Stabilizing a bell state of two superconducting qubits by
  dissipation engineering},\ }\href
  {https://doi.org/10.1103/PhysRevA.88.023849} {\bibfield  {journal} {\bibinfo
  {journal} {Phys. Rev. A}\ }\textbf {\bibinfo {volume} {88}},\ \bibinfo
  {pages} {023849} (\bibinfo {year} {2013})}\BibitemShut {NoStop}%
\bibitem [{\citenamefont {Reiter}\ \emph {et~al.}(2013)\citenamefont {Reiter},
  \citenamefont {Tornberg}, \citenamefont {Johansson},\ and\ \citenamefont
  {S\o{}rensen}}]{transmEnt16}%
  \BibitemOpen
  \bibfield  {author} {\bibinfo {author} {\bibfnamefont {F.}~\bibnamefont
  {Reiter}}, \bibinfo {author} {\bibfnamefont {L.}~\bibnamefont {Tornberg}},
  \bibinfo {author} {\bibfnamefont {G.}~\bibnamefont {Johansson}},\ and\
  \bibinfo {author} {\bibfnamefont {A.~S.}\ \bibnamefont {S\o{}rensen}},\
  }\bibfield  {title} {\bibinfo {title} {Steady-state entanglement of two
  superconducting qubits engineered by dissipation},\ }\href
  {https://doi.org/10.1103/PhysRevA.88.032317} {\bibfield  {journal} {\bibinfo
  {journal} {Phys. Rev. A}\ }\textbf {\bibinfo {volume} {88}},\ \bibinfo
  {pages} {032317} (\bibinfo {year} {2013})}\BibitemShut {NoStop}%
\bibitem [{\citenamefont {Egger}\ and\ \citenamefont
  {Wilhelm}(2013)}]{transmEnt17}%
  \BibitemOpen
  \bibfield  {author} {\bibinfo {author} {\bibfnamefont {D.~J.}\ \bibnamefont
  {Egger}}\ and\ \bibinfo {author} {\bibfnamefont {F.~K.}\ \bibnamefont
  {Wilhelm}},\ }\bibfield  {title} {\bibinfo {title} {Optimized controlled-z
  gates for two superconducting qubits coupled through a resonator},\ }\href
  {https://doi.org/10.1088/0953-2048/27/1/014001} {\bibfield  {journal}
  {\bibinfo  {journal} {Superconductor Science and Technology}\ }\textbf
  {\bibinfo {volume} {27}},\ \bibinfo {pages} {014001} (\bibinfo {year}
  {2013})}\BibitemShut {NoStop}%
\bibitem [{\citenamefont {Huang}\ and\ \citenamefont
  {Goan}(2014)}]{transmEnt18}%
  \BibitemOpen
  \bibfield  {author} {\bibinfo {author} {\bibfnamefont {S.-Y.}\ \bibnamefont
  {Huang}}\ and\ \bibinfo {author} {\bibfnamefont {H.-S.}\ \bibnamefont
  {Goan}},\ }\bibfield  {title} {\bibinfo {title} {Optimal control for fast and
  high-fidelity quantum gates in coupled superconducting flux qubits},\ }\href
  {https://doi.org/10.1103/PhysRevA.90.012318} {\bibfield  {journal} {\bibinfo
  {journal} {Phys. Rev. A}\ }\textbf {\bibinfo {volume} {90}},\ \bibinfo
  {pages} {012318} (\bibinfo {year} {2014})}\BibitemShut {NoStop}%
\bibitem [{\citenamefont {Chakhmakhchyan}\ \emph {et~al.}(2014)\citenamefont
  {Chakhmakhchyan}, \citenamefont {Leroy}, \citenamefont {Ananikian},\ and\
  \citenamefont {Gu\'erin}}]{transmEnt19}%
  \BibitemOpen
  \bibfield  {author} {\bibinfo {author} {\bibfnamefont {L.}~\bibnamefont
  {Chakhmakhchyan}}, \bibinfo {author} {\bibfnamefont {C.}~\bibnamefont
  {Leroy}}, \bibinfo {author} {\bibfnamefont {N.}~\bibnamefont {Ananikian}},\
  and\ \bibinfo {author} {\bibfnamefont {S.}~\bibnamefont {Gu\'erin}},\
  }\bibfield  {title} {\bibinfo {title} {Generation of entanglement in systems
  of intercoupled qubits},\ }\href {https://doi.org/10.1103/PhysRevA.90.042324}
  {\bibfield  {journal} {\bibinfo  {journal} {Phys. Rev. A}\ }\textbf {\bibinfo
  {volume} {90}},\ \bibinfo {pages} {042324} (\bibinfo {year}
  {2014})}\BibitemShut {NoStop}%
\bibitem [{\citenamefont {Yang}\ \emph {et~al.}(2016)\citenamefont {Yang},
  \citenamefont {Su}, \citenamefont {Zheng},\ and\ \citenamefont
  {Nori}}]{transmEnt20}%
  \BibitemOpen
  \bibfield  {author} {\bibinfo {author} {\bibfnamefont {C.-P.}\ \bibnamefont
  {Yang}}, \bibinfo {author} {\bibfnamefont {Q.-P.}\ \bibnamefont {Su}},
  \bibinfo {author} {\bibfnamefont {S.-B.}\ \bibnamefont {Zheng}},\ and\
  \bibinfo {author} {\bibfnamefont {F.}~\bibnamefont {Nori}},\ }\bibfield
  {title} {\bibinfo {title} {Entangling superconducting qubits in a
  multi-cavity system},\ }\href {https://doi.org/10.1088/1367-2630/18/1/013025}
  {\bibfield  {journal} {\bibinfo  {journal} {New Journal of Physics}\ }\textbf
  {\bibinfo {volume} {18}},\ \bibinfo {pages} {013025} (\bibinfo {year}
  {2016})}\BibitemShut {NoStop}%
\bibitem [{\citenamefont {Kirchhoff}\ \emph {et~al.}(2018)\citenamefont
  {Kirchhoff}, \citenamefont {Ke\ss{}ler}, \citenamefont {Liebermann},
  \citenamefont {Ass\'emat}, \citenamefont {Machnes}, \citenamefont {Motzoi},\
  and\ \citenamefont {Wilhelm}}]{transmEnt21}%
  \BibitemOpen
  \bibfield  {author} {\bibinfo {author} {\bibfnamefont {S.}~\bibnamefont
  {Kirchhoff}}, \bibinfo {author} {\bibfnamefont {T.}~\bibnamefont
  {Ke\ss{}ler}}, \bibinfo {author} {\bibfnamefont {P.~J.}\ \bibnamefont
  {Liebermann}}, \bibinfo {author} {\bibfnamefont {E.}~\bibnamefont
  {Ass\'emat}}, \bibinfo {author} {\bibfnamefont {S.}~\bibnamefont {Machnes}},
  \bibinfo {author} {\bibfnamefont {F.}~\bibnamefont {Motzoi}},\ and\ \bibinfo
  {author} {\bibfnamefont {F.~K.}\ \bibnamefont {Wilhelm}},\ }\bibfield
  {title} {\bibinfo {title} {Optimized cross-resonance gate for coupled
  transmon systems},\ }\href {https://doi.org/10.1103/PhysRevA.97.042348}
  {\bibfield  {journal} {\bibinfo  {journal} {Phys. Rev. A}\ }\textbf {\bibinfo
  {volume} {97}},\ \bibinfo {pages} {042348} (\bibinfo {year}
  {2018})}\BibitemShut {NoStop}%
\bibitem [{\citenamefont {Qi}\ \emph {et~al.}(2018)\citenamefont {Qi},
  \citenamefont {Xie}, \citenamefont {Shabani}, \citenamefont {Manucharyan},
  \citenamefont {Levchenko},\ and\ \citenamefont {Vavilov}}]{transmEnt22}%
  \BibitemOpen
  \bibfield  {author} {\bibinfo {author} {\bibfnamefont {Z.}~\bibnamefont
  {Qi}}, \bibinfo {author} {\bibfnamefont {H.-Y.}\ \bibnamefont {Xie}},
  \bibinfo {author} {\bibfnamefont {J.}~\bibnamefont {Shabani}}, \bibinfo
  {author} {\bibfnamefont {V.~E.}\ \bibnamefont {Manucharyan}}, \bibinfo
  {author} {\bibfnamefont {A.}~\bibnamefont {Levchenko}},\ and\ \bibinfo
  {author} {\bibfnamefont {M.~G.}\ \bibnamefont {Vavilov}},\ }\bibfield
  {title} {\bibinfo {title} {Controlled-z gate for transmon qubits coupled by
  semiconductor junctions},\ }\href
  {https://doi.org/10.1103/PhysRevB.97.134518} {\bibfield  {journal} {\bibinfo
  {journal} {Phys. Rev. B}\ }\textbf {\bibinfo {volume} {97}},\ \bibinfo
  {pages} {134518} (\bibinfo {year} {2018})}\BibitemShut {NoStop}%
\bibitem [{\citenamefont {Noh}\ \emph {et~al.}(2018)\citenamefont {Noh},
  \citenamefont {Park}, \citenamefont {Lee}, \citenamefont {Song},\ and\
  \citenamefont {Chong}}]{transmEnt23}%
  \BibitemOpen
  \bibfield  {author} {\bibinfo {author} {\bibfnamefont {T.}~\bibnamefont
  {Noh}}, \bibinfo {author} {\bibfnamefont {G.}~\bibnamefont {Park}}, \bibinfo
  {author} {\bibfnamefont {S.-G.}\ \bibnamefont {Lee}}, \bibinfo {author}
  {\bibfnamefont {W.}~\bibnamefont {Song}},\ and\ \bibinfo {author}
  {\bibfnamefont {Y.}~\bibnamefont {Chong}},\ }\bibfield  {title} {\bibinfo
  {title} {Construction of controlled-not gate based on microwave-activated
  phase (map) gate in two transmon system},\ }\href
  {https://doi.org/10.1038/s41598-018-31896-3} {\bibfield  {journal} {\bibinfo
  {journal} {Scientific Reports}\ }\textbf {\bibinfo {volume} {8}},\ \bibinfo
  {pages} {13598} (\bibinfo {year} {2018})}\BibitemShut {NoStop}%
\bibitem [{\citenamefont {Nesterov}\ \emph {et~al.}(2018)\citenamefont
  {Nesterov}, \citenamefont {Pechenezhskiy}, \citenamefont {Wang},
  \citenamefont {Manucharyan},\ and\ \citenamefont {Vavilov}}]{transmEnt24}%
  \BibitemOpen
  \bibfield  {author} {\bibinfo {author} {\bibfnamefont {K.~N.}\ \bibnamefont
  {Nesterov}}, \bibinfo {author} {\bibfnamefont {I.~V.}\ \bibnamefont
  {Pechenezhskiy}}, \bibinfo {author} {\bibfnamefont {C.}~\bibnamefont {Wang}},
  \bibinfo {author} {\bibfnamefont {V.~E.}\ \bibnamefont {Manucharyan}},\ and\
  \bibinfo {author} {\bibfnamefont {M.~G.}\ \bibnamefont {Vavilov}},\
  }\bibfield  {title} {\bibinfo {title} {Microwave-activated controlled-$z$
  gate for fixed-frequency fluxonium qubits},\ }\href
  {https://doi.org/10.1103/PhysRevA.98.030301} {\bibfield  {journal} {\bibinfo
  {journal} {Phys. Rev. A}\ }\textbf {\bibinfo {volume} {98}},\ \bibinfo
  {pages} {030301} (\bibinfo {year} {2018})}\BibitemShut {NoStop}%
\bibitem [{\citenamefont {Caldwell}\ \emph {et~al.}(2018)\citenamefont
  {Caldwell}, \citenamefont {Didier}, \citenamefont {Ryan}, \citenamefont
  {Sete}, \citenamefont {Hudson}, \citenamefont {Karalekas}, \citenamefont
  {Manenti}, \citenamefont {da~Silva}, \citenamefont {Sinclair}, \citenamefont
  {Acala}, \citenamefont {Alidoust}, \citenamefont {Angeles}, \citenamefont
  {Bestwick}, \citenamefont {Block}, \citenamefont {Bloom}, \citenamefont
  {Bradley}, \citenamefont {Bui}, \citenamefont {Capelluto}, \citenamefont
  {Chilcott}, \citenamefont {Cordova}, \citenamefont {Crossman}, \citenamefont
  {Curtis}, \citenamefont {Deshpande}, \citenamefont {Bouayadi}, \citenamefont
  {Girshovich}, \citenamefont {Hong}, \citenamefont {Kuang}, \citenamefont
  {Lenihan}, \citenamefont {Manning}, \citenamefont {Marchenkov}, \citenamefont
  {Marshall}, \citenamefont {Maydra}, \citenamefont {Mohan}, \citenamefont
  {O'Brien}, \citenamefont {Osborn}, \citenamefont {Otterbach}, \citenamefont
  {Papageorge}, \citenamefont {Paquette}, \citenamefont {Pelstring},
  \citenamefont {Polloreno}, \citenamefont {Prawiroatmodjo}, \citenamefont
  {Rawat}, \citenamefont {Reagor}, \citenamefont {Renzas}, \citenamefont
  {Rubin}, \citenamefont {Russell}, \citenamefont {Rust}, \citenamefont
  {Scarabelli}, \citenamefont {Scheer}, \citenamefont {Selvanayagam},
  \citenamefont {Smith}, \citenamefont {Staley}, \citenamefont {Suska},
  \citenamefont {Tezak}, \citenamefont {Thompson}, \citenamefont {To},
  \citenamefont {Vahidpour}, \citenamefont {Vodrahalli}, \citenamefont
  {Whyland}, \citenamefont {Yadav}, \citenamefont {Zeng},\ and\ \citenamefont
  {Rigetti}}]{transmEnt25}%
  \BibitemOpen
  \bibfield  {author} {\bibinfo {author} {\bibfnamefont {S.~A.}\ \bibnamefont
  {Caldwell}}, \bibinfo {author} {\bibfnamefont {N.}~\bibnamefont {Didier}},
  \bibinfo {author} {\bibfnamefont {C.~A.}\ \bibnamefont {Ryan}}, \bibinfo
  {author} {\bibfnamefont {E.~A.}\ \bibnamefont {Sete}}, \bibinfo {author}
  {\bibfnamefont {A.}~\bibnamefont {Hudson}}, \bibinfo {author} {\bibfnamefont
  {P.}~\bibnamefont {Karalekas}}, \bibinfo {author} {\bibfnamefont
  {R.}~\bibnamefont {Manenti}}, \bibinfo {author} {\bibfnamefont {M.~P.}\
  \bibnamefont {da~Silva}}, \bibinfo {author} {\bibfnamefont {R.}~\bibnamefont
  {Sinclair}}, \bibinfo {author} {\bibfnamefont {E.}~\bibnamefont {Acala}},
  \bibinfo {author} {\bibfnamefont {N.}~\bibnamefont {Alidoust}}, \bibinfo
  {author} {\bibfnamefont {J.}~\bibnamefont {Angeles}}, \bibinfo {author}
  {\bibfnamefont {A.}~\bibnamefont {Bestwick}}, \bibinfo {author}
  {\bibfnamefont {M.}~\bibnamefont {Block}}, \bibinfo {author} {\bibfnamefont
  {B.}~\bibnamefont {Bloom}}, \bibinfo {author} {\bibfnamefont
  {A.}~\bibnamefont {Bradley}}, \bibinfo {author} {\bibfnamefont
  {C.}~\bibnamefont {Bui}}, \bibinfo {author} {\bibfnamefont {L.}~\bibnamefont
  {Capelluto}}, \bibinfo {author} {\bibfnamefont {R.}~\bibnamefont {Chilcott}},
  \bibinfo {author} {\bibfnamefont {J.}~\bibnamefont {Cordova}}, \bibinfo
  {author} {\bibfnamefont {G.}~\bibnamefont {Crossman}}, \bibinfo {author}
  {\bibfnamefont {M.}~\bibnamefont {Curtis}}, \bibinfo {author} {\bibfnamefont
  {S.}~\bibnamefont {Deshpande}}, \bibinfo {author} {\bibfnamefont {T.~E.}\
  \bibnamefont {Bouayadi}}, \bibinfo {author} {\bibfnamefont {D.}~\bibnamefont
  {Girshovich}}, \bibinfo {author} {\bibfnamefont {S.}~\bibnamefont {Hong}},
  \bibinfo {author} {\bibfnamefont {K.}~\bibnamefont {Kuang}}, \bibinfo
  {author} {\bibfnamefont {M.}~\bibnamefont {Lenihan}}, \bibinfo {author}
  {\bibfnamefont {T.}~\bibnamefont {Manning}}, \bibinfo {author} {\bibfnamefont
  {A.}~\bibnamefont {Marchenkov}}, \bibinfo {author} {\bibfnamefont
  {J.}~\bibnamefont {Marshall}}, \bibinfo {author} {\bibfnamefont
  {R.}~\bibnamefont {Maydra}}, \bibinfo {author} {\bibfnamefont
  {Y.}~\bibnamefont {Mohan}}, \bibinfo {author} {\bibfnamefont
  {W.}~\bibnamefont {O'Brien}}, \bibinfo {author} {\bibfnamefont
  {C.}~\bibnamefont {Osborn}}, \bibinfo {author} {\bibfnamefont
  {J.}~\bibnamefont {Otterbach}}, \bibinfo {author} {\bibfnamefont
  {A.}~\bibnamefont {Papageorge}}, \bibinfo {author} {\bibfnamefont {J.-P.}\
  \bibnamefont {Paquette}}, \bibinfo {author} {\bibfnamefont {M.}~\bibnamefont
  {Pelstring}}, \bibinfo {author} {\bibfnamefont {A.}~\bibnamefont
  {Polloreno}}, \bibinfo {author} {\bibfnamefont {G.}~\bibnamefont
  {Prawiroatmodjo}}, \bibinfo {author} {\bibfnamefont {V.}~\bibnamefont
  {Rawat}}, \bibinfo {author} {\bibfnamefont {M.}~\bibnamefont {Reagor}},
  \bibinfo {author} {\bibfnamefont {R.}~\bibnamefont {Renzas}}, \bibinfo
  {author} {\bibfnamefont {N.}~\bibnamefont {Rubin}}, \bibinfo {author}
  {\bibfnamefont {D.}~\bibnamefont {Russell}}, \bibinfo {author} {\bibfnamefont
  {M.}~\bibnamefont {Rust}}, \bibinfo {author} {\bibfnamefont {D.}~\bibnamefont
  {Scarabelli}}, \bibinfo {author} {\bibfnamefont {M.}~\bibnamefont {Scheer}},
  \bibinfo {author} {\bibfnamefont {M.}~\bibnamefont {Selvanayagam}}, \bibinfo
  {author} {\bibfnamefont {R.}~\bibnamefont {Smith}}, \bibinfo {author}
  {\bibfnamefont {A.}~\bibnamefont {Staley}}, \bibinfo {author} {\bibfnamefont
  {M.}~\bibnamefont {Suska}}, \bibinfo {author} {\bibfnamefont
  {N.}~\bibnamefont {Tezak}}, \bibinfo {author} {\bibfnamefont {D.~C.}\
  \bibnamefont {Thompson}}, \bibinfo {author} {\bibfnamefont {T.-W.}\
  \bibnamefont {To}}, \bibinfo {author} {\bibfnamefont {M.}~\bibnamefont
  {Vahidpour}}, \bibinfo {author} {\bibfnamefont {N.}~\bibnamefont
  {Vodrahalli}}, \bibinfo {author} {\bibfnamefont {T.}~\bibnamefont {Whyland}},
  \bibinfo {author} {\bibfnamefont {K.}~\bibnamefont {Yadav}}, \bibinfo
  {author} {\bibfnamefont {W.}~\bibnamefont {Zeng}},\ and\ \bibinfo {author}
  {\bibfnamefont {C.}~\bibnamefont {Rigetti}},\ }\bibfield  {title} {\bibinfo
  {title} {Parametrically activated entangling gates using transmon qubits},\
  }\href {https://doi.org/10.1103/PhysRevApplied.10.034050} {\bibfield
  {journal} {\bibinfo  {journal} {Phys. Rev. Applied}\ }\textbf {\bibinfo
  {volume} {10}},\ \bibinfo {pages} {034050} (\bibinfo {year}
  {2018})}\BibitemShut {NoStop}%
\bibitem [{\citenamefont {Yan}\ \emph {et~al.}(2018)\citenamefont {Yan},
  \citenamefont {Krantz}, \citenamefont {Sung}, \citenamefont {Kjaergaard},
  \citenamefont {Campbell}, \citenamefont {Orlando}, \citenamefont
  {Gustavsson},\ and\ \citenamefont {Oliver}}]{transmEnt26}%
  \BibitemOpen
  \bibfield  {author} {\bibinfo {author} {\bibfnamefont {F.}~\bibnamefont
  {Yan}}, \bibinfo {author} {\bibfnamefont {P.}~\bibnamefont {Krantz}},
  \bibinfo {author} {\bibfnamefont {Y.}~\bibnamefont {Sung}}, \bibinfo {author}
  {\bibfnamefont {M.}~\bibnamefont {Kjaergaard}}, \bibinfo {author}
  {\bibfnamefont {D.~L.}\ \bibnamefont {Campbell}}, \bibinfo {author}
  {\bibfnamefont {T.~P.}\ \bibnamefont {Orlando}}, \bibinfo {author}
  {\bibfnamefont {S.}~\bibnamefont {Gustavsson}},\ and\ \bibinfo {author}
  {\bibfnamefont {W.~D.}\ \bibnamefont {Oliver}},\ }\bibfield  {title}
  {\bibinfo {title} {Tunable coupling scheme for implementing high-fidelity
  two-qubit gates},\ }\href {https://doi.org/10.1103/PhysRevApplied.10.054062}
  {\bibfield  {journal} {\bibinfo  {journal} {Phys. Rev. Applied}\ }\textbf
  {\bibinfo {volume} {10}},\ \bibinfo {pages} {054062} (\bibinfo {year}
  {2018})}\BibitemShut {NoStop}%
\bibitem [{\citenamefont {Premaratne}\ \emph {et~al.}(2019)\citenamefont
  {Premaratne}, \citenamefont {Yeh}, \citenamefont {Wellstood},\ and\
  \citenamefont {Palmer}}]{transmEnt27}%
  \BibitemOpen
  \bibfield  {author} {\bibinfo {author} {\bibfnamefont {S.~P.}\ \bibnamefont
  {Premaratne}}, \bibinfo {author} {\bibfnamefont {J.-H.}\ \bibnamefont {Yeh}},
  \bibinfo {author} {\bibfnamefont {F.~C.}\ \bibnamefont {Wellstood}},\ and\
  \bibinfo {author} {\bibfnamefont {B.~S.}\ \bibnamefont {Palmer}},\ }\bibfield
   {title} {\bibinfo {title} {Implementation of a generalized controlled-not
  gate between fixed-frequency transmons},\ }\href
  {https://doi.org/10.1103/PhysRevA.99.012317} {\bibfield  {journal} {\bibinfo
  {journal} {Phys. Rev. A}\ }\textbf {\bibinfo {volume} {99}},\ \bibinfo
  {pages} {012317} (\bibinfo {year} {2019})}\BibitemShut {NoStop}%
\bibitem [{\citenamefont {Rasmussen}\ \emph {et~al.}(2019)\citenamefont
  {Rasmussen}, \citenamefont {Christensen},\ and\ \citenamefont
  {Zinner}}]{transmEnt28}%
  \BibitemOpen
  \bibfield  {author} {\bibinfo {author} {\bibfnamefont {S.~E.}\ \bibnamefont
  {Rasmussen}}, \bibinfo {author} {\bibfnamefont {K.~S.}\ \bibnamefont
  {Christensen}},\ and\ \bibinfo {author} {\bibfnamefont {N.~T.}\ \bibnamefont
  {Zinner}},\ }\bibfield  {title} {\bibinfo {title} {Controllable two-qubit
  swapping gate using superconducting circuits},\ }\href
  {https://doi.org/10.1103/PhysRevB.99.134508} {\bibfield  {journal} {\bibinfo
  {journal} {Phys. Rev. B}\ }\textbf {\bibinfo {volume} {99}},\ \bibinfo
  {pages} {134508} (\bibinfo {year} {2019})}\BibitemShut {NoStop}%
\bibitem [{\citenamefont {Tripathi}\ \emph {et~al.}(2019)\citenamefont
  {Tripathi}, \citenamefont {Khezri},\ and\ \citenamefont
  {Korotkov}}]{transmEnt29}%
  \BibitemOpen
  \bibfield  {author} {\bibinfo {author} {\bibfnamefont {V.}~\bibnamefont
  {Tripathi}}, \bibinfo {author} {\bibfnamefont {M.}~\bibnamefont {Khezri}},\
  and\ \bibinfo {author} {\bibfnamefont {A.~N.}\ \bibnamefont {Korotkov}},\
  }\bibfield  {title} {\bibinfo {title} {Operation and intrinsic error budget
  of a two-qubit cross-resonance gate},\ }\href
  {https://doi.org/10.1103/PhysRevA.100.012301} {\bibfield  {journal} {\bibinfo
   {journal} {Phys. Rev. A}\ }\textbf {\bibinfo {volume} {100}},\ \bibinfo
  {pages} {012301} (\bibinfo {year} {2019})}\BibitemShut {NoStop}%
\bibitem [{\citenamefont {Patterson}\ \emph {et~al.}(2019)\citenamefont
  {Patterson}, \citenamefont {Rahamim}, \citenamefont {Tsunoda}, \citenamefont
  {Spring}, \citenamefont {Jebari}, \citenamefont {Ratter}, \citenamefont
  {Mergenthaler}, \citenamefont {Tancredi}, \citenamefont {Vlastakis},
  \citenamefont {Esposito},\ and\ \citenamefont {Leek}}]{transmEnt30}%
  \BibitemOpen
  \bibfield  {author} {\bibinfo {author} {\bibfnamefont {A.}~\bibnamefont
  {Patterson}}, \bibinfo {author} {\bibfnamefont {J.}~\bibnamefont {Rahamim}},
  \bibinfo {author} {\bibfnamefont {T.}~\bibnamefont {Tsunoda}}, \bibinfo
  {author} {\bibfnamefont {P.}~\bibnamefont {Spring}}, \bibinfo {author}
  {\bibfnamefont {S.}~\bibnamefont {Jebari}}, \bibinfo {author} {\bibfnamefont
  {K.}~\bibnamefont {Ratter}}, \bibinfo {author} {\bibfnamefont
  {M.}~\bibnamefont {Mergenthaler}}, \bibinfo {author} {\bibfnamefont
  {G.}~\bibnamefont {Tancredi}}, \bibinfo {author} {\bibfnamefont
  {B.}~\bibnamefont {Vlastakis}}, \bibinfo {author} {\bibfnamefont
  {M.}~\bibnamefont {Esposito}},\ and\ \bibinfo {author} {\bibfnamefont
  {P.}~\bibnamefont {Leek}},\ }\bibfield  {title} {\bibinfo {title}
  {Calibration of a cross-resonance two-qubit gate between directly coupled
  transmons},\ }\href {https://doi.org/10.1103/PhysRevApplied.12.064013}
  {\bibfield  {journal} {\bibinfo  {journal} {Phys. Rev. Applied}\ }\textbf
  {\bibinfo {volume} {12}},\ \bibinfo {pages} {064013} (\bibinfo {year}
  {2019})}\BibitemShut {NoStop}%
\bibitem [{\citenamefont {Hong}\ \emph {et~al.}(2020)\citenamefont {Hong},
  \citenamefont {Papageorge}, \citenamefont {Sivarajah}, \citenamefont
  {Crossman}, \citenamefont {Didier}, \citenamefont {Polloreno}, \citenamefont
  {Sete}, \citenamefont {Turkowski}, \citenamefont {da~Silva},\ and\
  \citenamefont {Johnson}}]{transmEnt31}%
  \BibitemOpen
  \bibfield  {author} {\bibinfo {author} {\bibfnamefont {S.~S.}\ \bibnamefont
  {Hong}}, \bibinfo {author} {\bibfnamefont {A.~T.}\ \bibnamefont
  {Papageorge}}, \bibinfo {author} {\bibfnamefont {P.}~\bibnamefont
  {Sivarajah}}, \bibinfo {author} {\bibfnamefont {G.}~\bibnamefont {Crossman}},
  \bibinfo {author} {\bibfnamefont {N.}~\bibnamefont {Didier}}, \bibinfo
  {author} {\bibfnamefont {A.~M.}\ \bibnamefont {Polloreno}}, \bibinfo {author}
  {\bibfnamefont {E.~A.}\ \bibnamefont {Sete}}, \bibinfo {author}
  {\bibfnamefont {S.~W.}\ \bibnamefont {Turkowski}}, \bibinfo {author}
  {\bibfnamefont {M.~P.}\ \bibnamefont {da~Silva}},\ and\ \bibinfo {author}
  {\bibfnamefont {B.~R.}\ \bibnamefont {Johnson}},\ }\bibfield  {title}
  {\bibinfo {title} {Demonstration of a parametrically activated entangling
  gate protected from flux noise},\ }\href
  {https://doi.org/10.1103/PhysRevA.101.012302} {\bibfield  {journal} {\bibinfo
   {journal} {Phys. Rev. A}\ }\textbf {\bibinfo {volume} {101}},\ \bibinfo
  {pages} {012302} (\bibinfo {year} {2020})}\BibitemShut {NoStop}%
\bibitem [{\citenamefont {Barron}\ \emph {et~al.}(2020)\citenamefont {Barron},
  \citenamefont {Calderon-Vargas}, \citenamefont {Long}, \citenamefont
  {Pappas},\ and\ \citenamefont {Economou}}]{transmEnt32}%
  \BibitemOpen
  \bibfield  {author} {\bibinfo {author} {\bibfnamefont {G.~S.}\ \bibnamefont
  {Barron}}, \bibinfo {author} {\bibfnamefont {F.~A.}\ \bibnamefont
  {Calderon-Vargas}}, \bibinfo {author} {\bibfnamefont {J.}~\bibnamefont
  {Long}}, \bibinfo {author} {\bibfnamefont {D.~P.}\ \bibnamefont {Pappas}},\
  and\ \bibinfo {author} {\bibfnamefont {S.~E.}\ \bibnamefont {Economou}},\
  }\bibfield  {title} {\bibinfo {title} {Microwave-based arbitrary cphase gates
  for transmon qubits},\ }\href {https://doi.org/10.1103/PhysRevB.101.054508}
  {\bibfield  {journal} {\bibinfo  {journal} {Phys. Rev. B}\ }\textbf {\bibinfo
  {volume} {101}},\ \bibinfo {pages} {054508} (\bibinfo {year}
  {2020})}\BibitemShut {NoStop}%
\bibitem [{\citenamefont {Loft}\ \emph {et~al.}(2020)\citenamefont {Loft},
  \citenamefont {Kjaergaard}, \citenamefont {Kristensen}, \citenamefont
  {Andersen}, \citenamefont {Larsen}, \citenamefont {Gustavsson}, \citenamefont
  {Oliver},\ and\ \citenamefont {Zinner}}]{transmEnt33}%
  \BibitemOpen
  \bibfield  {author} {\bibinfo {author} {\bibfnamefont {N.~J.~S.}\
  \bibnamefont {Loft}}, \bibinfo {author} {\bibfnamefont {M.}~\bibnamefont
  {Kjaergaard}}, \bibinfo {author} {\bibfnamefont {L.~B.}\ \bibnamefont
  {Kristensen}}, \bibinfo {author} {\bibfnamefont {C.~K.}\ \bibnamefont
  {Andersen}}, \bibinfo {author} {\bibfnamefont {T.~W.}\ \bibnamefont
  {Larsen}}, \bibinfo {author} {\bibfnamefont {S.}~\bibnamefont {Gustavsson}},
  \bibinfo {author} {\bibfnamefont {W.~D.}\ \bibnamefont {Oliver}},\ and\
  \bibinfo {author} {\bibfnamefont {N.~T.}\ \bibnamefont {Zinner}},\ }\bibfield
   {title} {\bibinfo {title} {Quantum interference device for controlled
  two-qubit operations},\ }\href {https://doi.org/10.1038/s41534-020-0275-3}
  {\bibfield  {journal} {\bibinfo  {journal} {npj Quantum Information}\
  }\textbf {\bibinfo {volume} {6}},\ \bibinfo {pages} {47} (\bibinfo {year}
  {2020})}\BibitemShut {NoStop}%
\bibitem [{\citenamefont {Rasmussen}\ and\ \citenamefont
  {Zinner}(2020)}]{transmEnt34}%
  \BibitemOpen
  \bibfield  {author} {\bibinfo {author} {\bibfnamefont {S.~E.}\ \bibnamefont
  {Rasmussen}}\ and\ \bibinfo {author} {\bibfnamefont {N.~T.}\ \bibnamefont
  {Zinner}},\ }\bibfield  {title} {\bibinfo {title} {Simple implementation of
  high fidelity controlled-$i$swap gates and quantum circuit exponentiation of
  non-hermitian gates},\ }\href
  {https://doi.org/10.1103/PhysRevResearch.2.033097} {\bibfield  {journal}
  {\bibinfo  {journal} {Phys. Rev. Research}\ }\textbf {\bibinfo {volume}
  {2}},\ \bibinfo {pages} {033097} (\bibinfo {year} {2020})}\BibitemShut
  {NoStop}%
\bibitem [{\citenamefont {Zhang}\ \emph {et~al.}(2020)\citenamefont {Zhang},
  \citenamefont {Liu},\ and\ \citenamefont {Lin}}]{transmEnt35}%
  \BibitemOpen
  \bibfield  {author} {\bibinfo {author} {\bibfnamefont {C.-L.}\ \bibnamefont
  {Zhang}}, \bibinfo {author} {\bibfnamefont {W.-W.}\ \bibnamefont {Liu}},\
  and\ \bibinfo {author} {\bibfnamefont {X.-M.}\ \bibnamefont {Lin}},\
  }\bibfield  {title} {\bibinfo {title} {One-step implementation of a robust
  fredkin gate based on path engineering},\ }\href
  {https://doi.org/10.1007/s11128-020-02767-6} {\bibfield  {journal} {\bibinfo
  {journal} {Quantum Information Processing}\ }\textbf {\bibinfo {volume}
  {19}},\ \bibinfo {pages} {265} (\bibinfo {year} {2020})}\BibitemShut
  {NoStop}%
\bibitem [{\citenamefont {Ganzhorn}\ \emph {et~al.}(2020)\citenamefont
  {Ganzhorn}, \citenamefont {Salis}, \citenamefont {Egger}, \citenamefont
  {Fuhrer}, \citenamefont {Mergenthaler}, \citenamefont {M\"uller},
  \citenamefont {M\"uller}, \citenamefont {Paredes}, \citenamefont {Pechal},
  \citenamefont {Werninghaus},\ and\ \citenamefont {Filipp}}]{transmEnt36}%
  \BibitemOpen
  \bibfield  {author} {\bibinfo {author} {\bibfnamefont {M.}~\bibnamefont
  {Ganzhorn}}, \bibinfo {author} {\bibfnamefont {G.}~\bibnamefont {Salis}},
  \bibinfo {author} {\bibfnamefont {D.~J.}\ \bibnamefont {Egger}}, \bibinfo
  {author} {\bibfnamefont {A.}~\bibnamefont {Fuhrer}}, \bibinfo {author}
  {\bibfnamefont {M.}~\bibnamefont {Mergenthaler}}, \bibinfo {author}
  {\bibfnamefont {C.}~\bibnamefont {M\"uller}}, \bibinfo {author}
  {\bibfnamefont {P.}~\bibnamefont {M\"uller}}, \bibinfo {author}
  {\bibfnamefont {S.}~\bibnamefont {Paredes}}, \bibinfo {author} {\bibfnamefont
  {M.}~\bibnamefont {Pechal}}, \bibinfo {author} {\bibfnamefont
  {M.}~\bibnamefont {Werninghaus}},\ and\ \bibinfo {author} {\bibfnamefont
  {S.}~\bibnamefont {Filipp}},\ }\bibfield  {title} {\bibinfo {title}
  {Benchmarking the noise sensitivity of different parametric two-qubit gates
  in a single superconducting quantum computing platform},\ }\href
  {https://doi.org/10.1103/PhysRevResearch.2.033447} {\bibfield  {journal}
  {\bibinfo  {journal} {Phys. Rev. Research}\ }\textbf {\bibinfo {volume}
  {2}},\ \bibinfo {pages} {033447} (\bibinfo {year} {2020})}\BibitemShut
  {NoStop}%
\bibitem [{\citenamefont {Fano}(1995)}]{densitymatI}%
  \BibitemOpen
  \bibfield  {author} {\bibinfo {author} {\bibfnamefont {U.}~\bibnamefont
  {Fano}},\ }\bibfield  {title} {\bibinfo {title} {Matrici di densit{\`a} come
  vettori di polarizzazione},\ }\href {https://doi.org/10.1007/BF03001661}
  {\bibfield  {journal} {\bibinfo  {journal} {Rendiconti Lincei}\ }\textbf
  {\bibinfo {volume} {6}},\ \bibinfo {pages} {123} (\bibinfo {year}
  {1995})}\BibitemShut {NoStop}%
\bibitem [{den(1965)}]{densitymatII}%
  \BibitemOpen
  \bibfield  {title} {\bibinfo {title} {2 - the damping problem in wave
  mechanics},\ }in\ \href
  {https://doi.org/https://doi.org/10.1016/B978-0-08-010586-4.50007-9} {\emph
  {\bibinfo {booktitle} {Collected Papers of L.D. Landau}}},\ \bibinfo {editor}
  {edited by\ \bibinfo {editor} {\bibfnamefont {D.~T.}\ \bibnamefont {HAAR}}}\
  (\bibinfo  {publisher} {Pergamon},\ \bibinfo {year} {1965})\ pp.\ \bibinfo
  {pages} {8 -- 18}\BibitemShut {NoStop}%
\bibitem [{\citenamefont {Neumann}(1927)}]{densitymatIII}%
  \BibitemOpen
  \bibfield  {author} {\bibinfo {author} {\bibfnamefont {J.~v.}\ \bibnamefont
  {Neumann}},\ }\bibfield  {title} {\bibinfo {title}
  {Wahrscheinlichkeitstheoretischer aufbau der quantenmechanik},\ }\href
  {https://eudml.org/doc/59230} {\bibfield  {journal} {\bibinfo  {journal}
  {Nachrichten von der Gesellschaft der Wissenschaften zu G{\"o}ttingen,
  Mathematisch-Physikalische Klasse}\ }\textbf {\bibinfo {volume} {1927}},\
  \bibinfo {pages} {245} (\bibinfo {year} {1927})}\BibitemShut {NoStop}%
\bibitem [{\citenamefont {Choi}(1975)}]{superOp}%
  \BibitemOpen
  \bibfield  {author} {\bibinfo {author} {\bibfnamefont {M.-D.}\ \bibnamefont
  {Choi}},\ }\bibfield  {title} {\bibinfo {title} {Completely positive linear
  maps on complex matrices},\ }\href
  {https://doi.org/https://doi.org/10.1016/0024-3795(75)90075-0} {\bibfield
  {journal} {\bibinfo  {journal} {Linear Algebra and its Applications}\
  }\textbf {\bibinfo {volume} {10}},\ \bibinfo {pages} {285 } (\bibinfo {year}
  {1975})}\BibitemShut {NoStop}%
\bibitem [{\citenamefont {Weedbrook}\ \emph {et~al.}(2012)\citenamefont
  {Weedbrook}, \citenamefont {Pirandola}, \citenamefont {Garc\'{\i}a-Patr\'on},
  \citenamefont {Cerf}, \citenamefont {Ralph}, \citenamefont {Shapiro},\ and\
  \citenamefont {Lloyd}}]{densitymatIV}%
  \BibitemOpen
  \bibfield  {author} {\bibinfo {author} {\bibfnamefont {C.}~\bibnamefont
  {Weedbrook}}, \bibinfo {author} {\bibfnamefont {S.}~\bibnamefont
  {Pirandola}}, \bibinfo {author} {\bibfnamefont {R.}~\bibnamefont
  {Garc\'{\i}a-Patr\'on}}, \bibinfo {author} {\bibfnamefont {N.~J.}\
  \bibnamefont {Cerf}}, \bibinfo {author} {\bibfnamefont {T.~C.}\ \bibnamefont
  {Ralph}}, \bibinfo {author} {\bibfnamefont {J.~H.}\ \bibnamefont {Shapiro}},\
  and\ \bibinfo {author} {\bibfnamefont {S.}~\bibnamefont {Lloyd}},\ }\bibfield
   {title} {\bibinfo {title} {Gaussian quantum information},\ }\href
  {https://doi.org/10.1103/RevModPhys.84.621} {\bibfield  {journal} {\bibinfo
  {journal} {Rev. Mod. Phys.}\ }\textbf {\bibinfo {volume} {84}},\ \bibinfo
  {pages} {621} (\bibinfo {year} {2012})}\BibitemShut {NoStop}%
\bibitem [{\citenamefont {Novikov}(1970)}]{gibbsstate}%
  \BibitemOpen
  \bibfield  {author} {\bibinfo {author} {\bibfnamefont {I.~D.}\ \bibnamefont
  {Novikov}},\ }\bibfield  {title} {\bibinfo {title} {Gibbs state in quantum
  statistical physics},\ }\href {https://doi.org/10.1007/BF01075980} {\bibfield
   {journal} {\bibinfo  {journal} {Functional Analysis and Its Applications}\
  }\textbf {\bibinfo {volume} {4}},\ \bibinfo {pages} {334} (\bibinfo {year}
  {1970})}\BibitemShut {NoStop}%
\bibitem [{\citenamefont {Ticozzi}\ and\ \citenamefont
  {Viola}(2017)}]{qpuMOD1}%
  \BibitemOpen
  \bibfield  {author} {\bibinfo {author} {\bibfnamefont {F.}~\bibnamefont
  {Ticozzi}}\ and\ \bibinfo {author} {\bibfnamefont {L.}~\bibnamefont
  {Viola}},\ }\bibfield  {title} {\bibinfo {title} {Quantum and classical
  resources for unitary design of open-system evolutions},\ }\href
  {https://doi.org/10.1088/2058-9565/aa722a} {\bibfield  {journal} {\bibinfo
  {journal} {Quantum Science and Technology}\ }\textbf {\bibinfo {volume}
  {2}},\ \bibinfo {pages} {034001} (\bibinfo {year} {2017})}\BibitemShut
  {NoStop}%
\bibitem [{\citenamefont {Stinespring}(1955)}]{stinespring}%
  \BibitemOpen
  \bibfield  {author} {\bibinfo {author} {\bibfnamefont {W.~F.}\ \bibnamefont
  {Stinespring}},\ }\bibfield  {title} {\bibinfo {title} {Positive functions on
  c*-algebras},\ }\href {http://www.jstor.org/stable/2032342} {\bibfield
  {journal} {\bibinfo  {journal} {Proceedings of the American Mathematical
  Society}\ }\textbf {\bibinfo {volume} {6}},\ \bibinfo {pages} {211} (\bibinfo
  {year} {1955})}\BibitemShut {NoStop}%
\bibitem [{\citenamefont {Nielsen}(2002)}]{gwnm1}%
  \BibitemOpen
  \bibfield  {author} {\bibinfo {author} {\bibfnamefont {M.~A.}\ \bibnamefont
  {Nielsen}},\ }\bibfield  {title} {\bibinfo {title} {A simple formula for the
  average gate fidelity of a quantum dynamical operation},\ }\href
  {https://doi.org/https://doi.org/10.1016/S0375-9601(02)01272-0} {\bibfield
  {journal} {\bibinfo  {journal} {Physics Letters A}\ }\textbf {\bibinfo
  {volume} {303}},\ \bibinfo {pages} {249 } (\bibinfo {year}
  {2002})}\BibitemShut {NoStop}%
\bibitem [{\citenamefont {Sobiczewski}\ and\ \citenamefont
  {Litvinov}(2014)}]{mbextrap0}%
  \BibitemOpen
  \bibfield  {author} {\bibinfo {author} {\bibfnamefont {A.}~\bibnamefont
  {Sobiczewski}}\ and\ \bibinfo {author} {\bibfnamefont {Y.~A.}\ \bibnamefont
  {Litvinov}},\ }\bibfield  {title} {\bibinfo {title} {Predictive power of
  nuclear-mass models},\ }\href {https://doi.org/10.1103/PhysRevC.90.017302}
  {\bibfield  {journal} {\bibinfo  {journal} {Phys. Rev. C}\ }\textbf {\bibinfo
  {volume} {90}},\ \bibinfo {pages} {017302} (\bibinfo {year}
  {2014})}\BibitemShut {NoStop}%
\bibitem [{\citenamefont {Neufcourt}\ \emph {et~al.}(2018)\citenamefont
  {Neufcourt}, \citenamefont {Cao}, \citenamefont {Nazarewicz},\ and\
  \citenamefont {Viens}}]{mbextrap1}%
  \BibitemOpen
  \bibfield  {author} {\bibinfo {author} {\bibfnamefont {L.}~\bibnamefont
  {Neufcourt}}, \bibinfo {author} {\bibfnamefont {Y.}~\bibnamefont {Cao}},
  \bibinfo {author} {\bibfnamefont {W.}~\bibnamefont {Nazarewicz}},\ and\
  \bibinfo {author} {\bibfnamefont {F.}~\bibnamefont {Viens}},\ }\bibfield
  {title} {\bibinfo {title} {Bayesian approach to model-based extrapolation of
  nuclear observables},\ }\href {https://doi.org/10.1103/PhysRevC.98.034318}
  {\bibfield  {journal} {\bibinfo  {journal} {Phys. Rev. C}\ }\textbf {\bibinfo
  {volume} {98}},\ \bibinfo {pages} {034318} (\bibinfo {year}
  {2018})}\BibitemShut {NoStop}%
\bibitem [{\citenamefont {Negoita}\ \emph {et~al.}(2019)\citenamefont
  {Negoita}, \citenamefont {Vary}, \citenamefont {Luecke}, \citenamefont
  {Maris}, \citenamefont {Shirokov}, \citenamefont {Shin}, \citenamefont {Kim},
  \citenamefont {Ng}, \citenamefont {Yang}, \citenamefont {Lockner},\ and\
  \citenamefont {Prabhu}}]{mbextrap2}%
  \BibitemOpen
  \bibfield  {author} {\bibinfo {author} {\bibfnamefont {G.~A.}\ \bibnamefont
  {Negoita}}, \bibinfo {author} {\bibfnamefont {J.~P.}\ \bibnamefont {Vary}},
  \bibinfo {author} {\bibfnamefont {G.~R.}\ \bibnamefont {Luecke}}, \bibinfo
  {author} {\bibfnamefont {P.}~\bibnamefont {Maris}}, \bibinfo {author}
  {\bibfnamefont {A.~M.}\ \bibnamefont {Shirokov}}, \bibinfo {author}
  {\bibfnamefont {I.~J.}\ \bibnamefont {Shin}}, \bibinfo {author}
  {\bibfnamefont {Y.}~\bibnamefont {Kim}}, \bibinfo {author} {\bibfnamefont
  {E.~G.}\ \bibnamefont {Ng}}, \bibinfo {author} {\bibfnamefont
  {C.}~\bibnamefont {Yang}}, \bibinfo {author} {\bibfnamefont {M.}~\bibnamefont
  {Lockner}},\ and\ \bibinfo {author} {\bibfnamefont {G.~M.}\ \bibnamefont
  {Prabhu}},\ }\bibfield  {title} {\bibinfo {title} {Deep learning:
  Extrapolation tool for ab initio nuclear theory},\ }\href
  {https://doi.org/10.1103/PhysRevC.99.054308} {\bibfield  {journal} {\bibinfo
  {journal} {Phys. Rev. C}\ }\textbf {\bibinfo {volume} {99}},\ \bibinfo
  {pages} {054308} (\bibinfo {year} {2019})}\BibitemShut {NoStop}%
\bibitem [{\citenamefont {Jiang}\ \emph {et~al.}(2019)\citenamefont {Jiang},
  \citenamefont {Hagen},\ and\ \citenamefont {Papenbrock}}]{mbextrap3}%
  \BibitemOpen
  \bibfield  {author} {\bibinfo {author} {\bibfnamefont {W.~G.}\ \bibnamefont
  {Jiang}}, \bibinfo {author} {\bibfnamefont {G.}~\bibnamefont {Hagen}},\ and\
  \bibinfo {author} {\bibfnamefont {T.}~\bibnamefont {Papenbrock}},\ }\bibfield
   {title} {\bibinfo {title} {Extrapolation of nuclear structure observables
  with artificial neural networks},\ }\href
  {https://doi.org/10.1103/PhysRevC.100.054326} {\bibfield  {journal} {\bibinfo
   {journal} {Phys. Rev. C}\ }\textbf {\bibinfo {volume} {100}},\ \bibinfo
  {pages} {054326} (\bibinfo {year} {2019})}\BibitemShut {NoStop}%
\bibitem [{\citenamefont {Temme}\ \emph {et~al.}(2017)\citenamefont {Temme},
  \citenamefont {Bravyi},\ and\ \citenamefont {Gambetta}}]{zeronoiseIII}%
  \BibitemOpen
  \bibfield  {author} {\bibinfo {author} {\bibfnamefont {K.}~\bibnamefont
  {Temme}}, \bibinfo {author} {\bibfnamefont {S.}~\bibnamefont {Bravyi}},\ and\
  \bibinfo {author} {\bibfnamefont {J.~M.}\ \bibnamefont {Gambetta}},\
  }\bibfield  {title} {\bibinfo {title} {Error mitigation for short-depth
  quantum circuits},\ }\href {https://doi.org/10.1103/PhysRevLett.119.180509}
  {\bibfield  {journal} {\bibinfo  {journal} {Phys. Rev. Lett.}\ }\textbf
  {\bibinfo {volume} {119}},\ \bibinfo {pages} {180509} (\bibinfo {year}
  {2017})}\BibitemShut {NoStop}%
\bibitem [{\citenamefont {Li}\ and\ \citenamefont
  {Benjamin}(2017)}]{zeronoise}%
  \BibitemOpen
  \bibfield  {author} {\bibinfo {author} {\bibfnamefont {Y.}~\bibnamefont
  {Li}}\ and\ \bibinfo {author} {\bibfnamefont {S.~C.}\ \bibnamefont
  {Benjamin}},\ }\bibfield  {title} {\bibinfo {title} {Efficient variational
  quantum simulator incorporating active error minimization},\ }\href
  {https://doi.org/10.1103/PhysRevX.7.021050} {\bibfield  {journal} {\bibinfo
  {journal} {Phys. Rev. X}\ }\textbf {\bibinfo {volume} {7}},\ \bibinfo {pages}
  {021050} (\bibinfo {year} {2017})}\BibitemShut {NoStop}%
\bibitem [{\citenamefont {Shaw}(2020{\natexlab{b}})}]{PNR0}%
  \BibitemOpen
  \bibfield  {author} {\bibinfo {author} {\bibfnamefont {A.}~\bibnamefont
  {Shaw}},\ }\bibfield  {title} {\bibinfo {title}
  {\href{http://bit.ly/BenchNonAb}{Benchmarking Non-Abelian Lattice Gauge
  Theories With NISQ Algorithms}}} (\bibinfo {year} {October 31,
  2020}{\natexlab{b}}),\ \bibinfo {note} {{2020} Fall Meeting of the APS
  Division of Nuclear Physics}\BibitemShut {NoStop}%
\bibitem [{\citenamefont {{Yamada}}(1962)}]{complexi}%
  \BibitemOpen
  \bibfield  {author} {\bibinfo {author} {\bibfnamefont {H.}~\bibnamefont
  {{Yamada}}},\ }\bibfield  {title} {\bibinfo {title} {Real-time computation
  and recursive functions not real-time computable},\ }\href
  {https://doi.org/10.1109/TEC.1962.5219459} {\bibfield  {journal} {\bibinfo
  {journal} {IRE Transactions on Electronic Computers}\ }\textbf {\bibinfo
  {volume} {EC-11}},\ \bibinfo {pages} {753} (\bibinfo {year}
  {1962})}\BibitemShut {NoStop}%
\bibitem [{\citenamefont {Smale}(1997)}]{complexii}%
  \BibitemOpen
  \bibfield  {author} {\bibinfo {author} {\bibfnamefont {S.}~\bibnamefont
  {Smale}},\ }\bibfield  {title} {\bibinfo {title} {Complexity theory and
  numerical analysis},\ }\href {https://doi.org/10.1017/S0962492900002774}
  {\bibfield  {journal} {\bibinfo  {journal} {Acta Numerica}\ }\textbf
  {\bibinfo {volume} {6}},\ \bibinfo {pages} {523–551} (\bibinfo {year}
  {1997})}\BibitemShut {NoStop}%
\bibitem [{\citenamefont {Aaronson}(2013)}]{complexAaron0}%
  \BibitemOpen
  \bibfield  {author} {\bibinfo {author} {\bibfnamefont {S.~J.}\ \bibnamefont
  {Aaronson}},\ }\href@noop {} {\emph {\bibinfo {title}
  {\href{http://bit.ly/saaronpaleoc}{Quantum Computing Since Democritus}}}}\
  (\bibinfo  {publisher} {Cambridge University Press},\ \bibinfo {address}
  {USA},\ \bibinfo {year} {2013})\ Chap.~\bibinfo {chapter} {5}, pp.\ \bibinfo
  {pages} {44--50}\BibitemShut {NoStop}%
\bibitem [{\citenamefont {Scott}\ \emph {et~al.}(2020)\citenamefont {Scott},
  \citenamefont {{Kuperberg}},\ and\ \citenamefont
  {{Granade}}}]{complexAaron1}%
  \BibitemOpen
  \bibfield  {author} {\bibinfo {author} {\bibfnamefont {A.}~\bibnamefont
  {Scott}}, \bibinfo {author} {\bibfnamefont {G.}~\bibnamefont {{Kuperberg}}},\
  and\ \bibinfo {author} {\bibfnamefont {C.}~\bibnamefont {{Granade}}},\ }\href
  {https://complexityzoo.uwaterloo.ca/Complexity_Zoo} {\bibinfo {title}
  {Complexity zoo}},\ \bibinfo {howpublished} {February 18} (\bibinfo {year}
  {2020})\BibitemShut {NoStop}%
\bibitem [{\citenamefont {Cobham}(1965)}]{cobham}%
  \BibitemOpen
  \bibfield  {author} {\bibinfo {author} {\bibfnamefont {A.}~\bibnamefont
  {Cobham}},\ }\href {https://doi.org/10.2307/2270886} {\emph {\bibinfo {title}
  {Logic, Methodology and Philosophy of Science: Proceedings of the 1964
  International Congress}}}\ (\bibinfo  {publisher} {North-Holland
  Publishing},\ \bibinfo {year} {1965})\ Chap.\ \bibinfo {chapter} {The
  Intrinsic Computational Difficulty of Functions}, pp.\ \bibinfo {pages}
  {24--30}\BibitemShut {NoStop}%
\bibitem [{\citenamefont {Edmonds}(1965)}]{edmonds}%
  \BibitemOpen
  \bibfield  {author} {\bibinfo {author} {\bibfnamefont {J.}~\bibnamefont
  {Edmonds}},\ }\bibfield  {title} {\bibinfo {title} {Paths, trees, and
  flowers},\ }\href {https://doi.org/10.4153/CJM-1965-045-4} {\bibfield
  {journal} {\bibinfo  {journal} {Canadian Journal of Mathematics}\ }\textbf
  {\bibinfo {volume} {17}},\ \bibinfo {pages} {449} (\bibinfo {year}
  {1965})}\BibitemShut {NoStop}%
\bibitem [{\citenamefont {Goldreich}(2008)}]{complexOne}%
  \BibitemOpen
  \bibfield  {author} {\bibinfo {author} {\bibfnamefont {O.}~\bibnamefont
  {Goldreich}},\ }\href {https://doi.org/10.1017/CBO9780511804106} {\emph
  {\bibinfo {title} {Computational complexity. A conceptual perspective}}},\
  Vol.~\bibinfo {volume} {39}\ (\bibinfo {year} {2008})\ pp.\ \bibinfo {pages}
  {32--33}\BibitemShut {NoStop}%
\bibitem [{\citenamefont {Turing}(1937)}]{cturing0}%
  \BibitemOpen
  \bibfield  {author} {\bibinfo {author} {\bibfnamefont {A.~M.}\ \bibnamefont
  {Turing}},\ }\bibfield  {title} {\bibinfo {title} {On computable numbers,
  with an application to the entscheidungsproblem},\ }\href
  {https://doi.org/10.1112/plms/s2-42.1.230} {\bibfield  {journal} {\bibinfo
  {journal} {Proceedings of the London Mathematical Society}\ }\textbf
  {\bibinfo {volume} {s2-42}},\ \bibinfo {pages} {230} (\bibinfo {year}
  {1937})}\BibitemShut {NoStop}%
\bibitem [{\citenamefont {Turing}(1938)}]{cturing1}%
  \BibitemOpen
  \bibfield  {author} {\bibinfo {author} {\bibfnamefont {A.~M.}\ \bibnamefont
  {Turing}},\ }\bibfield  {title} {\bibinfo {title} {On computable numbers,
  with an application to the entscheidungsproblem. a correction},\ }\href
  {https://doi.org/10.1112/plms/s2-43.6.544} {\bibfield  {journal} {\bibinfo
  {journal} {Proceedings of the London Mathematical Society}\ }\textbf
  {\bibinfo {volume} {s2-43}},\ \bibinfo {pages} {544} (\bibinfo {year}
  {1938})}\BibitemShut {NoStop}%
\bibitem [{\citenamefont {Turing}(1996)}]{cturing2}%
  \BibitemOpen
  \bibfield  {author} {\bibinfo {author} {\bibfnamefont {A.~M.}\ \bibnamefont
  {Turing}},\ }\bibfield  {title} {\bibinfo {title} {{Intelligent Machinery, A
  Heretical Theory*}},\ }\href {https://doi.org/10.1093/philmat/4.3.256}
  {\bibfield  {journal} {\bibinfo  {journal} {Philosophia Mathematica}\
  }\textbf {\bibinfo {volume} {4}},\ \bibinfo {pages} {256} (\bibinfo {year}
  {1996})}\BibitemShut {NoStop}%
\bibitem [{\citenamefont {Benioff}(1980)}]{qturing0}%
  \BibitemOpen
  \bibfield  {author} {\bibinfo {author} {\bibfnamefont {P.}~\bibnamefont
  {Benioff}},\ }\bibfield  {title} {\bibinfo {title} {The computer as a
  physical system: A microscopic quantum mechanical hamiltonian model of
  computers as represented by turing machines},\ }\href
  {https://doi.org/10.1007/BF01011339} {\bibfield  {journal} {\bibinfo
  {journal} {Journal of Statistical Physics}\ }\textbf {\bibinfo {volume}
  {22}},\ \bibinfo {pages} {563} (\bibinfo {year} {1980})}\BibitemShut
  {NoStop}%
\bibitem [{\citenamefont {Benioff}(1982)}]{qturing1}%
  \BibitemOpen
  \bibfield  {author} {\bibinfo {author} {\bibfnamefont {P.}~\bibnamefont
  {Benioff}},\ }\bibfield  {title} {\bibinfo {title} {Quantum mechanical
  hamiltonian models of turing machines},\ }\href
  {https://doi.org/10.1007/BF01342185} {\bibfield  {journal} {\bibinfo
  {journal} {Journal of Statistical Physics}\ }\textbf {\bibinfo {volume}
  {29}},\ \bibinfo {pages} {515} (\bibinfo {year} {1982})}\BibitemShut
  {NoStop}%
\bibitem [{\citenamefont {Deutsch}\ and\ \citenamefont
  {Penrose}(1985)}]{qturing2}%
  \BibitemOpen
  \bibfield  {author} {\bibinfo {author} {\bibfnamefont {D.}~\bibnamefont
  {Deutsch}}\ and\ \bibinfo {author} {\bibfnamefont {R.}~\bibnamefont
  {Penrose}},\ }\bibfield  {title} {\bibinfo {title} {Quantum theory, the
  church-turing principle and the universal quantum computer},\ }\href
  {https://doi.org/10.1098/rspa.1985.0070} {\bibfield  {journal} {\bibinfo
  {journal} {Proceedings of the Royal Society of London. A. Mathematical and
  Physical Sciences}\ }\textbf {\bibinfo {volume} {400}},\ \bibinfo {pages}
  {97} (\bibinfo {year} {1985})}\BibitemShut {NoStop}%
\bibitem [{\citenamefont {Sipser}(2012)}]{complexThree}%
  \BibitemOpen
  \bibfield  {author} {\bibinfo {author} {\bibfnamefont {M.}~\bibnamefont
  {Sipser}},\ }\href {https://books.google.com/books?id=P3f6CAAAQBAJ} {\emph
  {\bibinfo {title} {Introduction to the Theory of Computation}}}\ (\bibinfo
  {publisher} {Cengage Learning},\ \bibinfo {year} {2012})\BibitemShut
  {NoStop}%
\bibitem [{\citenamefont {Bernstein}\ and\ \citenamefont
  {Vazirani}(1997)}]{complexFour}%
  \BibitemOpen
  \bibfield  {author} {\bibinfo {author} {\bibfnamefont {E.}~\bibnamefont
  {Bernstein}}\ and\ \bibinfo {author} {\bibfnamefont {U.}~\bibnamefont
  {Vazirani}},\ }\bibfield  {title} {\bibinfo {title} {Quantum complexity
  theory},\ }\href {https://doi.org/10.1137/S0097539796300921} {\bibfield
  {journal} {\bibinfo  {journal} {SIAM Journal on Computing}\ }\textbf
  {\bibinfo {volume} {26}},\ \bibinfo {pages} {1411} (\bibinfo {year}
  {1997})}\BibitemShut {NoStop}%
\bibitem [{\citenamefont {{Aaronson}}\ and\ \citenamefont
  {{Ben-David}}(2015)}]{qspeedII}%
  \BibitemOpen
  \bibfield  {author} {\bibinfo {author} {\bibfnamefont {S.}~\bibnamefont
  {{Aaronson}}}\ and\ \bibinfo {author} {\bibfnamefont {S.}~\bibnamefont
  {{Ben-David}}},\ }\bibfield  {title} {\bibinfo {title} {{Sculpting Quantum
  Speedups}},\ }\href@noop {} {\bibfield  {journal} {\bibinfo  {journal} {arXiv
  e-prints}\ ,\ \bibinfo {eid} {arXiv:1512.04016}} (\bibinfo {year} {2015})},\
  \Eprint {https://arxiv.org/abs/1512.04016} {arXiv:1512.04016 [quant-ph]}
  \BibitemShut {NoStop}%
\bibitem [{\citenamefont {Feynman}(1982)}]{feynmansupremacy}%
  \BibitemOpen
  \bibfield  {author} {\bibinfo {author} {\bibfnamefont {R.~P.}\ \bibnamefont
  {Feynman}},\ }\bibfield  {title} {\bibinfo {title} {Simulating physics with
  computers},\ }\href {https://doi.org/10.1007/BF02650179} {\bibfield
  {journal} {\bibinfo  {journal} {International Journal of Theoretical
  Physics}\ }\textbf {\bibinfo {volume} {21}},\ \bibinfo {pages} {467}
  (\bibinfo {year} {1982})}\BibitemShut {NoStop}%
\bibitem [{\citenamefont {Lloyd}(1996)}]{complexTwo}%
  \BibitemOpen
  \bibfield  {author} {\bibinfo {author} {\bibfnamefont {S.}~\bibnamefont
  {Lloyd}},\ }\bibfield  {title} {\bibinfo {title} {Universal quantum
  simulators},\ }\href {https://doi.org/10.1126/science.273.5278.1073}
  {\bibfield  {journal} {\bibinfo  {journal} {Science}\ }\textbf {\bibinfo
  {volume} {273}},\ \bibinfo {pages} {1073} (\bibinfo {year}
  {1996})}\BibitemShut {NoStop}%
\bibitem [{\citenamefont {Aharonov}\ and\ \citenamefont
  {Ta-Shma}(2003)}]{completeIII}%
  \BibitemOpen
  \bibfield  {author} {\bibinfo {author} {\bibfnamefont {D.}~\bibnamefont
  {Aharonov}}\ and\ \bibinfo {author} {\bibfnamefont {A.}~\bibnamefont
  {Ta-Shma}},\ }\bibfield  {title} {\bibinfo {title} {Adiabatic quantum state
  generation and statistical zero knowledge},\ }in\ \href
  {https://doi.org/10.1145/780542.780546} {\emph {\bibinfo {booktitle}
  {Proceedings of the Thirty-Fifth Annual ACM Symposium on Theory of
  Computing}}}\ (\bibinfo  {publisher} {Association for Computing Machinery},\
  \bibinfo {address} {New York, NY, USA},\ \bibinfo {year} {2003})\ pp.\
  \bibinfo {pages} {20--29}\BibitemShut {NoStop}%
\bibitem [{\citenamefont {{Nagaj}}(2010)}]{completeIV}%
  \BibitemOpen
  \bibfield  {author} {\bibinfo {author} {\bibfnamefont {D.}~\bibnamefont
  {{Nagaj}}},\ }\bibfield  {title} {\bibinfo {title} {{Fast universal quantum
  computation with railroad-switch local Hamiltonians}},\ }\href
  {https://doi.org/10.1063/1.3384661} {\bibfield  {journal} {\bibinfo
  {journal} {Journal of Mathematical Physics}\ }\textbf {\bibinfo {volume}
  {51}},\ \bibinfo {pages} {062201} (\bibinfo {year} {2010})},\ \Eprint
  {https://arxiv.org/abs/0908.4219} {arXiv:0908.4219 [quant-ph]} \BibitemShut
  {NoStop}%
\bibitem [{\citenamefont {{Berry}}\ \emph {et~al.}(2015)\citenamefont
  {{Berry}}, \citenamefont {{Childs}},\ and\ \citenamefont
  {{Kothari}}}]{completeI}%
  \BibitemOpen
  \bibfield  {author} {\bibinfo {author} {\bibfnamefont {D.~W.}\ \bibnamefont
  {{Berry}}}, \bibinfo {author} {\bibfnamefont {A.~M.}\ \bibnamefont
  {{Childs}}},\ and\ \bibinfo {author} {\bibfnamefont {R.}~\bibnamefont
  {{Kothari}}},\ }\bibfield  {title} {\bibinfo {title} {{Hamiltonian simulation
  with nearly optimal dependence on all parameters}},\ }\href@noop {}
  {\bibfield  {journal} {\bibinfo  {journal} {arXiv e-prints}\ ,\ \bibinfo
  {eid} {arXiv:1501.01715}} (\bibinfo {year} {2015})},\ \Eprint
  {https://arxiv.org/abs/1501.01715} {arXiv:1501.01715 [quant-ph]} \BibitemShut
  {NoStop}%
\bibitem [{\citenamefont {{Hao Low}}\ and\ \citenamefont
  {{Chuang}}(2016)}]{completeII}%
  \BibitemOpen
  \bibfield  {author} {\bibinfo {author} {\bibfnamefont {G.}~\bibnamefont {{Hao
  Low}}}\ and\ \bibinfo {author} {\bibfnamefont {I.~L.}\ \bibnamefont
  {{Chuang}}},\ }\bibfield  {title} {\bibinfo {title} {{Hamiltonian Simulation
  by Qubitization}},\ }\href@noop {} {\bibfield  {journal} {\bibinfo  {journal}
  {arXiv e-prints}\ ,\ \bibinfo {eid} {arXiv:1610.06546}} (\bibinfo {year}
  {2016})},\ \Eprint {https://arxiv.org/abs/1610.06546} {arXiv:1610.06546
  [quant-ph]} \BibitemShut {NoStop}%
\bibitem [{\citenamefont {Affleck}\ \emph {et~al.}(1987)\citenamefont
  {Affleck}, \citenamefont {Kennedy}, \citenamefont {Lieb},\ and\ \citenamefont
  {Tasaki}}]{MPSI}%
  \BibitemOpen
  \bibfield  {author} {\bibinfo {author} {\bibfnamefont {I.}~\bibnamefont
  {Affleck}}, \bibinfo {author} {\bibfnamefont {T.}~\bibnamefont {Kennedy}},
  \bibinfo {author} {\bibfnamefont {E.~H.}\ \bibnamefont {Lieb}},\ and\
  \bibinfo {author} {\bibfnamefont {H.}~\bibnamefont {Tasaki}},\ }\bibfield
  {title} {\bibinfo {title} {Rigorous results on valence-bond ground states in
  antiferromagnets},\ }\href {https://doi.org/10.1103/PhysRevLett.59.799}
  {\bibfield  {journal} {\bibinfo  {journal} {Phys. Rev. Lett.}\ }\textbf
  {\bibinfo {volume} {59}},\ \bibinfo {pages} {799} (\bibinfo {year}
  {1987})}\BibitemShut {NoStop}%
\bibitem [{\citenamefont {{Perez-Garcia}}\ \emph {et~al.}(2006)\citenamefont
  {{Perez-Garcia}}, \citenamefont {{Verstraete}}, \citenamefont {{Wolf}},\ and\
  \citenamefont {{Cirac}}}]{MPSII}%
  \BibitemOpen
  \bibfield  {author} {\bibinfo {author} {\bibfnamefont {D.}~\bibnamefont
  {{Perez-Garcia}}}, \bibinfo {author} {\bibfnamefont {F.}~\bibnamefont
  {{Verstraete}}}, \bibinfo {author} {\bibfnamefont {M.~M.}\ \bibnamefont
  {{Wolf}}},\ and\ \bibinfo {author} {\bibfnamefont {J.~I.}\ \bibnamefont
  {{Cirac}}},\ }\bibfield  {title} {\bibinfo {title} {{Matrix Product State
  Representations}},\ }\href@noop {} {\  (\bibinfo {year} {2006})},\ \Eprint
  {https://arxiv.org/abs/quant-ph/0608197} {arXiv:quant-ph/0608197 [quant-ph]}
  \BibitemShut {NoStop}%
\bibitem [{\citenamefont {Verstraete}\ \emph {et~al.}(2008)\citenamefont
  {Verstraete}, \citenamefont {Murg},\ and\ \citenamefont {Cirac}}]{MPSIII}%
  \BibitemOpen
  \bibfield  {author} {\bibinfo {author} {\bibfnamefont {F.}~\bibnamefont
  {Verstraete}}, \bibinfo {author} {\bibfnamefont {V.}~\bibnamefont {Murg}},\
  and\ \bibinfo {author} {\bibfnamefont {J.}~\bibnamefont {Cirac}},\ }\bibfield
   {title} {\bibinfo {title} {Matrix product states, projected entangled pair
  states, and variational renormalization group methods for quantum spin
  systems},\ }\href {https://doi.org/10.1080/14789940801912366} {\bibfield
  {journal} {\bibinfo  {journal} {Advances in Physics}\ }\textbf {\bibinfo
  {volume} {57}},\ \bibinfo {pages} {143} (\bibinfo {year} {2008})}\BibitemShut
  {NoStop}%
\bibitem [{\citenamefont {Ganahl}\ \emph {et~al.}(2015)\citenamefont {Ganahl},
  \citenamefont {Aichhorn}, \citenamefont {Evertz}, \citenamefont
  {Thunstr\"om}, \citenamefont {Held},\ and\ \citenamefont
  {Verstraete}}]{MPSIV}%
  \BibitemOpen
  \bibfield  {author} {\bibinfo {author} {\bibfnamefont {M.}~\bibnamefont
  {Ganahl}}, \bibinfo {author} {\bibfnamefont {M.}~\bibnamefont {Aichhorn}},
  \bibinfo {author} {\bibfnamefont {H.~G.}\ \bibnamefont {Evertz}}, \bibinfo
  {author} {\bibfnamefont {P.}~\bibnamefont {Thunstr\"om}}, \bibinfo {author}
  {\bibfnamefont {K.}~\bibnamefont {Held}},\ and\ \bibinfo {author}
  {\bibfnamefont {F.}~\bibnamefont {Verstraete}},\ }\bibfield  {title}
  {\bibinfo {title} {Efficient dmft impurity solver using real-time dynamics
  with matrix product states},\ }\href
  {https://doi.org/10.1103/PhysRevB.92.155132} {\bibfield  {journal} {\bibinfo
  {journal} {Phys. Rev. B}\ }\textbf {\bibinfo {volume} {92}},\ \bibinfo
  {pages} {155132} (\bibinfo {year} {2015})}\BibitemShut {NoStop}%
\bibitem [{\citenamefont {Buyens}\ \emph {et~al.}(2017)\citenamefont {Buyens},
  \citenamefont {Haegeman}, \citenamefont {Hebenstreit}, \citenamefont
  {Verstraete},\ and\ \citenamefont {Van~Acoleyen}}]{MPSV}%
  \BibitemOpen
  \bibfield  {author} {\bibinfo {author} {\bibfnamefont {B.}~\bibnamefont
  {Buyens}}, \bibinfo {author} {\bibfnamefont {J.}~\bibnamefont {Haegeman}},
  \bibinfo {author} {\bibfnamefont {F.}~\bibnamefont {Hebenstreit}}, \bibinfo
  {author} {\bibfnamefont {F.}~\bibnamefont {Verstraete}},\ and\ \bibinfo
  {author} {\bibfnamefont {K.}~\bibnamefont {Van~Acoleyen}},\ }\bibfield
  {title} {\bibinfo {title} {Real-time simulation of the schwinger effect with
  matrix product states},\ }\href {https://doi.org/10.1103/PhysRevD.96.114501}
  {\bibfield  {journal} {\bibinfo  {journal} {Phys. Rev. D}\ }\textbf {\bibinfo
  {volume} {96}},\ \bibinfo {pages} {114501} (\bibinfo {year}
  {2017})}\BibitemShut {NoStop}%
\bibitem [{\citenamefont {Paeckel}\ \emph {et~al.}(2019)\citenamefont
  {Paeckel}, \citenamefont {Köhler}, \citenamefont {Swoboda}, \citenamefont
  {Manmana}, \citenamefont {Schollwöck},\ and\ \citenamefont {Hubig}}]{MPSVI}%
  \BibitemOpen
  \bibfield  {author} {\bibinfo {author} {\bibfnamefont {S.}~\bibnamefont
  {Paeckel}}, \bibinfo {author} {\bibfnamefont {T.}~\bibnamefont {Köhler}},
  \bibinfo {author} {\bibfnamefont {A.}~\bibnamefont {Swoboda}}, \bibinfo
  {author} {\bibfnamefont {S.~R.}\ \bibnamefont {Manmana}}, \bibinfo {author}
  {\bibfnamefont {U.}~\bibnamefont {Schollwöck}},\ and\ \bibinfo {author}
  {\bibfnamefont {C.}~\bibnamefont {Hubig}},\ }\bibfield  {title} {\bibinfo
  {title} {Time-evolution methods for matrix-product states},\ }\href
  {https://doi.org/https://doi.org/10.1016/j.aop.2019.167998} {\bibfield
  {journal} {\bibinfo  {journal} {Annals of Physics}\ }\textbf {\bibinfo
  {volume} {411}},\ \bibinfo {pages} {167998} (\bibinfo {year}
  {2019})}\BibitemShut {NoStop}%
\bibitem [{\citenamefont {White}(1992)}]{DMRGI}%
  \BibitemOpen
  \bibfield  {author} {\bibinfo {author} {\bibfnamefont {S.~R.}\ \bibnamefont
  {White}},\ }\bibfield  {title} {\bibinfo {title} {Density matrix formulation
  for quantum renormalization groups},\ }\href
  {https://doi.org/10.1103/PhysRevLett.69.2863} {\bibfield  {journal} {\bibinfo
   {journal} {Phys. Rev. Lett.}\ }\textbf {\bibinfo {volume} {69}},\ \bibinfo
  {pages} {2863} (\bibinfo {year} {1992})}\BibitemShut {NoStop}%
\bibitem [{\citenamefont {Gobert}\ \emph {et~al.}(2005)\citenamefont {Gobert},
  \citenamefont {Kollath}, \citenamefont {Schollw\"ock},\ and\ \citenamefont
  {Sch\"utz}}]{DMRGII}%
  \BibitemOpen
  \bibfield  {author} {\bibinfo {author} {\bibfnamefont {D.}~\bibnamefont
  {Gobert}}, \bibinfo {author} {\bibfnamefont {C.}~\bibnamefont {Kollath}},
  \bibinfo {author} {\bibfnamefont {U.}~\bibnamefont {Schollw\"ock}},\ and\
  \bibinfo {author} {\bibfnamefont {G.}~\bibnamefont {Sch\"utz}},\ }\bibfield
  {title} {\bibinfo {title} {Real-time dynamics in spin-$\frac{1}{2}$ chains
  with adaptive time-dependent density matrix renormalization group},\ }\href
  {https://doi.org/10.1103/PhysRevE.71.036102} {\bibfield  {journal} {\bibinfo
  {journal} {Phys. Rev. E}\ }\textbf {\bibinfo {volume} {71}},\ \bibinfo
  {pages} {036102} (\bibinfo {year} {2005})}\BibitemShut {NoStop}%
\bibitem [{\citenamefont {Feiguin}\ and\ \citenamefont
  {White}(2005)}]{DMRGIII}%
  \BibitemOpen
  \bibfield  {author} {\bibinfo {author} {\bibfnamefont {A.~E.}\ \bibnamefont
  {Feiguin}}\ and\ \bibinfo {author} {\bibfnamefont {S.~R.}\ \bibnamefont
  {White}},\ }\bibfield  {title} {\bibinfo {title} {Time-step targeting methods
  for real-time dynamics using the density matrix renormalization group},\
  }\href {https://doi.org/10.1103/PhysRevB.72.020404} {\bibfield  {journal}
  {\bibinfo  {journal} {Phys. Rev. B}\ }\textbf {\bibinfo {volume} {72}},\
  \bibinfo {pages} {020404} (\bibinfo {year} {2005})}\BibitemShut {NoStop}%
\bibitem [{\citenamefont {Schollwöck}(2011)}]{DMRGIV}%
  \BibitemOpen
  \bibfield  {author} {\bibinfo {author} {\bibfnamefont {U.}~\bibnamefont
  {Schollwöck}},\ }\bibfield  {title} {\bibinfo {title} {The density-matrix
  renormalization group in the age of matrix product states},\ }\href
  {https://doi.org/https://doi.org/10.1016/j.aop.2010.09.012} {\bibfield
  {journal} {\bibinfo  {journal} {Annals of Physics}\ }\textbf {\bibinfo
  {volume} {326}},\ \bibinfo {pages} {96 } (\bibinfo {year} {2011})},\ \bibinfo
  {note} {january 2011 Special Issue}\BibitemShut {NoStop}%
\bibitem [{\citenamefont {{Haah}}\ \emph {et~al.}(2018)\citenamefont {{Haah}},
  \citenamefont {{Hastings}}, \citenamefont {{Kothari}},\ and\ \citenamefont
  {{Hao Low}}}]{interI}%
  \BibitemOpen
  \bibfield  {author} {\bibinfo {author} {\bibfnamefont {J.}~\bibnamefont
  {{Haah}}}, \bibinfo {author} {\bibfnamefont {M.~B.}\ \bibnamefont
  {{Hastings}}}, \bibinfo {author} {\bibfnamefont {R.}~\bibnamefont
  {{Kothari}}},\ and\ \bibinfo {author} {\bibfnamefont {G.}~\bibnamefont {{Hao
  Low}}},\ }\bibfield  {title} {\bibinfo {title} {{Quantum algorithm for
  simulating real time evolution of lattice Hamiltonians}},\ }\href@noop {}
  {\bibfield  {journal} {\bibinfo  {journal} {arXiv e-prints}\ ,\ \bibinfo
  {eid} {arXiv:1801.03922}} (\bibinfo {year} {2018})},\ \Eprint
  {https://arxiv.org/abs/1801.03922} {arXiv:1801.03922 [quant-ph]} \BibitemShut
  {NoStop}%
\bibitem [{\citenamefont {Trotter}(1959)}]{trotterI}%
  \BibitemOpen
  \bibfield  {author} {\bibinfo {author} {\bibfnamefont {H.~F.}\ \bibnamefont
  {Trotter}},\ }\bibfield  {title} {\bibinfo {title} {On the product of
  semi-groups of operators},\ }\href {http://www.jstor.org/stable/2033649}
  {\bibfield  {journal} {\bibinfo  {journal} {Proceedings of the American
  Mathematical Society}\ }\textbf {\bibinfo {volume} {10}},\ \bibinfo {pages}
  {545} (\bibinfo {year} {1959})}\BibitemShut {NoStop}%
\bibitem [{\citenamefont {Kato}(1974)}]{trotterII}%
  \BibitemOpen
  \bibfield  {author} {\bibinfo {author} {\bibfnamefont {T.}~\bibnamefont
  {Kato}},\ }\bibfield  {title} {\bibinfo {title} {On the trotter-lie product
  formula},\ }\href {https://doi.org/10.3792/pja/1195518790} {\bibfield
  {journal} {\bibinfo  {journal} {Proc. Japan Acad.}\ }\textbf {\bibinfo
  {volume} {50}},\ \bibinfo {pages} {694} (\bibinfo {year} {1974})}\BibitemShut
  {NoStop}%
\bibitem [{\citenamefont {Hatano}\ and\ \citenamefont
  {Suzuki}(2005)}]{trotterIII}%
  \BibitemOpen
  \bibfield  {author} {\bibinfo {author} {\bibfnamefont {N.}~\bibnamefont
  {Hatano}}\ and\ \bibinfo {author} {\bibfnamefont {M.}~\bibnamefont
  {Suzuki}},\ }\href {https://doi.org/10.1007/11526216_2} {\emph {\bibinfo
  {title} {Quantum Annealing and Other Optimization Methods}}},\ edited by\
  \bibinfo {editor} {\bibfnamefont {A.}~\bibnamefont {Das}}\ and\ \bibinfo
  {editor} {\bibfnamefont {B.}~\bibnamefont {K.~Chakrabarti}}\ (\bibinfo
  {publisher} {Springer Berlin Heidelberg},\ \bibinfo {address} {Berlin,
  Heidelberg},\ \bibinfo {year} {2005})\ pp.\ \bibinfo {pages}
  {37--68}\BibitemShut {NoStop}%
\bibitem [{\citenamefont {Abraham}\ \emph {et~al.}(2019)\citenamefont
  {Abraham}, \citenamefont {AduOffei}, \citenamefont {Akhalwaya}, \citenamefont
  {Aleksandrowicz}, \citenamefont {Alexander}, \citenamefont {Alexandrowics},
  \citenamefont {Arbel}, \citenamefont {Asfaw}, \citenamefont {Azaustre},
  \citenamefont {AzizNgoueya}, \citenamefont {Barkoutsos}, \citenamefont
  {Barron}, \citenamefont {Bello}, \citenamefont {Ben-Haim}, \citenamefont
  {Bevenius}, \citenamefont {Bishop}, \citenamefont {Bolos}, \citenamefont
  {Bosch}, \citenamefont {Bravyi}, \citenamefont {Bucher}, \citenamefont
  {Burov}, \citenamefont {Cabrera}, \citenamefont {Calpin}, \citenamefont
  {Capelluto}, \citenamefont {Carballo}, \citenamefont {Carrascal},
  \citenamefont {Chen}, \citenamefont {Chen}, \citenamefont {Chen},
  \citenamefont {Chow}, \citenamefont {Claus}, \citenamefont {Clauss},
  \citenamefont {Cross}, \citenamefont {Cross}, \citenamefont {Cross},
  \citenamefont {Cruz-Benito}, \citenamefont {Culver}, \citenamefont
  {C{\'o}rcoles-Gonzales}, \citenamefont {Dague}, \citenamefont {Dandachi},
  \citenamefont {Dartiailh}, \citenamefont {DavideFrr}, \citenamefont {Davila},
  \citenamefont {Dekusar}, \citenamefont {Ding}, \citenamefont {Doi},
  \citenamefont {Drechsler}, \citenamefont {Drew}, \citenamefont {Dumitrescu},
  \citenamefont {Dumon}, \citenamefont {Duran}, \citenamefont {EL-Safty},
  \citenamefont {Eastman}, \citenamefont {Eendebak}, \citenamefont {Egger},
  \citenamefont {Everitt}, \citenamefont {Fern{\'a}ndez}, \citenamefont
  {Ferrera}, \citenamefont {Frisch}, \citenamefont {Fuhrer}, \citenamefont
  {GEORGE}, \citenamefont {Gacon}, \citenamefont {Gadi}, \citenamefont {Gago},
  \citenamefont {Gambella}, \citenamefont {Gambetta}, \citenamefont
  {Gammanpila}, \citenamefont {Garcia}, \citenamefont {Garion}, \citenamefont
  {Gilliam}, \citenamefont {Gomez-Mosquera}, \citenamefont {de~la
  Puente~Gonz{\'a}lez}, \citenamefont {Gorzinski}, \citenamefont {Gould},
  \citenamefont {Greenberg}, \citenamefont {Grinko}, \citenamefont {Guan},
  \citenamefont {Gunnels}, \citenamefont {Haglund}, \citenamefont {Haide},
  \citenamefont {Hamamura}, \citenamefont {Havlicek}, \citenamefont {Hellmers},
  \citenamefont {Herok}, \citenamefont {Hillmich}, \citenamefont {Horii},
  \citenamefont {Howington}, \citenamefont {Hu}, \citenamefont {Hu},
  \citenamefont {Imai}, \citenamefont {Imamichi}, \citenamefont {Ishizaki},
  \citenamefont {Iten}, \citenamefont {Itoko}, \citenamefont {JamesSeaward},
  \citenamefont {Javadi}, \citenamefont {Javadi-Abhari}, \citenamefont
  {Jessica}, \citenamefont {Johns}, \citenamefont {Kachmann}, \citenamefont
  {Kanazawa}, \citenamefont {Kang-Bae}, \citenamefont {Karazeev}, \citenamefont
  {Kassebaum}, \citenamefont {King}, \citenamefont {Knabberjoe}, \citenamefont
  {Kovyrshin}, \citenamefont {Krishnakumar}, \citenamefont {Krishnan},
  \citenamefont {Krsulich}, \citenamefont {Kus}, \citenamefont {LaRose},
  \citenamefont {Lambert}, \citenamefont {Latone}, \citenamefont {Lawrence},
  \citenamefont {Liu}, \citenamefont {Liu}, \citenamefont {Maeng},
  \citenamefont {Malyshev}, \citenamefont {Marecek}, \citenamefont {Marques},
  \citenamefont {Mathews}, \citenamefont {Matsuo}, \citenamefont {McClure},
  \citenamefont {McGarry}, \citenamefont {McKay}, \citenamefont {McPherson},
  \citenamefont {Meesala}, \citenamefont {Mevissen}, \citenamefont {Mezzacapo},
  \citenamefont {Midha}, \citenamefont {Minev}, \citenamefont {Mitchell},
  \citenamefont {Moll}, \citenamefont {Mooring}, \citenamefont {Morales},
  \citenamefont {Moran}, \citenamefont {MrF}, \citenamefont {Murali},
  \citenamefont {M{\"u}ggenburg}, \citenamefont {Nadlinger}, \citenamefont
  {Nakanishi}, \citenamefont {Nannicini}, \citenamefont {Nation}, \citenamefont
  {Navarro}, \citenamefont {Naveh}, \citenamefont {Neagle}, \citenamefont
  {Neuweiler}, \citenamefont {Niroula}, \citenamefont {Norlen}, \citenamefont
  {O'Riordan}, \citenamefont {Ogunbayo}, \citenamefont {Ollitrault},
  \citenamefont {Oud}, \citenamefont {Padilha}, \citenamefont {Paik},
  \citenamefont {Perriello}, \citenamefont {Phan}, \citenamefont {Piro},
  \citenamefont {Pistoia}, \citenamefont {Pozas-iKerstjens}, \citenamefont
  {Prutyanov}, \citenamefont {Puzzuoli}, \citenamefont {P{\'e}rez},
  \citenamefont {Quintiii}, \citenamefont {Raymond}, \citenamefont {Redondo},
  \citenamefont {Reuter}, \citenamefont {Rice}, \citenamefont {Rodr{\'\i}guez},
  \citenamefont {RohithKarur}, \citenamefont {Rossmannek}, \citenamefont {Ryu},
  \citenamefont {SAPV}, \citenamefont {SamFerracin}, \citenamefont {Sandberg},
  \citenamefont {Sargsyan}, \citenamefont {Sathaye}, \citenamefont {Schmitt},
  \citenamefont {Schnabel}, \citenamefont {Schoenfeld}, \citenamefont
  {Scholten}, \citenamefont {Schoute}, \citenamefont {Schwarm}, \citenamefont
  {Sertage}, \citenamefont {Setia}, \citenamefont {Shammah}, \citenamefont
  {Shi}, \citenamefont {Silva}, \citenamefont {Simonetto}, \citenamefont
  {Singstock}, \citenamefont {Siraichi}, \citenamefont {Sitdikov},
  \citenamefont {Sivarajah}, \citenamefont {Sletfjerding}, \citenamefont
  {Smolin}, \citenamefont {Soeken}, \citenamefont {Sokolov}, \citenamefont
  {SooluThomas}, \citenamefont {Steenken}, \citenamefont {Stypulkoski},
  \citenamefont {Suen}, \citenamefont {Sun}, \citenamefont {Sung},
  \citenamefont {Takahashi}, \citenamefont {Tavernelli}, \citenamefont
  {Taylor}, \citenamefont {Taylour}, \citenamefont {Thomas}, \citenamefont
  {Tillet}, \citenamefont {Tod}, \citenamefont {de~la Torre}, \citenamefont
  {Trabing}, \citenamefont {Treinish}, \citenamefont {TrishaPe}, \citenamefont
  {Turner}, \citenamefont {Vaknin}, \citenamefont {Valcarce}, \citenamefont
  {Varchon}, \citenamefont {Vazquez}, \citenamefont {Vogt-Lee}, \citenamefont
  {Vuillot}, \citenamefont {Weaver}, \citenamefont {Wieczorek}, \citenamefont
  {Wildstrom}, \citenamefont {Wille}, \citenamefont {Winston}, \citenamefont
  {Woehr}, \citenamefont {Woerner}, \citenamefont {Woo}, \citenamefont {Wood},
  \citenamefont {Wood}, \citenamefont {Wood}, \citenamefont {Wood},
  \citenamefont {Wootton}, \citenamefont {Yeralin}, \citenamefont {Young},
  \citenamefont {Yu}, \citenamefont {Zachow}, \citenamefont {Zdanski},
  \citenamefont {Zoufal}, \citenamefont {Zoufalc}, \citenamefont {a~matsuo},
  \citenamefont {adekusar drl}, \citenamefont {azulehner}, \citenamefont
  {bcamorrison}, \citenamefont {brandhsn}, \citenamefont {chlorophyll zz},
  \citenamefont {dan1pal}, \citenamefont {dime10}, \citenamefont {drholmie},
  \citenamefont {elfrocampeador}, \citenamefont {faisaldebouni}, \citenamefont
  {fanizzamarco}, \citenamefont {gadial}, \citenamefont {gruu}, \citenamefont
  {jliu45}, \citenamefont {kanejess}, \citenamefont {klinvill}, \citenamefont
  {kurarrr}, \citenamefont {lerongil}, \citenamefont {ma5x}, \citenamefont
  {merav aharoni}, \citenamefont {michelle4654}, \citenamefont {ordmoj},
  \citenamefont {sethmerkel}, \citenamefont {strickroman}, \citenamefont
  {sumitpuri}, \citenamefont {tigerjack}, \citenamefont {toural}, \citenamefont
  {vvilpas}, \citenamefont {welien}, \citenamefont {willhbang}, \citenamefont
  {yang.luh}, \citenamefont {yelojakit},\ and\ \citenamefont
  {yotamvakninibm}}]{QisKit}%
  \BibitemOpen
  \bibfield  {author} {\bibinfo {author} {\bibfnamefont {H.}~\bibnamefont
  {Abraham}}, \bibinfo {author} {\bibnamefont {AduOffei}}, \bibinfo {author}
  {\bibfnamefont {I.~Y.}\ \bibnamefont {Akhalwaya}}, \bibinfo {author}
  {\bibfnamefont {G.}~\bibnamefont {Aleksandrowicz}}, \bibinfo {author}
  {\bibfnamefont {T.}~\bibnamefont {Alexander}}, \bibinfo {author}
  {\bibfnamefont {G.}~\bibnamefont {Alexandrowics}}, \bibinfo {author}
  {\bibfnamefont {E.}~\bibnamefont {Arbel}}, \bibinfo {author} {\bibfnamefont
  {A.}~\bibnamefont {Asfaw}}, \bibinfo {author} {\bibfnamefont
  {C.}~\bibnamefont {Azaustre}}, \bibinfo {author} {\bibnamefont
  {AzizNgoueya}}, \bibinfo {author} {\bibfnamefont {P.}~\bibnamefont
  {Barkoutsos}}, \bibinfo {author} {\bibfnamefont {G.}~\bibnamefont {Barron}},
  \bibinfo {author} {\bibfnamefont {L.}~\bibnamefont {Bello}}, \bibinfo
  {author} {\bibfnamefont {Y.}~\bibnamefont {Ben-Haim}}, \bibinfo {author}
  {\bibfnamefont {D.}~\bibnamefont {Bevenius}}, \bibinfo {author}
  {\bibfnamefont {L.~S.}\ \bibnamefont {Bishop}}, \bibinfo {author}
  {\bibfnamefont {S.}~\bibnamefont {Bolos}}, \bibinfo {author} {\bibfnamefont
  {S.}~\bibnamefont {Bosch}}, \bibinfo {author} {\bibfnamefont
  {S.}~\bibnamefont {Bravyi}}, \bibinfo {author} {\bibfnamefont
  {D.}~\bibnamefont {Bucher}}, \bibinfo {author} {\bibfnamefont
  {A.}~\bibnamefont {Burov}}, \bibinfo {author} {\bibfnamefont
  {F.}~\bibnamefont {Cabrera}}, \bibinfo {author} {\bibfnamefont
  {P.}~\bibnamefont {Calpin}}, \bibinfo {author} {\bibfnamefont
  {L.}~\bibnamefont {Capelluto}}, \bibinfo {author} {\bibfnamefont
  {J.}~\bibnamefont {Carballo}}, \bibinfo {author} {\bibfnamefont
  {G.}~\bibnamefont {Carrascal}}, \bibinfo {author} {\bibfnamefont
  {A.}~\bibnamefont {Chen}}, \bibinfo {author} {\bibfnamefont {C.-F.}\
  \bibnamefont {Chen}}, \bibinfo {author} {\bibfnamefont {R.}~\bibnamefont
  {Chen}}, \bibinfo {author} {\bibfnamefont {J.~M.}\ \bibnamefont {Chow}},
  \bibinfo {author} {\bibfnamefont {C.}~\bibnamefont {Claus}}, \bibinfo
  {author} {\bibfnamefont {C.}~\bibnamefont {Clauss}}, \bibinfo {author}
  {\bibfnamefont {A.~J.}\ \bibnamefont {Cross}}, \bibinfo {author}
  {\bibfnamefont {A.~W.}\ \bibnamefont {Cross}}, \bibinfo {author}
  {\bibfnamefont {S.}~\bibnamefont {Cross}}, \bibinfo {author} {\bibfnamefont
  {J.}~\bibnamefont {Cruz-Benito}}, \bibinfo {author} {\bibfnamefont
  {C.}~\bibnamefont {Culver}}, \bibinfo {author} {\bibfnamefont {A.~D.}\
  \bibnamefont {C{\'o}rcoles-Gonzales}}, \bibinfo {author} {\bibfnamefont
  {S.}~\bibnamefont {Dague}}, \bibinfo {author} {\bibfnamefont {T.~E.}\
  \bibnamefont {Dandachi}}, \bibinfo {author} {\bibfnamefont {M.}~\bibnamefont
  {Dartiailh}}, \bibinfo {author} {\bibnamefont {DavideFrr}}, \bibinfo {author}
  {\bibfnamefont {A.~R.}\ \bibnamefont {Davila}}, \bibinfo {author}
  {\bibfnamefont {A.}~\bibnamefont {Dekusar}}, \bibinfo {author} {\bibfnamefont
  {D.}~\bibnamefont {Ding}}, \bibinfo {author} {\bibfnamefont {J.}~\bibnamefont
  {Doi}}, \bibinfo {author} {\bibfnamefont {E.}~\bibnamefont {Drechsler}},
  \bibinfo {author} {\bibnamefont {Drew}}, \bibinfo {author} {\bibfnamefont
  {E.}~\bibnamefont {Dumitrescu}}, \bibinfo {author} {\bibfnamefont
  {K.}~\bibnamefont {Dumon}}, \bibinfo {author} {\bibfnamefont
  {I.}~\bibnamefont {Duran}}, \bibinfo {author} {\bibfnamefont
  {K.}~\bibnamefont {EL-Safty}}, \bibinfo {author} {\bibfnamefont
  {E.}~\bibnamefont {Eastman}}, \bibinfo {author} {\bibfnamefont
  {P.}~\bibnamefont {Eendebak}}, \bibinfo {author} {\bibfnamefont
  {D.}~\bibnamefont {Egger}}, \bibinfo {author} {\bibfnamefont
  {M.}~\bibnamefont {Everitt}}, \bibinfo {author} {\bibfnamefont {P.~M.}\
  \bibnamefont {Fern{\'a}ndez}}, \bibinfo {author} {\bibfnamefont {A.~H.}\
  \bibnamefont {Ferrera}}, \bibinfo {author} {\bibfnamefont {A.}~\bibnamefont
  {Frisch}}, \bibinfo {author} {\bibfnamefont {A.}~\bibnamefont {Fuhrer}},
  \bibinfo {author} {\bibfnamefont {M.}~\bibnamefont {GEORGE}}, \bibinfo
  {author} {\bibfnamefont {J.}~\bibnamefont {Gacon}}, \bibinfo {author}
  {\bibnamefont {Gadi}}, \bibinfo {author} {\bibfnamefont {B.~G.}\ \bibnamefont
  {Gago}}, \bibinfo {author} {\bibfnamefont {C.}~\bibnamefont {Gambella}},
  \bibinfo {author} {\bibfnamefont {J.~M.}\ \bibnamefont {Gambetta}}, \bibinfo
  {author} {\bibfnamefont {A.}~\bibnamefont {Gammanpila}}, \bibinfo {author}
  {\bibfnamefont {L.}~\bibnamefont {Garcia}}, \bibinfo {author} {\bibfnamefont
  {S.}~\bibnamefont {Garion}}, \bibinfo {author} {\bibfnamefont
  {A.}~\bibnamefont {Gilliam}}, \bibinfo {author} {\bibfnamefont
  {J.}~\bibnamefont {Gomez-Mosquera}}, \bibinfo {author} {\bibfnamefont
  {S.}~\bibnamefont {de~la Puente~Gonz{\'a}lez}}, \bibinfo {author}
  {\bibfnamefont {J.}~\bibnamefont {Gorzinski}}, \bibinfo {author}
  {\bibfnamefont {I.}~\bibnamefont {Gould}}, \bibinfo {author} {\bibfnamefont
  {D.}~\bibnamefont {Greenberg}}, \bibinfo {author} {\bibfnamefont
  {D.}~\bibnamefont {Grinko}}, \bibinfo {author} {\bibfnamefont
  {W.}~\bibnamefont {Guan}}, \bibinfo {author} {\bibfnamefont {J.~A.}\
  \bibnamefont {Gunnels}}, \bibinfo {author} {\bibfnamefont {M.}~\bibnamefont
  {Haglund}}, \bibinfo {author} {\bibfnamefont {I.}~\bibnamefont {Haide}},
  \bibinfo {author} {\bibfnamefont {I.}~\bibnamefont {Hamamura}}, \bibinfo
  {author} {\bibfnamefont {V.}~\bibnamefont {Havlicek}}, \bibinfo {author}
  {\bibfnamefont {J.}~\bibnamefont {Hellmers}}, \bibinfo {author}
  {\bibfnamefont {{\L}.}~\bibnamefont {Herok}}, \bibinfo {author}
  {\bibfnamefont {S.}~\bibnamefont {Hillmich}}, \bibinfo {author}
  {\bibfnamefont {H.}~\bibnamefont {Horii}}, \bibinfo {author} {\bibfnamefont
  {C.}~\bibnamefont {Howington}}, \bibinfo {author} {\bibfnamefont
  {S.}~\bibnamefont {Hu}}, \bibinfo {author} {\bibfnamefont {W.}~\bibnamefont
  {Hu}}, \bibinfo {author} {\bibfnamefont {H.}~\bibnamefont {Imai}}, \bibinfo
  {author} {\bibfnamefont {T.}~\bibnamefont {Imamichi}}, \bibinfo {author}
  {\bibfnamefont {K.}~\bibnamefont {Ishizaki}}, \bibinfo {author}
  {\bibfnamefont {R.}~\bibnamefont {Iten}}, \bibinfo {author} {\bibfnamefont
  {T.}~\bibnamefont {Itoko}}, \bibinfo {author} {\bibnamefont {JamesSeaward}},
  \bibinfo {author} {\bibfnamefont {A.}~\bibnamefont {Javadi}}, \bibinfo
  {author} {\bibfnamefont {A.}~\bibnamefont {Javadi-Abhari}}, \bibinfo {author}
  {\bibnamefont {Jessica}}, \bibinfo {author} {\bibfnamefont {K.}~\bibnamefont
  {Johns}}, \bibinfo {author} {\bibfnamefont {T.}~\bibnamefont {Kachmann}},
  \bibinfo {author} {\bibfnamefont {N.}~\bibnamefont {Kanazawa}}, \bibinfo
  {author} {\bibnamefont {Kang-Bae}}, \bibinfo {author} {\bibfnamefont
  {A.}~\bibnamefont {Karazeev}}, \bibinfo {author} {\bibfnamefont
  {P.}~\bibnamefont {Kassebaum}}, \bibinfo {author} {\bibfnamefont
  {S.}~\bibnamefont {King}}, \bibinfo {author} {\bibnamefont {Knabberjoe}},
  \bibinfo {author} {\bibfnamefont {A.}~\bibnamefont {Kovyrshin}}, \bibinfo
  {author} {\bibfnamefont {R.}~\bibnamefont {Krishnakumar}}, \bibinfo {author}
  {\bibfnamefont {V.}~\bibnamefont {Krishnan}}, \bibinfo {author}
  {\bibfnamefont {K.}~\bibnamefont {Krsulich}}, \bibinfo {author}
  {\bibfnamefont {G.}~\bibnamefont {Kus}}, \bibinfo {author} {\bibfnamefont
  {R.}~\bibnamefont {LaRose}}, \bibinfo {author} {\bibfnamefont
  {R.}~\bibnamefont {Lambert}}, \bibinfo {author} {\bibfnamefont
  {J.}~\bibnamefont {Latone}}, \bibinfo {author} {\bibfnamefont
  {S.}~\bibnamefont {Lawrence}}, \bibinfo {author} {\bibfnamefont
  {D.}~\bibnamefont {Liu}}, \bibinfo {author} {\bibfnamefont {P.}~\bibnamefont
  {Liu}}, \bibinfo {author} {\bibfnamefont {Y.}~\bibnamefont {Maeng}}, \bibinfo
  {author} {\bibfnamefont {A.}~\bibnamefont {Malyshev}}, \bibinfo {author}
  {\bibfnamefont {J.}~\bibnamefont {Marecek}}, \bibinfo {author} {\bibfnamefont
  {M.}~\bibnamefont {Marques}}, \bibinfo {author} {\bibfnamefont
  {D.}~\bibnamefont {Mathews}}, \bibinfo {author} {\bibfnamefont
  {A.}~\bibnamefont {Matsuo}}, \bibinfo {author} {\bibfnamefont {D.~T.}\
  \bibnamefont {McClure}}, \bibinfo {author} {\bibfnamefont {C.}~\bibnamefont
  {McGarry}}, \bibinfo {author} {\bibfnamefont {D.}~\bibnamefont {McKay}},
  \bibinfo {author} {\bibfnamefont {D.}~\bibnamefont {McPherson}}, \bibinfo
  {author} {\bibfnamefont {S.}~\bibnamefont {Meesala}}, \bibinfo {author}
  {\bibfnamefont {M.}~\bibnamefont {Mevissen}}, \bibinfo {author}
  {\bibfnamefont {A.}~\bibnamefont {Mezzacapo}}, \bibinfo {author}
  {\bibfnamefont {R.}~\bibnamefont {Midha}}, \bibinfo {author} {\bibfnamefont
  {Z.}~\bibnamefont {Minev}}, \bibinfo {author} {\bibfnamefont
  {A.}~\bibnamefont {Mitchell}}, \bibinfo {author} {\bibfnamefont
  {N.}~\bibnamefont {Moll}}, \bibinfo {author} {\bibfnamefont {M.~D.}\
  \bibnamefont {Mooring}}, \bibinfo {author} {\bibfnamefont {R.}~\bibnamefont
  {Morales}}, \bibinfo {author} {\bibfnamefont {N.}~\bibnamefont {Moran}},
  \bibinfo {author} {\bibnamefont {MrF}}, \bibinfo {author} {\bibfnamefont
  {P.}~\bibnamefont {Murali}}, \bibinfo {author} {\bibfnamefont
  {J.}~\bibnamefont {M{\"u}ggenburg}}, \bibinfo {author} {\bibfnamefont
  {D.}~\bibnamefont {Nadlinger}}, \bibinfo {author} {\bibfnamefont
  {K.}~\bibnamefont {Nakanishi}}, \bibinfo {author} {\bibfnamefont
  {G.}~\bibnamefont {Nannicini}}, \bibinfo {author} {\bibfnamefont
  {P.}~\bibnamefont {Nation}}, \bibinfo {author} {\bibfnamefont
  {E.}~\bibnamefont {Navarro}}, \bibinfo {author} {\bibfnamefont
  {Y.}~\bibnamefont {Naveh}}, \bibinfo {author} {\bibfnamefont {S.~W.}\
  \bibnamefont {Neagle}}, \bibinfo {author} {\bibfnamefont {P.}~\bibnamefont
  {Neuweiler}}, \bibinfo {author} {\bibfnamefont {P.}~\bibnamefont {Niroula}},
  \bibinfo {author} {\bibfnamefont {H.}~\bibnamefont {Norlen}}, \bibinfo
  {author} {\bibfnamefont {L.~J.}\ \bibnamefont {O'Riordan}}, \bibinfo {author}
  {\bibfnamefont {O.}~\bibnamefont {Ogunbayo}}, \bibinfo {author}
  {\bibfnamefont {P.}~\bibnamefont {Ollitrault}}, \bibinfo {author}
  {\bibfnamefont {S.}~\bibnamefont {Oud}}, \bibinfo {author} {\bibfnamefont
  {D.}~\bibnamefont {Padilha}}, \bibinfo {author} {\bibfnamefont
  {H.}~\bibnamefont {Paik}}, \bibinfo {author} {\bibfnamefont {S.}~\bibnamefont
  {Perriello}}, \bibinfo {author} {\bibfnamefont {A.}~\bibnamefont {Phan}},
  \bibinfo {author} {\bibfnamefont {F.}~\bibnamefont {Piro}}, \bibinfo {author}
  {\bibfnamefont {M.}~\bibnamefont {Pistoia}}, \bibinfo {author} {\bibfnamefont
  {A.}~\bibnamefont {Pozas-iKerstjens}}, \bibinfo {author} {\bibfnamefont
  {V.}~\bibnamefont {Prutyanov}}, \bibinfo {author} {\bibfnamefont
  {D.}~\bibnamefont {Puzzuoli}}, \bibinfo {author} {\bibfnamefont
  {J.}~\bibnamefont {P{\'e}rez}}, \bibinfo {author} {\bibnamefont {Quintiii}},
  \bibinfo {author} {\bibfnamefont {R.}~\bibnamefont {Raymond}}, \bibinfo
  {author} {\bibfnamefont {R.~M.-C.}\ \bibnamefont {Redondo}}, \bibinfo
  {author} {\bibfnamefont {M.}~\bibnamefont {Reuter}}, \bibinfo {author}
  {\bibfnamefont {J.}~\bibnamefont {Rice}}, \bibinfo {author} {\bibfnamefont
  {D.~M.}\ \bibnamefont {Rodr{\'\i}guez}}, \bibinfo {author} {\bibnamefont
  {RohithKarur}}, \bibinfo {author} {\bibfnamefont {M.}~\bibnamefont
  {Rossmannek}}, \bibinfo {author} {\bibfnamefont {M.}~\bibnamefont {Ryu}},
  \bibinfo {author} {\bibfnamefont {T.}~\bibnamefont {SAPV}}, \bibinfo {author}
  {\bibnamefont {SamFerracin}}, \bibinfo {author} {\bibfnamefont
  {M.}~\bibnamefont {Sandberg}}, \bibinfo {author} {\bibfnamefont
  {H.}~\bibnamefont {Sargsyan}}, \bibinfo {author} {\bibfnamefont
  {N.}~\bibnamefont {Sathaye}}, \bibinfo {author} {\bibfnamefont
  {B.}~\bibnamefont {Schmitt}}, \bibinfo {author} {\bibfnamefont
  {C.}~\bibnamefont {Schnabel}}, \bibinfo {author} {\bibfnamefont
  {Z.}~\bibnamefont {Schoenfeld}}, \bibinfo {author} {\bibfnamefont {T.~L.}\
  \bibnamefont {Scholten}}, \bibinfo {author} {\bibfnamefont {E.}~\bibnamefont
  {Schoute}}, \bibinfo {author} {\bibfnamefont {J.}~\bibnamefont {Schwarm}},
  \bibinfo {author} {\bibfnamefont {I.~F.}\ \bibnamefont {Sertage}}, \bibinfo
  {author} {\bibfnamefont {K.}~\bibnamefont {Setia}}, \bibinfo {author}
  {\bibfnamefont {N.}~\bibnamefont {Shammah}}, \bibinfo {author} {\bibfnamefont
  {Y.}~\bibnamefont {Shi}}, \bibinfo {author} {\bibfnamefont {A.}~\bibnamefont
  {Silva}}, \bibinfo {author} {\bibfnamefont {A.}~\bibnamefont {Simonetto}},
  \bibinfo {author} {\bibfnamefont {N.}~\bibnamefont {Singstock}}, \bibinfo
  {author} {\bibfnamefont {Y.}~\bibnamefont {Siraichi}}, \bibinfo {author}
  {\bibfnamefont {I.}~\bibnamefont {Sitdikov}}, \bibinfo {author}
  {\bibfnamefont {S.}~\bibnamefont {Sivarajah}}, \bibinfo {author}
  {\bibfnamefont {M.~B.}\ \bibnamefont {Sletfjerding}}, \bibinfo {author}
  {\bibfnamefont {J.~A.}\ \bibnamefont {Smolin}}, \bibinfo {author}
  {\bibfnamefont {M.}~\bibnamefont {Soeken}}, \bibinfo {author} {\bibfnamefont
  {I.~O.}\ \bibnamefont {Sokolov}}, \bibinfo {author} {\bibnamefont
  {SooluThomas}}, \bibinfo {author} {\bibfnamefont {D.}~\bibnamefont
  {Steenken}}, \bibinfo {author} {\bibfnamefont {M.}~\bibnamefont
  {Stypulkoski}}, \bibinfo {author} {\bibfnamefont {J.}~\bibnamefont {Suen}},
  \bibinfo {author} {\bibfnamefont {S.}~\bibnamefont {Sun}}, \bibinfo {author}
  {\bibfnamefont {K.~J.}\ \bibnamefont {Sung}}, \bibinfo {author}
  {\bibfnamefont {H.}~\bibnamefont {Takahashi}}, \bibinfo {author}
  {\bibfnamefont {I.}~\bibnamefont {Tavernelli}}, \bibinfo {author}
  {\bibfnamefont {C.}~\bibnamefont {Taylor}}, \bibinfo {author} {\bibfnamefont
  {P.}~\bibnamefont {Taylour}}, \bibinfo {author} {\bibfnamefont
  {S.}~\bibnamefont {Thomas}}, \bibinfo {author} {\bibfnamefont
  {M.}~\bibnamefont {Tillet}}, \bibinfo {author} {\bibfnamefont
  {M.}~\bibnamefont {Tod}}, \bibinfo {author} {\bibfnamefont {E.}~\bibnamefont
  {de~la Torre}}, \bibinfo {author} {\bibfnamefont {K.}~\bibnamefont
  {Trabing}}, \bibinfo {author} {\bibfnamefont {M.}~\bibnamefont {Treinish}},
  \bibinfo {author} {\bibnamefont {TrishaPe}}, \bibinfo {author} {\bibfnamefont
  {W.}~\bibnamefont {Turner}}, \bibinfo {author} {\bibfnamefont
  {Y.}~\bibnamefont {Vaknin}}, \bibinfo {author} {\bibfnamefont {C.~R.}\
  \bibnamefont {Valcarce}}, \bibinfo {author} {\bibfnamefont {F.}~\bibnamefont
  {Varchon}}, \bibinfo {author} {\bibfnamefont {A.~C.}\ \bibnamefont
  {Vazquez}}, \bibinfo {author} {\bibfnamefont {D.}~\bibnamefont {Vogt-Lee}},
  \bibinfo {author} {\bibfnamefont {C.}~\bibnamefont {Vuillot}}, \bibinfo
  {author} {\bibfnamefont {J.}~\bibnamefont {Weaver}}, \bibinfo {author}
  {\bibfnamefont {R.}~\bibnamefont {Wieczorek}}, \bibinfo {author}
  {\bibfnamefont {J.~A.}\ \bibnamefont {Wildstrom}}, \bibinfo {author}
  {\bibfnamefont {R.}~\bibnamefont {Wille}}, \bibinfo {author} {\bibfnamefont
  {E.}~\bibnamefont {Winston}}, \bibinfo {author} {\bibfnamefont {J.~J.}\
  \bibnamefont {Woehr}}, \bibinfo {author} {\bibfnamefont {S.}~\bibnamefont
  {Woerner}}, \bibinfo {author} {\bibfnamefont {R.}~\bibnamefont {Woo}},
  \bibinfo {author} {\bibfnamefont {C.~J.}\ \bibnamefont {Wood}}, \bibinfo
  {author} {\bibfnamefont {R.}~\bibnamefont {Wood}}, \bibinfo {author}
  {\bibfnamefont {S.}~\bibnamefont {Wood}}, \bibinfo {author} {\bibfnamefont
  {S.}~\bibnamefont {Wood}}, \bibinfo {author} {\bibfnamefont {J.}~\bibnamefont
  {Wootton}}, \bibinfo {author} {\bibfnamefont {D.}~\bibnamefont {Yeralin}},
  \bibinfo {author} {\bibfnamefont {R.}~\bibnamefont {Young}}, \bibinfo
  {author} {\bibfnamefont {J.}~\bibnamefont {Yu}}, \bibinfo {author}
  {\bibfnamefont {C.}~\bibnamefont {Zachow}}, \bibinfo {author} {\bibfnamefont
  {L.}~\bibnamefont {Zdanski}}, \bibinfo {author} {\bibfnamefont
  {C.}~\bibnamefont {Zoufal}}, \bibinfo {author} {\bibnamefont {Zoufalc}},
  \bibinfo {author} {\bibnamefont {a~matsuo}}, \bibinfo {author} {\bibnamefont
  {adekusar drl}}, \bibinfo {author} {\bibnamefont {azulehner}}, \bibinfo
  {author} {\bibnamefont {bcamorrison}}, \bibinfo {author} {\bibnamefont
  {brandhsn}}, \bibinfo {author} {\bibnamefont {chlorophyll zz}}, \bibinfo
  {author} {\bibnamefont {dan1pal}}, \bibinfo {author} {\bibnamefont {dime10}},
  \bibinfo {author} {\bibnamefont {drholmie}}, \bibinfo {author} {\bibnamefont
  {elfrocampeador}}, \bibinfo {author} {\bibnamefont {faisaldebouni}}, \bibinfo
  {author} {\bibnamefont {fanizzamarco}}, \bibinfo {author} {\bibnamefont
  {gadial}}, \bibinfo {author} {\bibnamefont {gruu}}, \bibinfo {author}
  {\bibnamefont {jliu45}}, \bibinfo {author} {\bibnamefont {kanejess}},
  \bibinfo {author} {\bibnamefont {klinvill}}, \bibinfo {author} {\bibnamefont
  {kurarrr}}, \bibinfo {author} {\bibnamefont {lerongil}}, \bibinfo {author}
  {\bibnamefont {ma5x}}, \bibinfo {author} {\bibnamefont {merav aharoni}},
  \bibinfo {author} {\bibnamefont {michelle4654}}, \bibinfo {author}
  {\bibnamefont {ordmoj}}, \bibinfo {author} {\bibnamefont {sethmerkel}},
  \bibinfo {author} {\bibnamefont {strickroman}}, \bibinfo {author}
  {\bibnamefont {sumitpuri}}, \bibinfo {author} {\bibnamefont {tigerjack}},
  \bibinfo {author} {\bibnamefont {toural}}, \bibinfo {author} {\bibnamefont
  {vvilpas}}, \bibinfo {author} {\bibnamefont {welien}}, \bibinfo {author}
  {\bibnamefont {willhbang}}, \bibinfo {author} {\bibnamefont {yang.luh}},
  \bibinfo {author} {\bibnamefont {yelojakit}},\ and\ \bibinfo {author}
  {\bibnamefont {yotamvakninibm}},\ }\href
  {https://doi.org/10.5281/zenodo.2562110} {\bibinfo {title} {Qiskit: An
  open-source framework for quantum computing}} (\bibinfo {year}
  {2019})\BibitemShut {NoStop}%
\bibitem [{\citenamefont {Efron}(1979)}]{bootstrapI}%
  \BibitemOpen
  \bibfield  {author} {\bibinfo {author} {\bibfnamefont {B.}~\bibnamefont
  {Efron}},\ }\bibfield  {title} {\bibinfo {title} {Bootstrap methods: Another
  look at the jackknife},\ }\href {https://doi.org/10.1214/aos/1176344552}
  {\bibfield  {journal} {\bibinfo  {journal} {Ann. Statist.}\ }\textbf
  {\bibinfo {volume} {7}},\ \bibinfo {pages} {1} (\bibinfo {year}
  {1979})}\BibitemShut {NoStop}%
\bibitem [{\citenamefont {Kunsch}(1989)}]{bootstrapII}%
  \BibitemOpen
  \bibfield  {author} {\bibinfo {author} {\bibfnamefont {H.~R.}\ \bibnamefont
  {Kunsch}},\ }\bibfield  {title} {\bibinfo {title} {The jackknife and the
  bootstrap for general stationary observations},\ }\href
  {https://doi.org/10.1214/aos/1176347265} {\bibfield  {journal} {\bibinfo
  {journal} {Ann. Statist.}\ }\textbf {\bibinfo {volume} {17}},\ \bibinfo
  {pages} {1217} (\bibinfo {year} {1989})}\BibitemShut {NoStop}%
\bibitem [{\citenamefont {Politis}\ and\ \citenamefont
  {Romano}(1994)}]{bootstrapIII}%
  \BibitemOpen
  \bibfield  {author} {\bibinfo {author} {\bibfnamefont {D.~N.}\ \bibnamefont
  {Politis}}\ and\ \bibinfo {author} {\bibfnamefont {J.~P.}\ \bibnamefont
  {Romano}},\ }\bibfield  {title} {\bibinfo {title} {The stationary
  bootstrap},\ }\href {https://doi.org/10.1080/01621459.1994.10476870}
  {\bibfield  {journal} {\bibinfo  {journal} {Journal of the American
  Statistical Association}\ }\textbf {\bibinfo {volume} {89}},\ \bibinfo
  {pages} {1303} (\bibinfo {year} {1994})}\BibitemShut {NoStop}%
\bibitem [{\citenamefont {Cincio}\ \emph {et~al.}(2018)\citenamefont {Cincio},
  \citenamefont {Suba{\c{s}}{\i}}, \citenamefont {Sornborger},\ and\
  \citenamefont {Coles}}]{reyniXI}%
  \BibitemOpen
  \bibfield  {author} {\bibinfo {author} {\bibfnamefont {L.}~\bibnamefont
  {Cincio}}, \bibinfo {author} {\bibfnamefont {Y.}~\bibnamefont
  {Suba{\c{s}}{\i}}}, \bibinfo {author} {\bibfnamefont {A.~T.}\ \bibnamefont
  {Sornborger}},\ and\ \bibinfo {author} {\bibfnamefont {P.~J.}\ \bibnamefont
  {Coles}},\ }\bibfield  {title} {\bibinfo {title} {Learning the quantum
  algorithm for state overlap},\ }\href
  {https://doi.org/10.1088/1367-2630/aae94a} {\bibfield  {journal} {\bibinfo
  {journal} {New Journal of Physics}\ }\textbf {\bibinfo {volume} {20}},\
  \bibinfo {pages} {113022} (\bibinfo {year} {2018})}\BibitemShut {NoStop}%
\bibitem [{\citenamefont {Kern}\ \emph {et~al.}(2005)\citenamefont {Kern},
  \citenamefont {Alber},\ and\ \citenamefont {L.~Shepelyansky}}]{PAREC}%
  \BibitemOpen
  \bibfield  {author} {\bibinfo {author} {\bibfnamefont {O.}~\bibnamefont
  {Kern}}, \bibinfo {author} {\bibfnamefont {G.}~\bibnamefont {Alber}},\ and\
  \bibinfo {author} {\bibfnamefont {D.}~\bibnamefont {L.~Shepelyansky}},\
  }\bibfield  {title} {\bibinfo {title} {Quantum error correction of coherent
  errors by randomization},\ }\href
  {https://doi.org/10.1140/epjd/e2004-00196-9} {\bibfield  {journal} {\bibinfo
  {journal} {The European Physical Journal D}\ }\textbf {\bibinfo {volume}
  {32}},\ \bibinfo {pages} {153} (\bibinfo {year} {2005})}\BibitemShut
  {NoStop}%
\bibitem [{\citenamefont {Hastings}(2017)}]{randomization}%
  \BibitemOpen
  \bibfield  {author} {\bibinfo {author} {\bibfnamefont {M.~B.}\ \bibnamefont
  {Hastings}},\ }\bibfield  {title} {\bibinfo {title} {Turning gate synthesis
  errors into incoherent errors},\ }\href
  {http://dl.acm.org/citation.cfm?id=3179543.3179550} {\bibfield  {journal}
  {\bibinfo  {journal} {Quantum Info. Comput.}\ }\textbf {\bibinfo {volume}
  {17}},\ \bibinfo {pages} {488} (\bibinfo {year} {2017})}\BibitemShut
  {NoStop}%
\bibitem [{\citenamefont {Wallman}\ and\ \citenamefont
  {Emerson}(2016)}]{randcomp}%
  \BibitemOpen
  \bibfield  {author} {\bibinfo {author} {\bibfnamefont {J.~J.}\ \bibnamefont
  {Wallman}}\ and\ \bibinfo {author} {\bibfnamefont {J.}~\bibnamefont
  {Emerson}},\ }\bibfield  {title} {\bibinfo {title} {Noise tailoring for
  scalable quantum computation via randomized compiling},\ }\href
  {https://doi.org/10.1103/PhysRevA.94.052325} {\bibfield  {journal} {\bibinfo
  {journal} {Phys. Rev. A}\ }\textbf {\bibinfo {volume} {94}},\ \bibinfo
  {pages} {052325} (\bibinfo {year} {2016})}\BibitemShut {NoStop}%
\bibitem [{\citenamefont {Zhang}\ \emph {et~al.}(2019)\citenamefont {Zhang},
  \citenamefont {Yu}, \citenamefont {Zhu},\ and\ \citenamefont
  {Pei}}]{randcompII}%
  \BibitemOpen
  \bibfield  {author} {\bibinfo {author} {\bibfnamefont {L.}~\bibnamefont
  {Zhang}}, \bibinfo {author} {\bibfnamefont {Y.}~\bibnamefont {Yu}}, \bibinfo
  {author} {\bibfnamefont {C.}~\bibnamefont {Zhu}},\ and\ \bibinfo {author}
  {\bibfnamefont {C.}~\bibnamefont {Pei}},\ }\bibfield  {title} {\bibinfo
  {title} {Noise tailoring for quantum circuits via unitary 2t-design},\ }\href
  {https://doi.org/10.1038/s41598-018-38158-2} {\bibfield  {journal} {\bibinfo
  {journal} {Scientific Reports}\ }\textbf {\bibinfo {volume} {9}},\ \bibinfo
  {pages} {1790} (\bibinfo {year} {2019})}\BibitemShut {NoStop}%
\bibitem [{\citenamefont {Viola}\ and\ \citenamefont {Lloyd}(1998)}]{DDI}%
  \BibitemOpen
  \bibfield  {author} {\bibinfo {author} {\bibfnamefont {L.}~\bibnamefont
  {Viola}}\ and\ \bibinfo {author} {\bibfnamefont {S.}~\bibnamefont {Lloyd}},\
  }\bibfield  {title} {\bibinfo {title} {Dynamical suppression of decoherence
  in two-state quantum systems},\ }\href
  {https://doi.org/10.1103/PhysRevA.58.2733} {\bibfield  {journal} {\bibinfo
  {journal} {Phys. Rev. A}\ }\textbf {\bibinfo {volume} {58}},\ \bibinfo
  {pages} {2733} (\bibinfo {year} {1998})}\BibitemShut {NoStop}%
\end{thebibliography}%


\providecommand{\noopsort}[1]{}\providecommand{\singleletter}[1]{#1}%
%

\end{document}